\documentclass{emulateapj}
\usepackage{natbib}

\newcommand{\los}{{\em los~}}

\slugcomment{Submitted 07 April 2010, Received 19 April 2010,
Accepted 18 July 2010}

\shorttitle{The Sgr Stream as traced by Red Clump stars}
\shortauthors{Correnti et al.}

\begin{document}

\title{The Northern wraps of the Sagittarius Stream as traced by Red Clump stars: distances, intrinsic widths and stellar densities.}

\author{M. Correnti}
\affil{Universit\'a di Bologna, Dipartimento di Astronomia, 
via Ranzani 1, 40127, Bologna, Italy}
\email{matteo.correnti@studio.unibo.it}

\author{M. Bellazzini}
\affil{INAF - Osservatorio Astronomico di Bologna, via Ranzani 1, 40127,
Bologna, Italy}

\author{R.A. Ibata}
\affil{Observatoire Astronomique, Universit\'e de Strasbourg, CNRS, 11, 
rue de l'Universit\'e, F-67000 Strasbourg, France}

\author{F.R. Ferraro}
\affil{Universit\'a di Bologna, Dipartimento di Astronomia,
via Ranzani 1, 40127, Bologna, Italy}

\and

\author{A. Varghese}
\affil{Observatoire Astronomique, Universit\'e de Strasbourg, CNRS, 11, 
rue de l'Universit\'e, F-67000 Strasbourg, France}

\begin{abstract}
We trace the tidal Stream of the Sagittarius dwarf spheroidal galaxy (Sgr dSph)
using Red Clump stars from the catalog of the Sloan Digital Sky Survey - Data
Release 6, in the range $150\degr\la RA\la 220\degr$, corresponding to the range
of orbital azimuth $220\degr \la \Lambda\la 290\degr$. Substructures along the
line of sight are identified as significant peaks in the differential star count
profiles (SCP) of candidate Red Clump stars. A proper modeling of the SCPs allows
us to obtain: (a) $\le 10$\% accurate, purely differential distances with
respect to the main body of Sgr, (b) estimates of the FWHM along the line of
sight, and (c) estimates of the local density, for each detected substructure.
In the range $255\degr\la \Lambda\la 290\degr$ we cleanly and continuously trace
various coherent structures that can be ascribed to the Stream, in particular:
the well known northern portion of the leading arm, running from  $d\simeq
43$~kpc at $\Lambda\simeq 290\degr$ to $d\simeq 30$~kpc at $\Lambda\simeq
255\degr$, and a more nearby coherent series of detections lying at constant
distance $d\simeq 25$~kpc, that can be identified with a wrap of the trailing
arm. The latter structure, predicted by several models of the disruption of Sgr
dSph, was never traced before; comparison with existing models indicates that
the difference in distance between these portions of the leading and trailing
arms may provide a powerful tool to discriminate between theoretical models
assuming different shapes of the Galactic potential. A further, more distant
wrap in the same portion of the sky is detected only along a couple of  lines of
sight. For $\Lambda\la 255\degr$ the detected structures are more complex and
less easily interpreted. We are confident to be able to trace the continuation
of the leading arm down to $\Lambda\simeq 220\degr$ and $d\simeq 20$~kpc; the
trailing arm is seen up to $\Lambda\simeq 240\degr$ where it is replaced by more
distant structures. Possible detections of more nearby wraps and of the Virgo
Stellar Stream are also discussed. These measured properties provide a coherent
set of observational constraints for the next generation of theoretical models
of the disruption of Sgr.
\end{abstract}


\section{Introduction}
\label{intro}

Stellar tidal streams as well as other substructures in the Milky Way (MW) halo
are generally interpreted as the relics of the process of hierarchical formation
of the MW, as envisaged by the currently accepted cosmological model
\citep[$\Lambda$-Cold Dark Matter, $\Lambda$-CDM hereafter, see][and references
therein]{bull,madau}. With the advent of large modern surveys, like the 2 Micron
All Sky Survey \citep[2MASS,][]{2mass} and the Sloan Digital Sky Survey
\citep[SDSS,][and references therein]{dr6}, our ability to detect stellar
systems and/or structures in the halo and in the disk of the MW has  increased
dramatically and several large-scale likely relics of the build-up of the
Galactic halo have been identified
\citep{iba01,heidi,yan03,maj03,cma,FoS,juric}. Also smaller tidal streams have
been found around disrupting globular clusters \citep[see, for
example][]{connie,GJ} or lacking an evident progenitor \citep[hereafter
Bel06]{gd06,FoS}.  The most spectacular example of the process of tidal
disruption and accretion of a dwarf satellite into our Galaxy is the Sagittarius
dwarf spheroidal galaxy (Sgr dSph), originally discovered by \citet{iba94}.  The
main body of the Sgr galaxy is located at $\sim 26$ kpc \citep{mon_dist} from
the Sun, beyond the Galactic bulge (Galactic coordinates $l,b
=+5.6\degr,-14.0\degr$). The stellar content of the Sgr dSph is dominated by an
intermediate-age relatively metal-rich population, with distributions peaking at
age $\sim$ 6-8 Gyr and $[Fe/H]\sim -0.5$, \citep[see][hereafter B06a, and
references therein]{B06a} but there is also clear evidence for the presence of
an older ($>10$ Gyr) and more metal-poor population as well, including Blue
Horizontal Branch \citep[BHB,][]{iba97,sdgs1,mon_age} and RR Lyrae stars
\citep{mateovar,alcock,cser}. All the available spectroscopic analyses indicate
that the Metallicity Distribution (MD) of Sgr stars is characterized by a broad
peak in the range $-1.0\la [Fe/H]\la 0.0$, with a weak tail likely extending
beyond $[Fe/H]\la -2.0$  \citep[see B06a,][]{cser,mon_met,andy,italboni,sbordo,nuc08,lagadec}.

The body of Sgr dSph appears tidally disturbed \citep{iba95}, and, soon after
its discovery, it was realized that there was some tidal debris surrounding the
galaxy \citep{mateostream,fahlman,alard96,iba97,maj99}. Indeed, it has been
subsequently established that there are two huge tidal tails emanating from
the edges of the galaxy and approximately tracing its orbital path, as expected
from N-body simulations \citep{kat95,ibamodel}. These tails  form a coherent and
dynamically cold filamentary structure (hereafter Sgr Stream) that extends for
tens of kpc from the parent galaxy and has been probed with many different
tracers. \citet{yan00} used SDSS first-year commissioning data to identify an
overdensity of blue A-type stars in two stripes located at ({\it l}, {\it b}, D)=
($341\degr,+57\degr$,46 kpc) and ($157^{\circ},-58^{\circ}$,33 kpc), which were
subsequently found to match the prediction of the available theoretical
models of the Sgr Stream \citep{iba01}. Similarly, \citet{ive} noticed that
clumps of RR Lyrae stars in SDSS commissioning data lay along the Sgr  orbit.
The thorough study of the structure of the halo as traced by F stars from the
SDSS, within a strip of $\pm 1.26\degr$ around the celestial equator, by
\citet{heidi} provided the first examples of Color Magnitude Diagrams (CMD) of
the Stream population toward ({\it l}, {\it b})= ($350\degr,50\degr$) and  ({\it
l}, {\it b})= ($167\degr,-54\degr$).  Other detections toward specific
directions were provided by \cite{david01,david}, \cite{sgrclus2} and \cite{vivas}. The first
panoramic view of the Sgr Stream was presented by \citet{cambresy}  using late M
giants selected from the Second Incremental Data Release of 2MASS. Subsequently,
\citet[][hereafter M03]{maj03}, having at disposal the final All Sky Data
Release of 2MASS, used a larger sample of M giants to provide a clearer view of
the whole complex, tracing very neatly the trailing tidal tail  all over the
Southern Galactic hemisphere, as well as the part of the leading arm closer to
the main body of the galaxy, up to $RA\simeq 190\degr$. In a very recent
analysis \citet{yan09} showed that M and K giants can be successfully selected
also from the SDSS and SEGUE datasets \citep{segue} and used to trace the
Stream; one main advantage of using giants as tracers is that they can be
(relatively) easily followed-up spectroscopically, thus providing crucial
kinematical and chemical information \citep{swope,mon_grad,chou,yan09}. 

Bel06 exploited the SDSS data release 5 (DR5) to provide a picture of the
leading arm of the Sgr Stream in the vicinity of the North  Galactic Cap with
remarkable clarity, using tracers (A-F dwarfs) that are intrinsically more
numerous than M giants, for a given space density and/or surface brightness
\footnote{In a stellar population of given age and chemical composition the
number of stars per units of sampled (integrated) luminosity in a given
evolutionary phase is proportional to the duration of the evolutionary phase
\cite[see][and references therein]{rbuz,rfp88,alviopix}. A-F stars are evolving
along the Main Sequence, a phase lasting several Gyr for these stars, while M
giants are in the latest phases of their evolution along the Red Giant Branch,
lasting $\la 10^8$ yr. Hence in any given field, independently of the absolute
density normalization, A-F dwarfs outnumber M giants by a factor of $\ga 10$.}.
In their Fig.~1 they show the density of (candidate) A-F dwarf stars (selected
with a simple color cut, $g-r \leq 0.4$, corresponding to $\sim B-V  \leq 0.6$) in
the portion of the sky covered by the SDSS. The Sgr Stream emerges very clearly
as a broad (and bifurcated) band going from 
($\alpha$,$\delta$)$\simeq$($220\degr ,0\degr$) to
($\alpha$,$\delta$)$\sim$($125\degr ,25\degr$), where  it plunges into the
Galactic Disk. The color cut adopted by Bel06 is very successful in tracing the
Stream structure as it takes advantage of the fact that Sgr stars are younger
than typical halo stars, hence they have a bluer Turn Off (TO) color with
respect to the halo population \cite[see][for another application of the same
principle]{unavane}. The density map by Bel06 shows evidence for a clear
distance gradient along the Stream, from the nearest part crossing the Disk at
$\alpha \approx 120\degr$, to the most distant part at $\alpha \approx
210\degr$, toward the North Galactic Pole (NGP).  More recently, in a pilot
project limited to a sub-sample of the SDSS (the so called Stripe 82)
\citet{cole08} described a more refined approach to the study of the spatial
structure of the Stream, using the same tracers as \citet{heidi}. Very recent
detections from different data and/or using different models can be found also
in \citet{dejong}, \citet{prior} and \citet{keller}.

In \citet[][hereafter B06c]{B06c} we demonstrated that yet another kind of
tracer can be efficiently used to study the Sgr Stream, i.e. core-He-burning
stars lying in the well populated Red Clump (RC) of the CMD of Sagittarius dSph.
In particular, we showed that it is possible to detect the RC  associated with a
given sub-structure as a peak in the differential Star Count Profiles (SCP) of
sub-samples of stars selected in a relatively narrow color range including the
RC.  The spatially localized RC population can be disentangled from the
fore/background contaminating population of the MW by subtracting the underlying
SCP, that is, in general, quite smooth and smoothly varying with position in the
sky. In B06c we used this technique to compare the Horizontal Branch (HB)
morphology in the Stream and in the main body of Sgr, finding an age/metallicity
gradient along the Sgr remnant \citep[see also][]{mon_grad,chou}, while in
\citet{boo3} we obtained an independent detection of the recently discovered
stellar system Bo\"otes~III \citep{grillbo,carlin}, providing new insight on its
nature, structure and stellar populations. \citet{carrell} recently presented
the results of a spectroscopic survey targeting RC stars in the Sgr Stream,
selected as in B06c.

The most natural and direct application of this technique is the determination
of accurate distance estimates from the magnitude of detected RC peaks, as the
RC is well known and widely used as a standard candle since long time
\citep[see][and references therein]{pacz,stagar,gs,babu,mrc}.  For 
intermediate/old-age populations, the luminosity of the RC peak shows relatively
modest variations as a function of age and metallicity, in particular when
measured in the reddest optical passband  \citep[as Cousins'$I$, see][]{gs}.  
When used differentially, i.e. looking at the same (or very
similar) stellar population in different places, the variations in the intrinsic
luminosity of the RC due to age/metallicity effects should vanish. Given also
the intrinsic narrowness of the feature in Sgr (see below), the RC seems the
ideal tool to accurately trace the run of the distance along the orbital azimuth
of the Sgr Stream, from the main body of the galaxy all over the portion of the
Stream sampled by the SDSS. The other large survey covering all the extent of
the Stream, 2MASS, cannot be used in this way as the associated photometry is
not sufficiently deep to reach the RC level. 

In this paper we will use the RC SCP method outlined above to take accurate {\em
purely differential measures} of the distance of the Northern arms of the Sgr
Stream with respect to the main body of the galaxy. This will provide strong
constraints for the models of the disruption of Sgr within the Galactic (dark)
halo, and, in turn on the physical properties of the dark halo itself
\citep{iba01b,aminashape,kat05}. The basic idea is the following: (a) we measure
the position of the RC peak in $V$ and $I$, with independent color selections
using $B-V$ and $V-I$ colors, in the main body of Sgr (from B06a photometry),
(b) we select SDSS fields projected onto the Sgr Stream as traced by Bel06, (c)
we transform the SDSS photometry into $B,V,I$ magnitudes, (d) we detect the RC
peak(s) in $V$ and $I$ SCPs from the SDSS on-Stream fields (adopting the same
color selections as in the main body), and (e) we obtain two independent
measures of the magnitude differences of the RC peaks between the main body and
the considered portion of the Stream. These are fully equivalent to differences
in distance modulus, that is, differences in distance. The whole set of
differential distances can be translated into a set of {\em absolute} distances
by adopting the preferred value of the distance modulus for the main body 
\cite[see, for example][]{alard96,ls00,mon_dist,kunder}. The detection of the
same peaks in both $V$ and $I$ SCPs provides a useful sanity check on the
interpretation of the SCPs and on the derived differential distances. As an
additional observational constraint to models of the disruption of Sgr, we
provide also an estimate of the  characteristic width of the Stream section
crossed by our fields (see Sect.~\ref{sec.sigma}). 

The plan of the paper is the following. In Sect.~2 we present the field of the
main body that we used as template and the fields of the Stream used for/in the
analysis. In Sect.~3 we describe the method used to analyze the SCPs and derive
the informations from the peaks. In Sect.~4 we present all the SCPs obtained from
each field and we discuss some special cases. In Sect.~5 we compare our results
with previous works in literature, with particular emphasis on the different
degree of uncertainty related to the distance estimates. In Sect.~6 we compare
our distance estimates with models that reproduce the three-dimensional shape of
the Stream. Finally, we summarize and discuss our results in Sect.~7.

Some preliminary reports on earlier phases of this project were presented in
\citet{mat1} and \citet{mat2}. 

\begin{figure*}
\plottwo{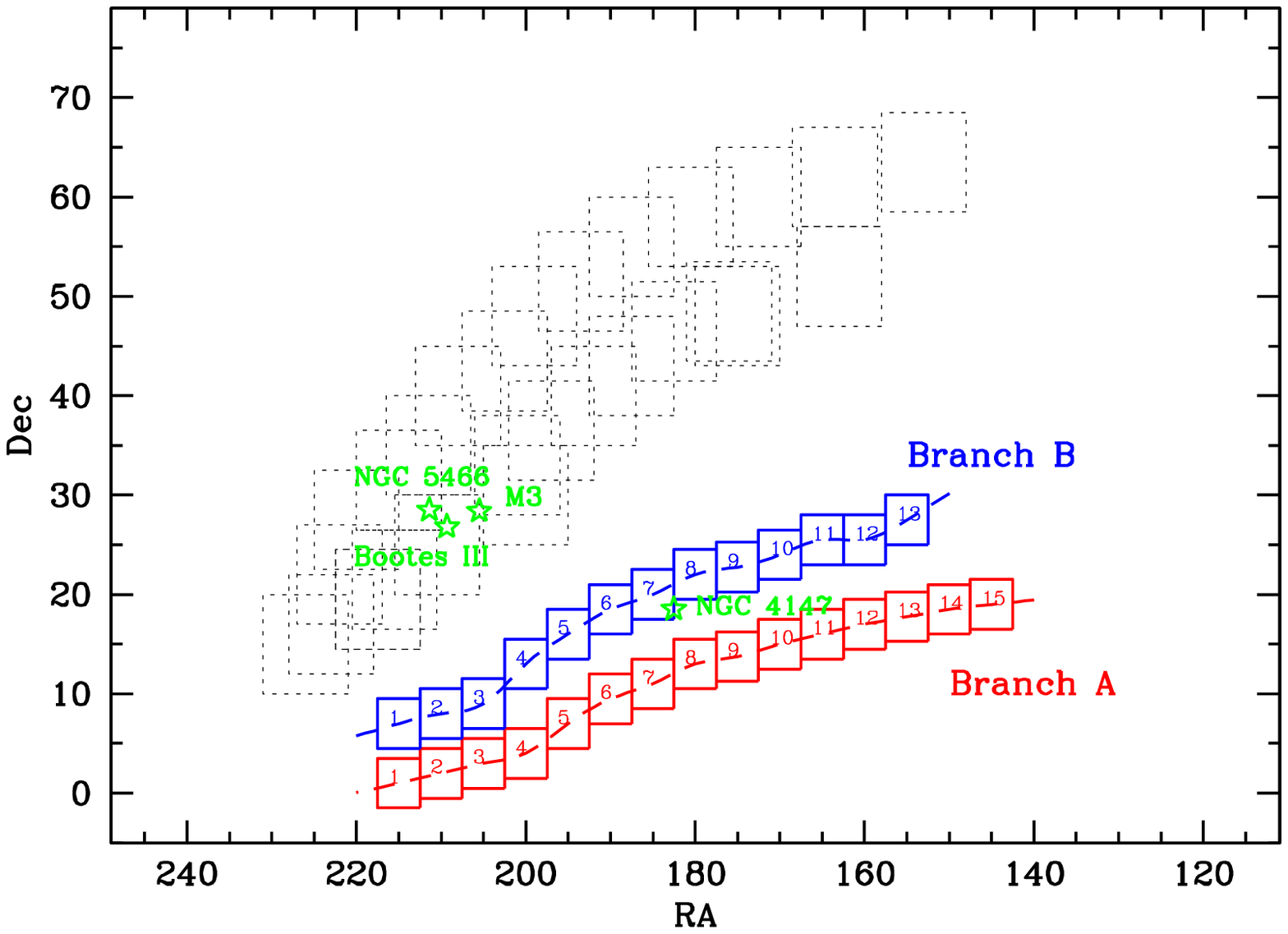}{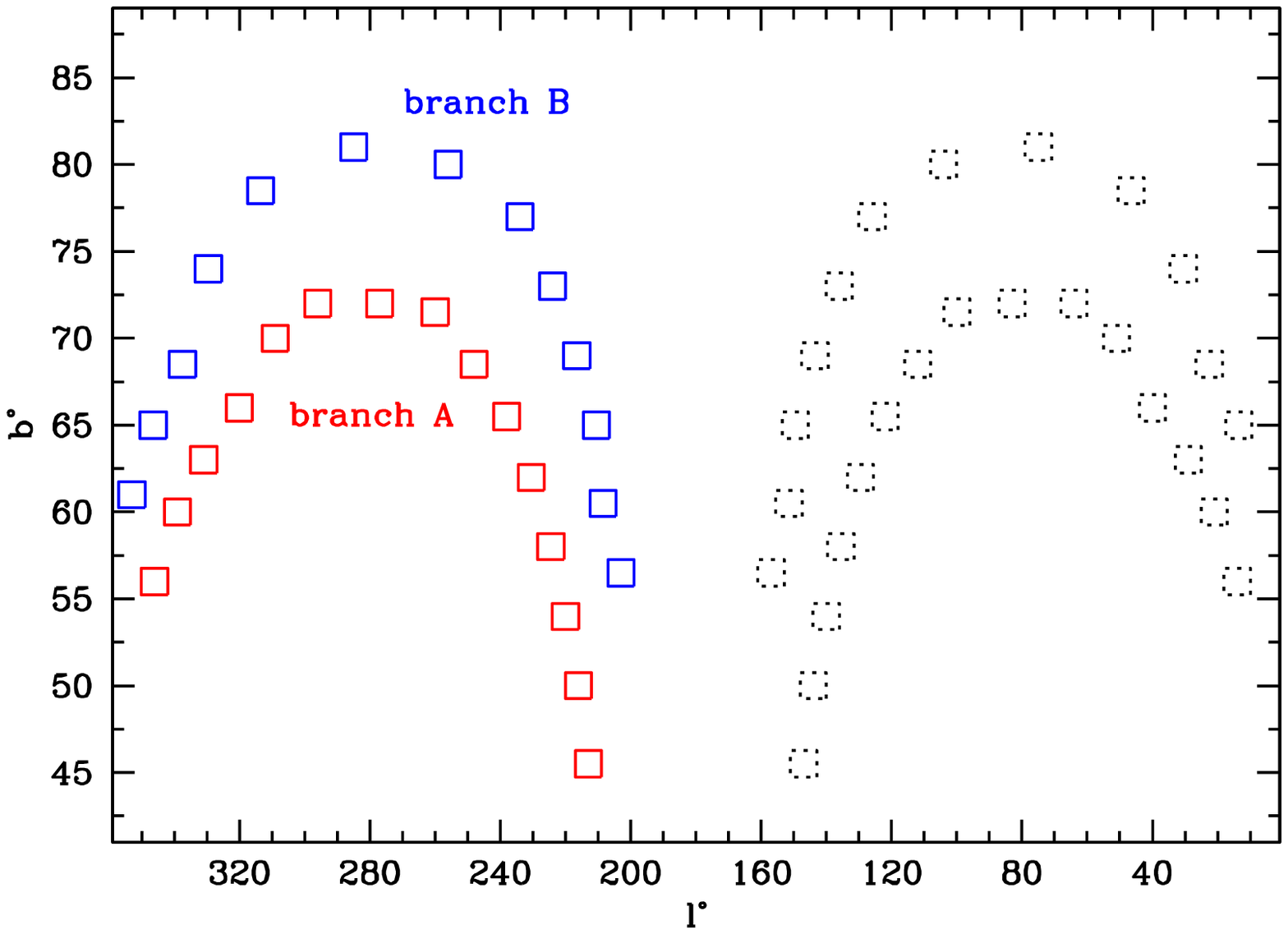}
\caption{Left panel: Distribution of the on-Stream fields of the branch A (red
squares) and the branch B (blue squares) of the Sgr Stream. The associated 
control fields are also plotted (black dashed squares). The plot is intended
to show the position of the fields and give a rough idea of their
dimension, projection effects are not taken into account. The shape of the two
branches is reproduced with a continuous line, following Fig.~1 of Bel06. The
positions of known stellar systems falling into the considered fields are 
indicated (green stars) and labeled. Right panel: 
Positions of the various fields in Galactic coordinates, to
highlight the symmetry (with respect to the Galactic Center and to the Galactic
Plane) of the adopted on-Stream fields and their corresponding Control Fields.}
\label{map}
\end{figure*}

\section{Data and Observables}
\label{data}

As a reference sample for the stellar population in the core of Sgr we take the
photometry of a $1^{\circ}\times 1^{\circ}$ wide field located  $\sim 2^{\circ}$
eastward of the galaxy center at ({\it l,b}) $\simeq$
($6.5^{\circ},-16.5^{\circ}$), presented in B06a and named {\it Sgr34}.  This
should be considered as fairly representative of the average population of the
Sgr galaxy \cite[see][]{sdgs1,sdgs2,giuf}, avoiding the youngest and most metal-rich
populations that appear to reside in the central nucleus \citep{siegel,nuc08}.
The strong similarity between the population of the Sgr main body and the Stream
has been shown by \citet{heidi} and Bel06, by direct comparison of
CMDs\footnote{In particular, Bel06 uses the same photometry of
{\it Sgr34} that is adopted here, as a reference.}. To sample the Galactic population at similar
angular distance from the Galactic Center  as for {\it Sgr34} we used the same control
field also presented in B06a: a $0.5^{\circ}\times 0.5^{\circ}$ field, named
{\it Gal\_Field}, at ({\it l,b}) $\simeq$  ($-6.0^{\circ},-14.5^{\circ}$), that
was used in B06a to perform the statistical decontamination of the {\it Sgr34}
CMD from the foreground/background Galactic stars.  Following B06a we adopted
the average reddening values $\langle E(B-V) \rangle = 0.116$ for {\it Sgr34}
and $\langle E(B-V) \rangle = 0.096$ for {\it Gal\_Field}, as derived from the
reddening maps of \citet[][hereafter SFD98]{ebv}.\\

\begin{table*}
\begin{center}
\caption{Position and reddening of the considered fields\tablenotemark{a}.}
\tiny
\begin{tabular}{ccccccc|cccccc}
\hline
field & $\alpha$ & $\delta$ & $l^{\circ}$ & $b^{\circ}$ & $\langle E(B-V) \rangle$
& $\sigma$ & $\alpha_{c}$ & $\delta_{c}$ & $l_{c}^{\circ}$ & 
$b_{c}^{\circ}$ & $\langle E(B-V)\rangle_{c}$ & $\sigma_{c}$ \\
\hline
  1A  & 215 & 1     & 346   & 56   & 0.038 & 0.007 & 226 & 15   & 14 & 56 
  & 0.034 & 0.009 \\
  2A  & 210 & 2     & 339   & 60   & 0.033 & 0.006 & 223 & 17   & 21 & 60 
  & 0.030 & 0.008 \\
  3A  & 205 & 3     & 331   & 63   & 0.026 & 0.003 & 222 & 22   & 29 & 63 
  & 0.034 & 0.008 \\
  4A  & 200 & 4     & 320   & 66   & 0.029 & 0.004 & 220 & 27.5 & 40 & 66 
  & 0.024 & 0.010 \\
  5A  & 195 & 7     & 309   & 70   & 0.031 & 0.005 & 215 & 31.5 & 51 & 70
  & 0.015 & 0.004 \\
  6A  & 190 & 9.5   & 296   & 72   & 0.024 & 0.006 & 211.5 & 35 & 64 & 72
  & 0.013 & 0.004 \\
  7A  & 185 & 11    & 277   & 72   & 0.027 & 0.007 & 208 & 40   & 83 & 72
  & 0.010 & 0.004 \\
  8A  & 180 & 13    & 260   & 71.5 & 0.030 & 0.006 & 202.5& 43.5& 100& 71.5
  & 0.014 & 0.006 \\
  9A  & 175 & 13.75 & 248   & 68.5 & 0.037 & 0.007 & 199 & 48   & 112& 68.5
  & 0.014 & 0.005 \\
  10A & 170 & 15    & 238   & 65.5 & 0.024 & 0.006 & 193.5& 51.5& 122& 65.5
  & 0.014 & 0.003 \\
  11A & 165 & 16    & 230.5 & 62   & 0.022 & 0.005 & 187.5& 55  &129.5 & 62
  & 0.015 & 0.004 \\
  12A & 160 & 17    & 224.5 & 58   & 0.030 & 0.006 & 180.5& 58  &135.5 & 58
  & 0.017 & 0.007 \\
  13A & 155 & 17.75 & 220   & 54   & 0.031 & 0.007 & 172.5& 60  & 140 & 54 
  & 0.014 & 0.007 \\
  14A & 150 & 18.5  & 216   & 50   & 0.030 & 0.004 & 163.5& 62  & 144 & 50 
  & 0.011 & 0.006 \\
  15A & 145 & 19    & 213   & 45.5 & 0.030 & 0.006 & 153 & 63.5 & 147 & 45.5
  & 0.017 & 0.016 \\
\hline
  1B  & 215 & 7     & 353   & 61   & 0.027 & 0.003 & 217.5 &19.5& 13.5& 65
  & 0.029 & 0.008 \\
  2B  & 210 & 8     & 346.5 & 65   & 0.026 & 0.003 & 217.5 &19.5& 13.5& 65
  & 0.029 & 0.008 \\
  3B  & 205 & 9     & 337.5 & 68.5 & 0.028 & 0.004 & 215.5 &21.5& 22.5& 68.5 
  & 0.028 & 0.008 \\
  4B  & 200 & 13    & 329.5 & 74   & 0.025 & 0.004 & 210.5 & 25 & 30.5& 74
  & 0.019 & 0.007 \\     
  5B  & 195 & 16    & 313.5 & 78.5 & 0.027 & 0.006 & 200  & 30  & 46.5& 78.5 
  & 0.012 & 0.002 \\
  6B  & 190 & 18.5  & 285   & 81   & 0.025 & 0.005 & 201  & 33  & 75 &  81
  & 0.012 & 0.002 \\
  7B  & 185 & 20    & 256   & 80   & 0.029 & 0.005 & 197  & 36.5& 104& 80 
  & 0.013 & 0.003 \\
  8B  & 180 & 22    & 234   & 77   & 0.028 & 0.006 & 192  & 40  & 126& 77 
  & 0.016 & 0.004 \\
  9B  & 175 & 22.75 & 224   & 73   & 0.023 & 0.004 & 187.5& 43  & 136& 73
  & 0.017 & 0.005 \\
 10B  & 170 & 24    & 216.5 & 69   & 0.017 & 0.002 & 182.5& 46.5& 143.5& 69 
  & 0.017 & 0.005 \\
 11B  & 165 & 25.5  & 210.5 & 65   & 0.022 & 0.008 & 176 & 48.5 & 149.5& 65 
  & 0.019 & 0.005 \\
 12B  & 160 & 25.5  & 208.5 & 60.5 & 0.025 & 0.007 & 175  & 48 & 151.5 & 65
  & 0.018 & 0.005 \\
 13B  & 155 & 27.5  & 203   & 56.5 & 0.028 & 0.007 & 163 & 52  & 157 & 56.5 
  & 0.013 & 0.004 \\  
\hline
\tablenotetext{a}{$\langle E(B-V) \rangle$ is the mean reddening of the field as extracted from \citet{ebv} maps and averaged over all the stars in the field; $\sigma$ is the corresponding standard deviation. The subscript $c$ refers to {\em control fields}.}
\label{coord}
\end{tabular}
\end{center}
\end{table*}

To study the Stream, we used the SDSS-DR6 photometry of objects classified as
stars \citep[extracted from the SDSS CasJobs query system,][]{dr6}\footnote{We
used a template SQL query provided in the Cas-SDSS web page called ``Clean Photometry'', in the version aimed at selecting stars. The corresponding SQL string can be found under the link
{\em Clean Photometry} at
{\tt http://cas.sdss.org/astrodr7/en/help/docs/realquery.asp}.}  for a series of
selected fields along the branch A and B, listed in Tab.~\ref{coord} and plotted
in Fig.~\ref{map}. We chose to follow the two branches separately, with
non-overlapping fields.
These on-Stream fields are similar, in position, to
those studied by Bel06, but are slightly smaller ($5^{\circ} \times 5^{\circ}$
instead of $6^{\circ} \times 6^{\circ}$), to avoid overlap between different
fields of the same branch. 
For each on-Stream field [located, for example, at (l,b)=($l_0,b_0$)] we
selected also a corresponding {\em control field}  (CF) located at the same latitude
and at the same angular distance from the  Galactic Center on the other side of
the Galaxy [i.e having (l,b)=($360^0-l_0, b_0$)]\footnote{Except in the case of
field F1B, for which the CF is the same adopted for the field F2B}. 
Assuming that the MW is symmetric about its center and its disk mid-plane
\cite[that should be a reasonable first-order approximation, at least at the Galactic latitudes considered here, $b\ge 45.5\degr$; but see][]{bell}, each control field should be fairly
representative of the typical Galactic population contaminating our on-Stream
fields. Following Bel06, to average out the effects of shot noise, the control
fields are larger than the on-Stream fields ($10^{\circ} \times  10^{\circ}$).
As shown in Fig.~\ref{map}, the globular clusters NGC5466 and M3, and the dwarf
galaxy remnant Bo\"otes~III are enclosed within some of our control fields
\citep{boo3}. To avoid any undesired contamination we excluded from the
corresponding samples the stars associated with these stellar systems by excising
areas of radius $1\degr$ (for the globulars) and $2\degr$ (for the dwarf galaxy)
around their centers. The only known stellar system that is (partially) enclosed
in one of our on-Stream fields is the globular cluster NGC4147 \cite[see Bel06
and][]{sgrclus1,sgrclus2}. Also in this case we excluded from the adopted sample
all the stars within $1\degr$ of the cluster center. In the following we will
use CFs only as a further observational check that the simple models we adopt to
account for the fore/background populations contaminating our SCPs are adequate
for our purposes (see Sect.~\ref{scp_cf}).

For our analysis we adopted the reddening-corrected $g,r,i,z$ magnitudes  as
provided by CasJobs. These magnitudes were also corrected using the 
SFD98 maps, hence the source of the reddening corrections is homogeneous for all
the datasets considered in the present analysis. The mean $E(B-V)$ and its
standard deviation for each field, averaged over all the stars included in the
field, are reported in Table~\ref{coord}. It is important to note that the
average reddening of our fields is remarkably low  ($0.010\le E(B-V)\le 0.038$)
and constant within each field  ($0.002\le \sigma_{E(B-V)}\le 0.010$), hence any
error in the adopted reddening correction would have only a minor impact on our
final {\em differential distance} estimates.  For brevity, in the following all
the reported magnitudes and colors are reddening-corrected
(i.e., for example, $V=V_0=$ extinction corrected $V$ magnitude).

The $g,r,i,z$ magnitudes in the SDSS system have been transformed to the
Johnson-Kron-Cousins $B,V,I$ system \cite[as defined by the standard stars
by][]{land92}  using robust empirical transformations that have been checked to
be particularly accurate in the color range typical of RC stars \citep[provided
by Lupton 2005\footnote{http://www.sdss.org/dr7/algorithms/sdssUBVRITransform.html},
derived from large samples of stars in common between SDSS and the extended
database of Landolt's standards by][]{stet}.  In particular, we obtain $B$ and
$V$ from $g$ and $r$, while $I$ is obtained from $i$ and $z$, adopting the
following equations:

\begin{eqnarray}
B = g + 0.3130(g-r) + 0.2271 ~~~~~\sigma=0.0107\\
V = g - 0.5784(g-r) - 0.0038 ~~~~~\sigma=0.0054\\
I = i - 0.3780(i-z) - 0.3974 ~~~~~~\sigma=0.0063
\end{eqnarray}

Note that the transformed $V$ and $I$ are {\em fully independent} as they are
obtained by independent couples of SDSS magnitudes. Consequently, measures of
the position of any significant peak detected in $V$ and $I$ SCPs will also be
independent, thus providing a powerful cross-check of any detection and
distance estimate.


\begin{figure*}
\plotone{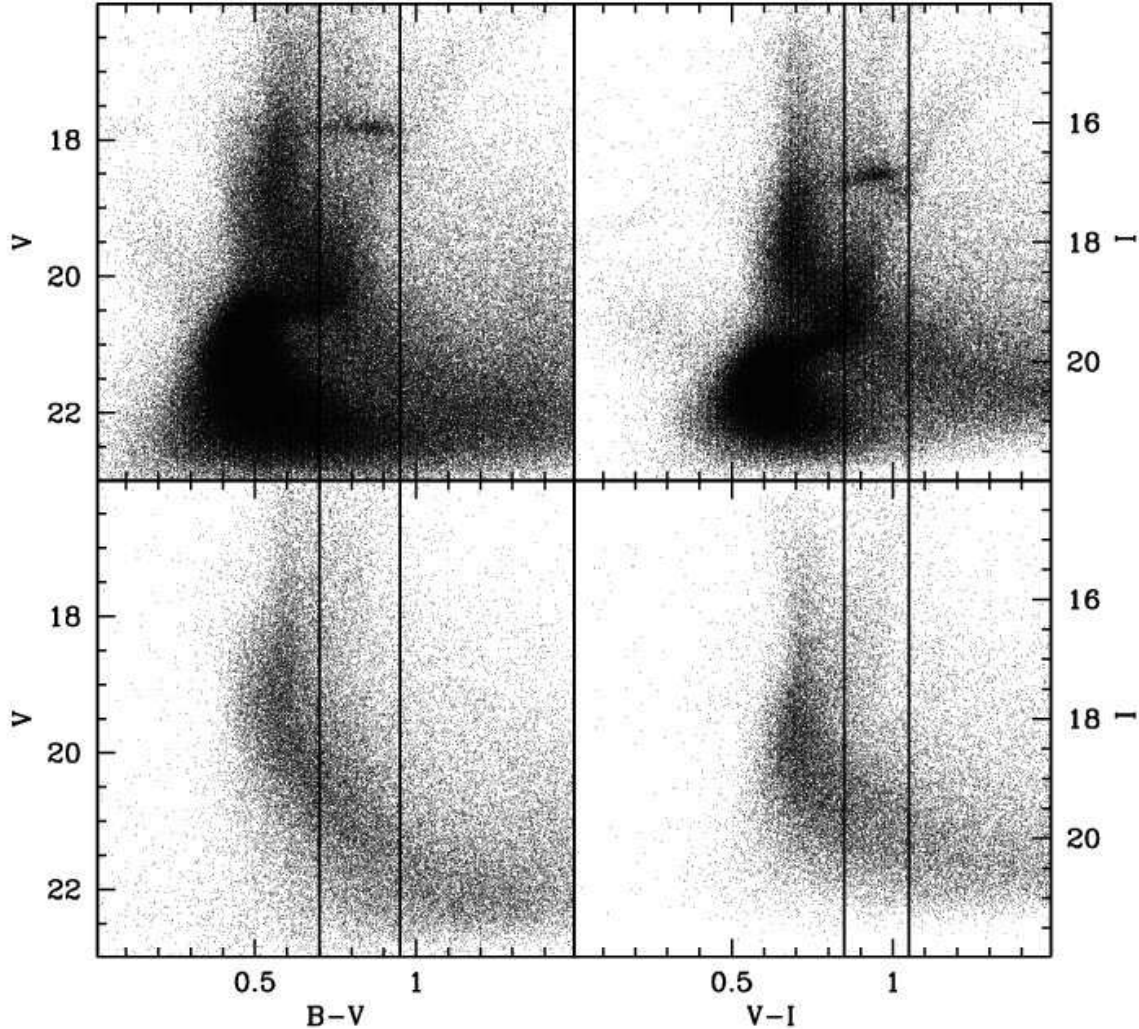}
\caption{Reddening corrected ({\it V,B-V}; left panels) and ({\it I,V-I};
right panels) CMDs, focused on the RC up to the upper part of the MS, of the
Sgr field  ({\it Sgr34}, upper panels) and of the control field ({\it
Gal\_Field}, lower  panels). The vertical lines enclose the RC populations and
are the color strips used to select the region where build the SCPs. The color
ranges are respectively, $0.70 \leq (B-V) \leq 0.95$ and $0.85 \leq (V-I)
\leq 1.05$.}
\label{cmd34}
\end{figure*}


\begin{figure}
\plotone{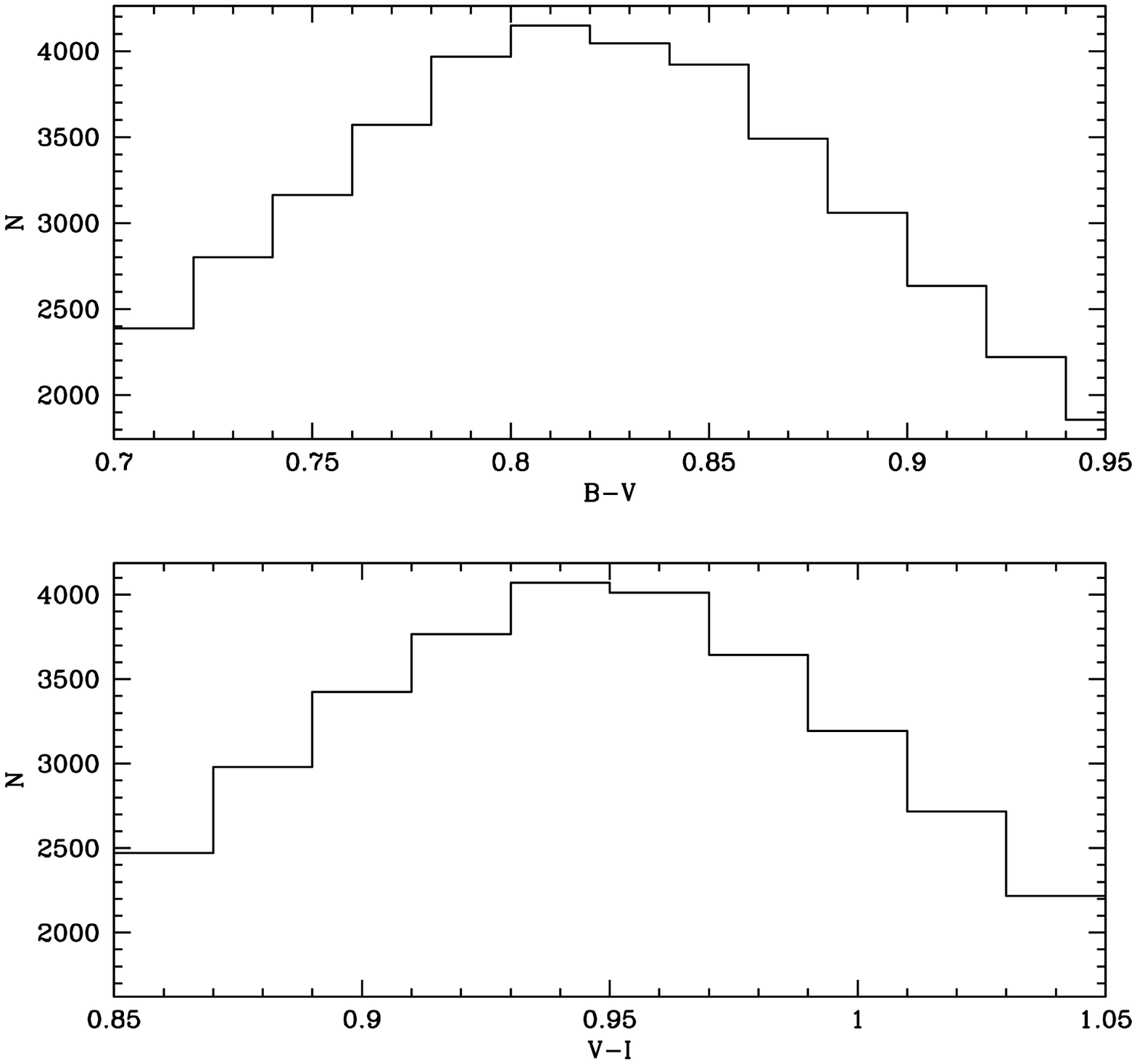}
\caption{Distribution of stars in {\it B-V} (upper panel) and {\it V-I} (lower
panel) color stripes for the {\it Sgr34} field. The selection must be optimized in
order to contain all the RC stars, but minimizing the presence of contaminating
stars. The RC is well confined inside the selected regions and peaks near the
middle in both; the magnitude limits of the selection are respectively, $17.3
\leq V \leq 18.3$ and $16.4 \leq I \leq 17.4$.}
\label{colorRC}
\end{figure}

\subsection{Selections on the Color Magnitude Diagram}
\label{SelCMD}

In Fig.~\ref{cmd34} we present reddening-corrected {\it V,B-V} and {\it I,V-I}
CMDs (focused on the RC features up to the upper region of the Main Sequence,
MS) of the main-body field {\it Sgr34} and of the corresponding control field 
{\it Gal\_Field}. The comparison between the CMDs of the two fields permits the
identification of the main features associated with Sgr and with the
fore/background Galactic populations. The RC of the Sagittarius dSph is a
prominent feature  in the CMDs of the {\it Sgr34} field (upper panels), around
$(I,V-I)\simeq(16.9, 0.9)$ and  $(V,B-V)\simeq(17.8, 0.8)$. The wide and inclined
Red Giant Branch (RGB) can be discerned over the background, going from
$(I,V-I)\simeq(16.9, 0.9)$ to $(I,V-I)\simeq(14.0, 1.5)$ [$(V,B-V)\simeq(19.5,
0.8)$ to $(V,B-V)\simeq(16.0, 1.4)$], and continuing beyond the limits of the
box. The RGB bump is apparent at $V\sim 18.2$ and $I\sim 17.2$, along the RGB
\cite[see][]{bump}. For $V-I\la 0.3$ ($B-V\la 0.2$) at $I\sim 19.0$ ($V\sim
17.9$) a portion of the Blue Horizontal Branch (BHB) is also visible
\citep{mon_age}; at  $V-I\la 0.5$ ($B-V\la 0.4$) and $I\ga 19.5$ ($V\ga 18.5$)
the Blue Plume \cite[BP,][B06a]{muskkk,sdgs1} population is visible. The
Sub Giant Branch (SGB, for $V-I\ga 0.8$ or $B-V\ga 0.7$) and the upper Main
Sequence (MS, to the blue of the above limits) appear for $I\ga 19.0$ ($V\ga
20.0$). For a more detailed description of the CMD of Sgr see B06a. The strong
vertical band around $V-I\sim 0.7$ ($B-V\sim 0.6$) running over the largest part
of the CMD, and bending to the red at $I\sim 19$, $V\sim 20$, is constituted by 
MS stars of the MW \cite[mostly from the Thick Disk, in this direction,
according to the Galactic  model by][]{besa}; the wide band running parallel, to
the red of the vertical portion of this feature is mainly populated by Galactic
giants, either in their RGB or RC/HB phase.  The majority of the stars redder
than $V-I=B-V\sim 1.0$ belongs to the vertical plume of local Galactic M dwarfs. 
 
The vertical lines in each panel of Fig.~\ref{cmd34} enclose the color stripes
that we adopted to select the RC population in the two colors, corresponding to
$0.70 \leq B-V \leq 0.95$ and $0.85 \le V-I \le 1.05$.  The choice of the color
limits was made in order to include the bulk of the RC population even if
small color shifts were present due to errors in the adopted reddening corrections and/or
population gradients, while keeping the contamination from other sources as low
as possible. The distribution in color within the selection windows (around the
magnitude of the observed RC of Sgr) shown in Fig.~\ref{colorRC} suggests that
color shifts of order $\pm 0.05$ mag would lead just to minor losses of the
signal (of the order of 10\% with respect to the number of stars obtained with our
choice in the selection window).

Fig.~\ref{cmd34} clearly shows that, in addition to Sgr RC stars, several
different contaminants are expected to enter the selection window. For $I\le
18.5$ ($V\le 19.5$) Galactic giants (mainly RC stars) should be the primary
source of contamination, while the sequence of Galactic MS stars crosses the
selection stripes at $I\ga 19$ and $V\ga 20$, boosting the star counts at faint
magnitudes. The RGB of the Sgr population, and in particular the RGB bump, are
also selected by the adopted windows. We will show below that this source of
contamination has a negligible effect on our SCPs. At $I\ga 19.0$ ($V\ga 20.0$)
the SGB stars of Sgr enter  the selection window; as they are much more numerous
than RGB and RC stars they may provide a serious contribution to the
``background'' in our SCPs, at faint magnitudes. Finally, the MS of
Sgr crosses the windows at $I\ga 21$ ($V\ga 22$). The actual structure of the
contamination entering the windows will obviously depend on (a) the Galactic
population encountered along the considered line of sight (\los, hereafter), and
(b) the distance of the wrap(s) of the Sgr Stream that is(are) crossed by the
considered \los. However the \los considered in this study are all at much
higher Galactic latitudes than {\it Sgr34}, hence the degree of contamination per unit
area of the sky should be lower, and the average distance of the encountered
stars should be higher, hence most of the contamination by Galactic dwarfs should
occur at fainter magnitudes than discussed above for {\it Sgr34}. Furthermore,
all the detections of the Stream presented here are at distances similar or
larger than the main body of the galaxy sampled by {\it Sgr34}; hence, in most cases
the contamination by the SGB of the Stream population will occur at fainter
magnitudes than in {\it Sgr34}\footnote{However it should be noted that the MS of
wraps of the Streams that are too nearby to have their RC detected with the
present technique may contribute to the contamination of our color-selected
samples of candidate RC stars. Moreover, other unknown substructures may
contribute to the contamination \cite[see][for example]{boo3}.}. In any case, to
limit the contribution by dwarf stars, independently of their origin, we limit our
analysis to the magnitude ranges $15.0\le I \le 19.5$ and $16.0\le V\le
20.5$\footnote{Except in the case of {\it Sgr34} where the limits are $15.0\le I \le
18.5$ and $16.0\le V\le 19.5$}. These limits approximately correspond to an
accessible range of heliocentric distances 12~kpc~$\la D\la$~70~kpc (see
Fig.~\ref{sint}, below).

While the surface brightness of Sgr at {\it Sgr34} is $\sim 25$ mag/arcsec$^2$,
typical values for the Stream are $\ga 30$ mag/arcsec$^2$ \cite[][and references
therein]{mario,sgrclus2,maj03}. It may be quite hard to identify the feeble
signal from such sparse populations even in the presence of low background. In fact,
even in the most favorable cases, the RC is barely visible in the CMDs of
on-Stream Fields \cite[see, e.g.][]{heidi}. The construction and modeling of
SCPs described below is very effective in extracting the
distance information in these cases \citep[B06c,][]{boo3}.

Finally, there are several indications that there is a sizable metallicity  (and
presumably age) gradient along the Stream, in the sense that the average
metallicity is lower in distant portion of the Stream with respect to the main
body of Sgr \citep[B06c,][]{mon_grad,chou}. This is generally interpreted as due
to a pre-existing population gradient within the original body of the Sgr
galaxy, as the tidal tails were preferentially populated by stars that resided
in the old and metal poor outskirts of Sgr \citep{chou}. It must be stressed
that the detected gradient means that the relative proportion of {\em
intermediate-age \& metal-rich stars} and of {\em old-age metal-poor stars}
changes along the Stream (and with respect to the main body). This, in turn,
changes the HB morphology, i.e. the relative abundance of RC and Blue HB stars
(as observed in B06c), but is it not expected to change the intrinsic luminosity
of the RC. Indeed, \citet{carrell} find that the mean metallicity of RC stars along the Stream is very similar to that found in the main body of Sgr.
Hence, while the population gradient may bias estimates of the
stellar density along the Stream obtained from RC stars, our distance estimates should be unaffected and our characteristic size estimates can be 
only marginally affected (see Sect.~\ref{sec.sigma} and
Fig.~\ref{viRC}, below, for further details and discussion).

\subsection{Detecting RC peaks in Star Count Profiles}  
\label{DetRC}

All the SCPs of color-selected RC samples presented in this paper are computed
as running histograms \cite[see][and references therein]{leo2}\footnote{Running
histograms are histograms in which the step is smaller than the bin width. The adoption of steps much smaller than the bin width removes the dependency from the starting position of the binning that affects classic histograms. Clearly, the values of adjacent bins are not statistically independent.}, as these couple the property of collecting the
signal from a wide bin with the ability of constraining with great accuracy the
location of density maxima, almost independently of the bin width. A bin width
of $\pm0.2$ mag\footnote{With rare exceptions in which bins of $\pm0.25$ mag
have been adopted to enhance the signal of a weak feature. All these cases are
clearly indicated in the following.} and step of 0.02 mag have been adopted
here. After different trials, they have been found to provide a good compromise
between the exigence of co-adding all the signal from a given RC population
(that requires larger bins) and the ability of distinguishing (resolving) nearby
peaks  (that is favored by smaller bins). The use of generalized histograms
\citep{laird} would have provided an higher degree of smoothing, possibly making
some of our SCPs easier to interpret. However we preferred running histograms as
they provide the reader a clearer idea of the local noise on the SCP as well as
a scale in real units ($\frac{stars}{mag~ bin\times FoV}$). The density scales
of the various fields have all been reported to unit standard area ($1\degr
\times 1\degr$) by applying the corrections due to spherical geometry that is
inherent to equatorial coordinates.

\begin{figure}
\plotone{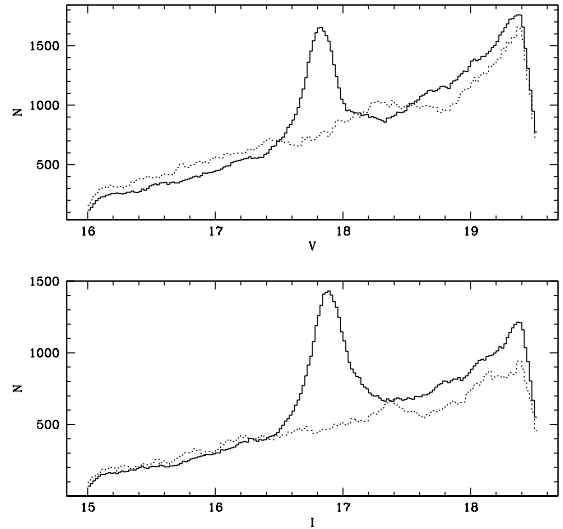}
\caption{De-reddened running histogram SCPs of color-selected RC candidates for
{\it Sgr34} (continuous line) and for the control field {\it Gal\_Field} (dashed
line) in $V$ and $I$ magnitude (respectively upper and bottom panel). The shape
of the SCPs is similar but, as expected, in the {\it Gal\_Field} the peak is
completely laking, while in the Sgr field it stands out very clearly.}
\label{lf34}
\end{figure}

To illustrate at best the case of the detection of the RC of a spatially
confined stellar system in a color-selected SCP, we show in Fig.~\ref{lf34} the
$V$ and $I$ SCPs for the {\it Sgr34} field (continuous lines), compared with those
obtained for the control field {\it Gal\_Field}, normalized by the ratio of
background densities  between the two fields\footnote{This ratio is dominated by
the ratio of the areas of the fields, {\it Sgr34} being $\simeq 4$ times larger
than {\it Gal\_Field}. However {\it Gal\_Field} sample a direction $\simeq
2\degr$ closer to the Galactic Plane and $\simeq 0.5\degr$ closer to the
Galactic Center than {\it Sgr34}, hence the (column) stellar density is
intrinsically larger in the former field. The actual density ratio, computed in
selected CMD boxes where the contribution from Sgr dSph stars is negligible, is 
$\simeq 3$, see \citet{sdgs2} and B06a.}.  

The shapes of the {\it Sgr34} and {\it Gal\_Field} SCPs are {\em remarkably similar}:  the
only exception is the very strong and well defined peak corresponding to the RC
of the Sgr galaxy seen in Fig.~\ref{cmd34}. It is interesting to note that
while  also other features related to Sgr are visible in the CMDs and (at least
partially) included in the selection windows, as for example the RGB bump, in
the SCPs the RC is the only signal emerging from the Sgr population. Independently of
the origin of the stellar mix actually selected, the SCP of the control field,
and, by analogy, the SCP of the contaminating population that is superposed on
the RC in the {\it Sgr34} SCP, are quite smooth and have a very simple behavior; in
B06c, \citet{boo3} and in Sect.~\ref{scp_cf} we show that this is the
general behavior of the SCP of the back/foreground population in the vast
majority of the considered \los, thus justifying the choice of a very simple
model for them, as described in  Sect.~\ref{modeling}, below.

\begin{figure*}
\plottwo{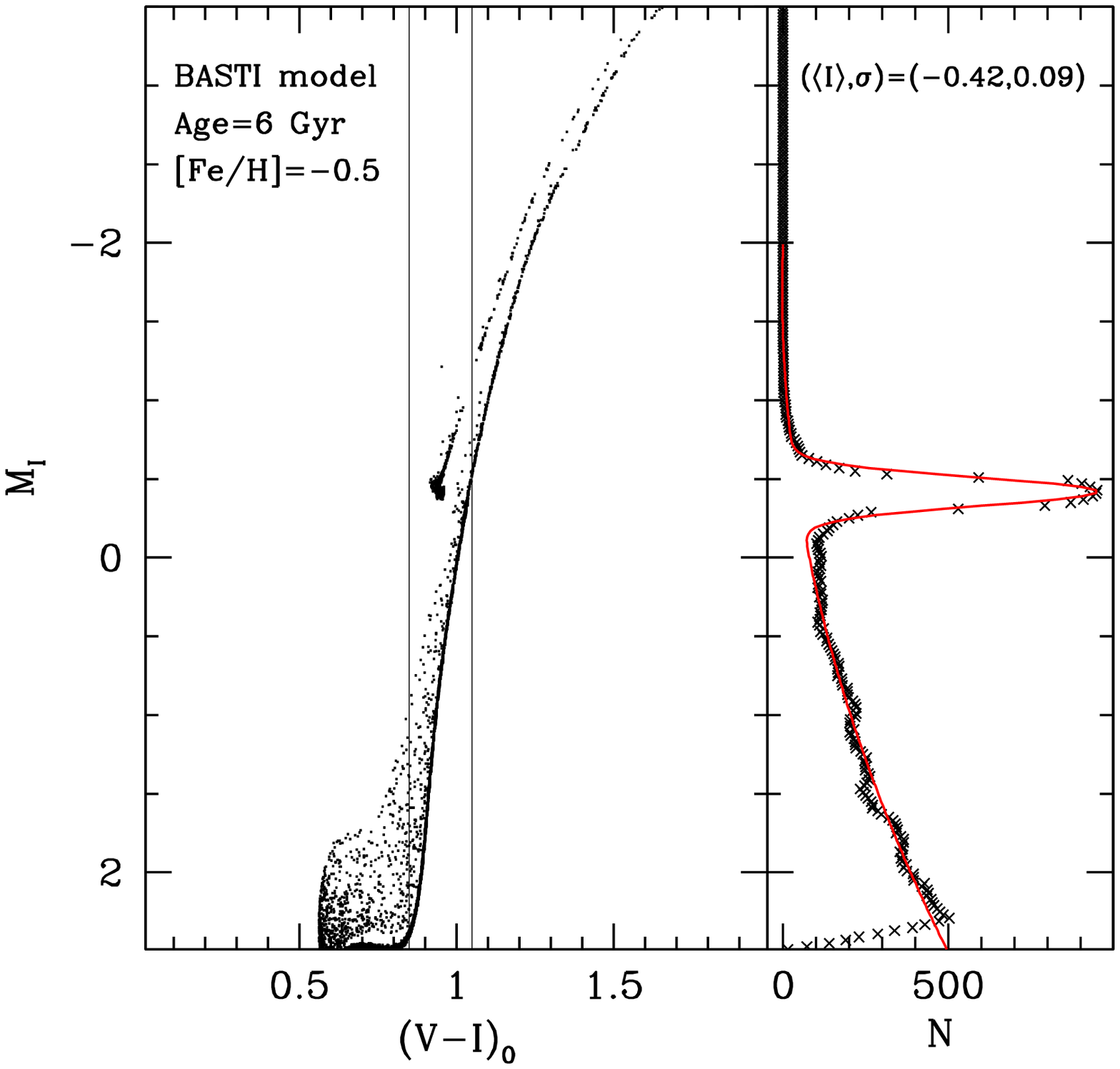}{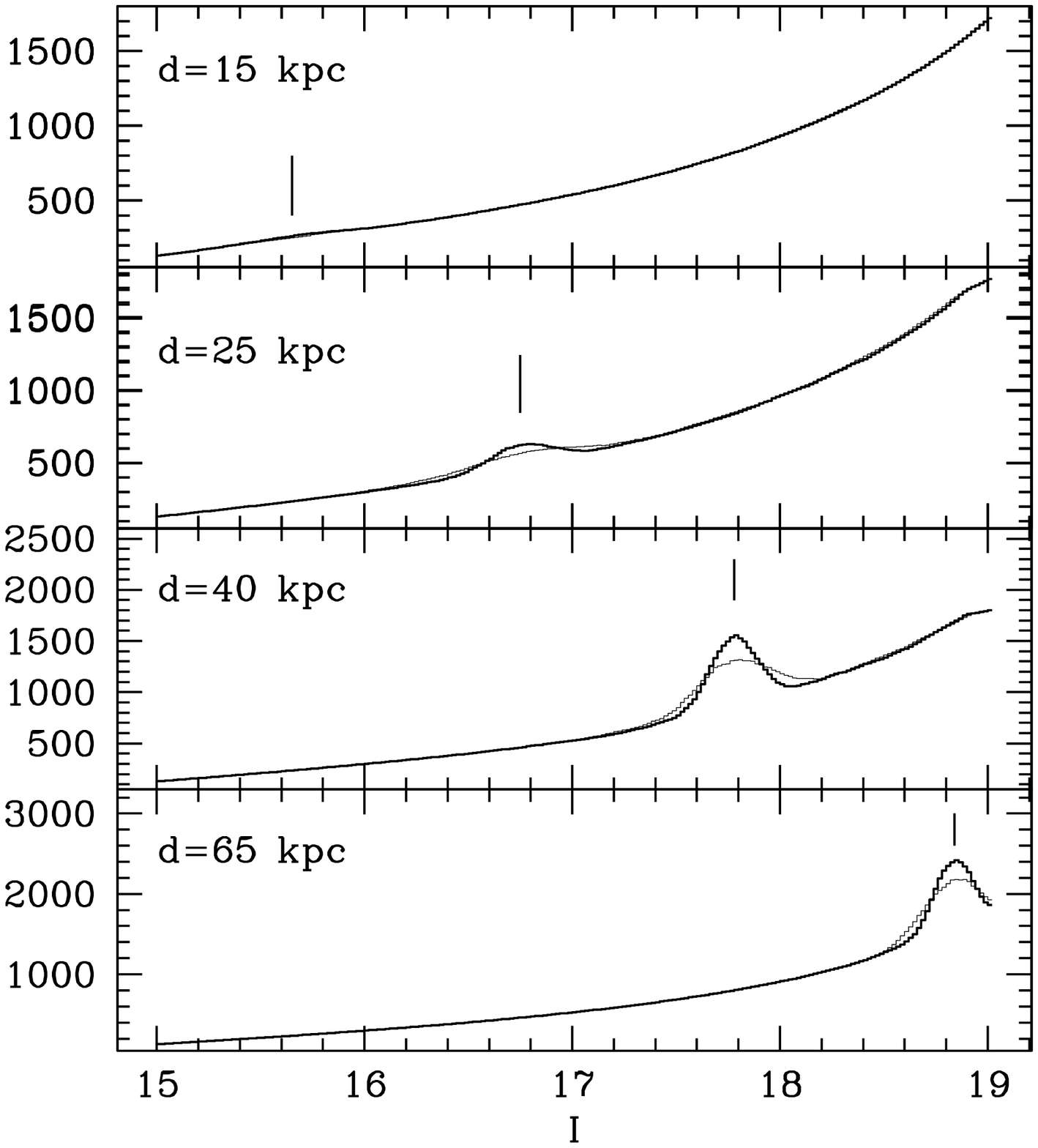}
\caption{{\em {(a).}} Left panel: CMD of a synthetic population composed of
45953 stars having age=6 Gyr,  $[Fe/H]=-0.5$, solar scaled abundance pattern,
10\% of binary systems, and Salpeter's Initial Mass Function, from the BASTI
database. The color window adopted to select RC candidates is enclosed within
the two thin parallel lines. Right panel: SCP of the color-selected RC
candidates obtained in the same way as Fig.~\ref{lf34}($\times$ symbols) and
fitted with the same kind of model (red continuous line). The mean and standard
deviation of the best-fitting Gaussian are reported within parentheses. {\em
{(b).}} The synthetic population shown in panel{\em (a)} has been distributed
along the line of sight according to gaussian distributions having various mean
distances (D=15, 25, 40 and 65 kpc, from the upper to the lower panel,
respectively) and Full Width at Half Maximum of 3.3 kpc (thick lines) or 6.6 kpc
(thin lines). The SCP of the resulting color-selected RC population has been
derived and superposed to the (model) SCP of the background of Fig.~\ref{lf34},
to simulate the detection of the same structure at different distances with the
method applied here. The vertical segments indicate the positions of the peaks
produced by the considered structures.}
\label{sint}
\end{figure*}

\subsection{Sensitivity of the technique}
\label{basti}

Before proceeding with the description of the method adopted to obtain the
actual differential distance estimates, it may be useful to study the
sensitivity of our SCPs to the various properties of any encountered substructure
(distance, density, etc.). To do that we used the dedicated web 
tool\footnote{\tt http://albione.oa-teramo.inaf.it} of the BASTI repository of
stellar models \citep{piet,basti} to produce a synthetic population of $\sim
45000$ stars having age and metallicity similar to the bulk of the Sgr
population (age=6 Gyr, $[Fe/H]=-0.5$). The CMD and the color-selected RC SCP of
the population are shown in the left panels of Fig.~\ref{sint}. The synthetic
stars have been distributed along the line of sight according to gaussian
distributions having various mean distances (D=15, 25, 40 and 65 kpc) and Full
Width at Half Maximum (FWHM) of 3.3 kpc or 6.6 kpc, to simulate the crossing of
a wrap of the Stream at various distances and with different characteristic
sizes along the \los.  A $FWHM\simeq 3.3$ kpc is quite typical of sections of
Stream wraps crossed perpendicularly by a given \los, as measured on the models
of the disruption of the Sgr galaxy by \citet{law}. The $FWHM\simeq 6.6$ kpc
has been considered to account for cases of sparser portions of the Stream
and/or non-perpendicular intersections with the \los. The SCPs of the resulting
color-selected RC population have been derived (properly including realistic
photometric errors\footnote{For each passband, we fitted the error curve derived from SDSS  photometric errors with exponential functions. Then we used the fitted functions to assign the proper average error to each synthetic star (according to its magnitude) and we added to each synthetic magnitude an error component extracted from a gaussian distribution having $\sigma$ equal to the average error.}) 
and added to the SCP of the background of Fig.~\ref{lf34}, as
modeled in Fig.~\ref{fitSgr}, below, to simulate the detection of the same
structure with the method applied here. The results of this simulation are shown
in the right panels of Fig.~\ref{sint}.

The most obvious effect shown in Fig.~\ref{sint} is the increase of sensitivity
with the distance of the structure. This is due to two factors: (1) the inherent
``compressive'' property of the magnitude scale, by which, for instance, a
difference in distance of 3.3 kpc corresponds to a difference in magnitudes of
0.43 mag at D=15 kpc and to just 0.14 mag at D=50 kpc, and (2) the relative
dimension of the considered structure and of the \los ~cone at the distance of
the structure; for nearby structures the fixed FoV adopted here may be smaller
than the structure itself, thus missing part of the signal that instead would be
included when more distant structures are encountered. This effect illustrates a
fundamental property of our method that should always be taken into account: a
structure that is very cleanly detected at, say, D=30 kpc may go completely
undetected if located  at D=15 kpc, instead. This implies that while significant
detections of RC peaks in our SCPs are robust and provide accurate distances, the
significance of non-detections must be evaluated with great care, on a case by
case basis, and, in general cannot be taken as a proof of the {\em absence} of a
given structure (that, for example, may be predicted by some model). One can
conceive various different techniques to mitigate this dependence of sensitivity
on distance, as, for example, to scale the bin width with magnitude to account
for the effect described at point (1), above, or to transform the magnitudes of
any color-selected RC candidate into distances, by assuming template values for
$M_I$ and $M_V$, and then to search for peaks in Distance Functions instead of
SCPs. However each of these possible solutions would have an impact on the
accuracy of the derived distance scale: for this reason we prefer to maintain an
approach that derives differential distances from the direct comparison of truly
{\em observable} quantities of strictly the same nature, i.e. the magnitudes of
peaks in SCPs that can be determined to within a few hundredths of mag, finally
providing differential distances with accuracies $<$10\%.   Other methods to
trace structures are intrinsically more powerful for other purposes (like the
{\em detection} of structures, for example; see Bel06). The technique adopted
here is best suited for distance measures and we decided to optimize it to this
task, at least in the present application.


\section{Modeling observed SCPs and measuring differential distances}\label{modeling}

\begin{figure*}
\plotone{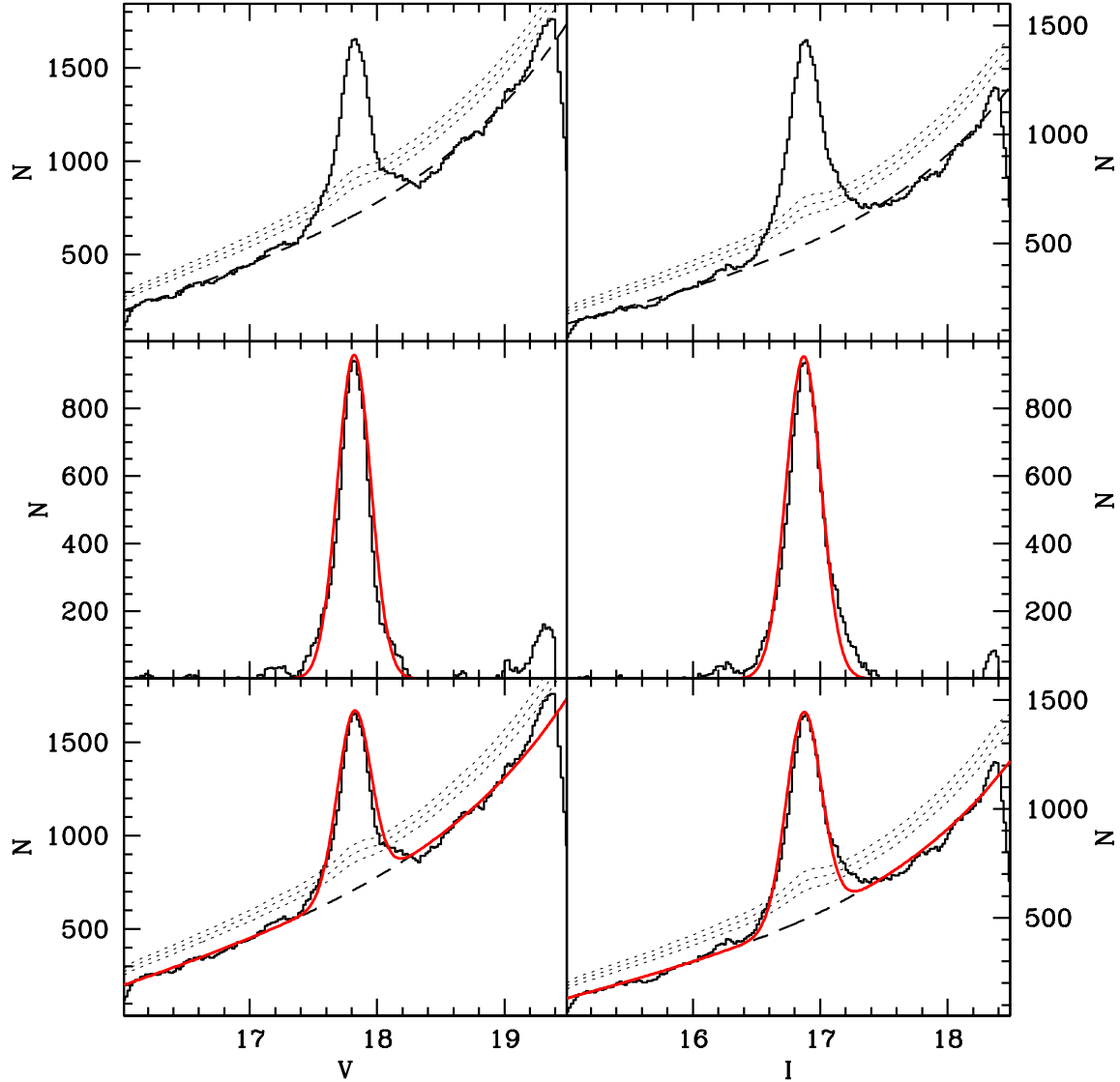}
\caption{Upper panels: de-reddened running histogram SCPs of color-selected RC 
candidates for the {\it Sgr34} field (continuous line) in $V$ and $I$ magnitude
(left and right panel, respectively); the dashed line represents the polynomial
fit of the background ($f(x)$). The dotted lines mark the 3,4 and 5$\sigma$
levels above the background, that is computed including both the Poisson noise
and the  uncertainty in the fit. Middle panels: Residuals of the subtraction
between the observed SCPs and the fit of the SCPs without the peaks; in red we
plotted the fit of the peaks, obtained with a gaussian ($G(x)$). Bottom panels:
same as the first ones, with added in red the total fit of the SCPs (polynomial
for the bkg + gaussian for the peaks, $f(x) + G(x)$).}
\label{fitSgr}
\end{figure*}

In the following we will describe the technique that we use to accurately
estimate the magnitude (and the statistical significance) of peaks in the SCP of
on-Stream fields.  To illustrate it we recur again to the case of the {\it
Sgr34} field (we will show below also two cases of on-Stream fields). The
approach is strictly the same for $V$ and $I$ SCPs and it is presented in parallel in
the left and right columns of panels of Fig.~\ref{fitSgr}:

\begin{enumerate}

\item Upper panels: a possible RC peak is detected in both SCPs. The underlying
smooth SCP (the back/foreground component) is fitted with a simple function of
the form $f(x) = Ae^{x} + B x + C$, once the points in the range enclosing the
peak are excluded \cite[see also B06a and][]{boo3}. This simple procedure is
very effective in all the cases considered here and it allows a reliable
interpolation of the background in the region of the peak. The dotted lines mark
the 3,4 and 5$\sigma$ levels above the background, that are computed including
both the Poisson noise and the  uncertainty in the fit of the background
component (that is conservatively assumed to be of the same order of the Poisson noise, i.e. $\sqrt N$, where N is the value of the model at a given position). 
To verify the validity of our assumption for running histograms (that have 
non-independent bins), we made several
trials (in different fields) using ``classical'' histograms, where Poisson statistics clearly apply. We found that the statistical significance lines are at the same level in both classical and running histograms, confirming that our simple assumptions lead to correct results \cite[see as another example of the same approach][]{boo3}.

\item Middle panels: the model of the background ($f(x)$) is subtracted from the
observed SCP. The only significant residual is the RC peak that, in general, has
a rather symmetric bell shape. The peak is fitted with a gaussian curve
($G(x)$)  by searching for the three parameters of $G(x)$ (mean, $\sigma$ and
normalization factor) that minimize the reduced $\chi^2$. The derived mean is
taken as the best estimate for the position of the considered peak.  As the bins
of the adopted running histogram are not independent, the resulting $\chi^2$
values can be considered only in a relative sense. After several trials on real
cases we found that increases of $\chi^2$ by a factor of 2 with respect to the
best solution (having $\chi^2=\chi^2_{min}$) always correspond to clearly
unsatisfactory fits (see Fig.~\ref{errfit}, for an example). For this reason we
adopt the difference between the best-fit mean and the mean of the solutions
having $\chi^2=2\chi^2_{min}$ as a robust  estimate of the accuracy of our
measures. As a sanity check we tested the assumed models against the
corresponding ordinary histogram (i.e. with independent bins) for nine flag=1
peaks, computing the actual $1\sigma$ uncertainty. It turned out that our
empirically defined error is always within a factor from 0.5 to 2 times the
statistically correct $1\sigma$ error, thus providing a realistic estimate of
the uncertainty of our measures.

\item Lower panels: the global model, obtained by summing $f(x)$ and $G(x)$, is
compared to the observed SCPs. This final form of the overall fit is what we
will show for all the considered fields in Sect.~\ref{detections}, below. In
Sect.~\ref{scp_cf} we will show that the adopted model of the fore/background
component of the considered SCPs ($f(x)$) provides an adequate representation of
what is observed in actual CFs and predicted by current Galactic models.

\end{enumerate} 

The application to the {\it Sgr34} field just described, provides also the Zero
Points of our differential distance scale, i.e. the magnitude of the RC in the
main body of Sgr, $V = 17.82 \pm 0.02$ and $I = 16.87 \pm 0.02$.  As a sanity
check, we verify if these numbers are compatible with theoretical stellar model
predictions\footnote{That, however, are quite uncertain and model dependent, in
an absolute sense, for stars in this evolutionary phase. For instance the
absolute $I$ magnitude of the peak for a age=6 Gyr, $[Fe/H]=-0.5$ model from the
BASTI dataset (shown in Fig.~\ref{sint}), is matched by a model of age=1.7 Gyr
and $[Fe/H]=-0.4$, from the set by \citet{gs}.}. Adopting the distance modulus
$(m-M)_0=17.10 \pm 0.15$ for Sgr \citep{mon_dist} we obtain $M_I=-0.23\pm 0.15$
and $M_V=+0.72\pm 0.15$. These correspond to ages in the range 5-7 Gyr for
$[Fe/H]=-0.4$ and 9-11 Gyr for $[Fe/H]=-0.7$ in the models by \citet{gs}, in
good agreement with all recent estimates of the typical age of the bulk of the
Sgr stars \cite[see][B06a]{ls00,bump}.

\begin{figure}
\plotone{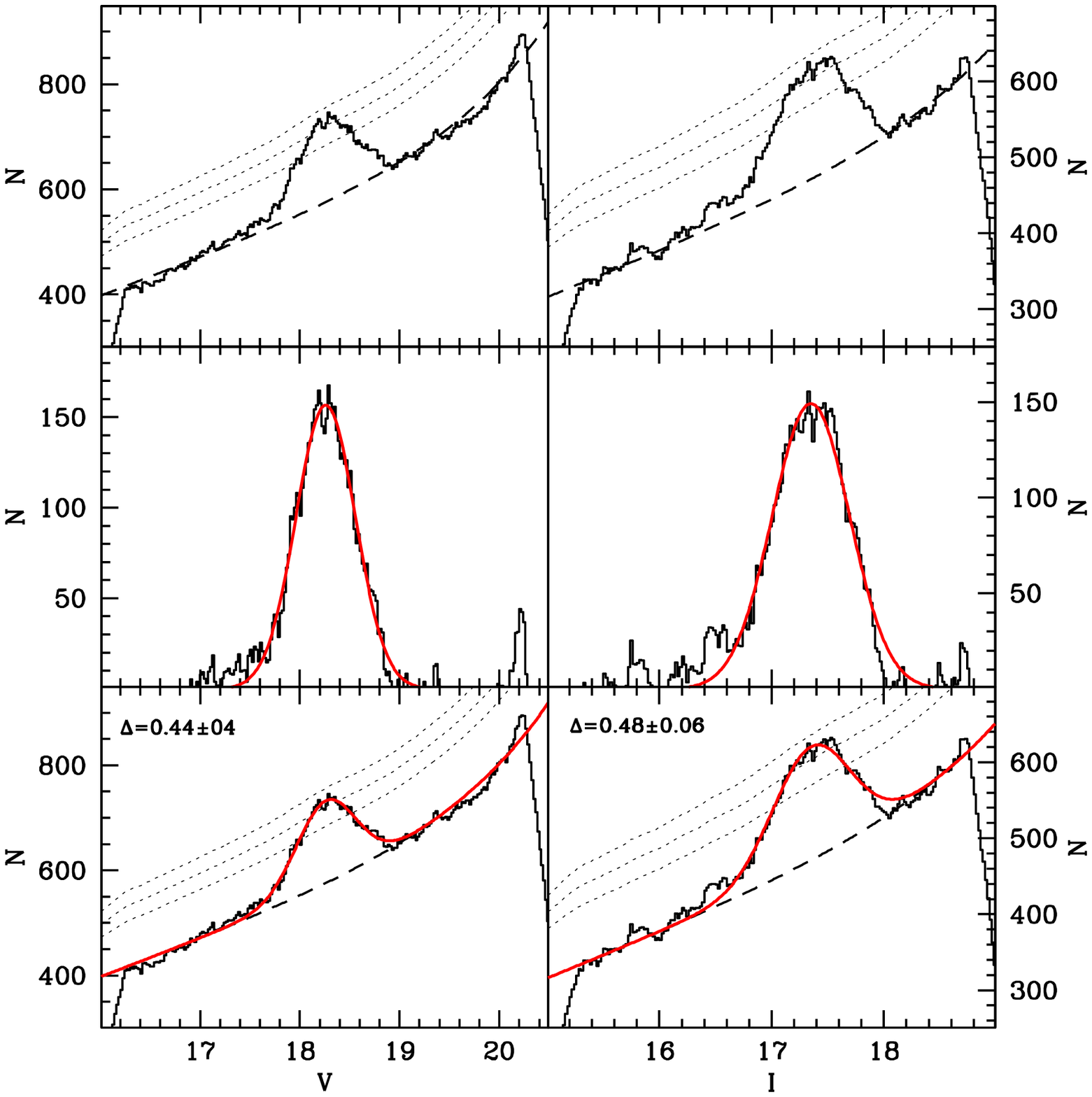}
\caption{Upper panels: de-reddened running histogram SCPs of color-selected RC 
candidates for a Stream field (field F7A, continuous line) in $V$ and $I$
magnitude (left and right panel, respectively); the dashed line represents the
polynomial fit of the background ($f(x)$). The dotted lines mark the 3,4 and
5$\sigma$ levels above the background, that is computed including both the
Poisson noise and the  uncertainty in the fit. Middle panels: Residuals of the
subtraction between the observed SCPs and the fit of the SCPs without the peaks;
in red we plotted the fit of the peaks, obtained with a gaussian ($G(x)$).
Bottom panels: same as the first ones, with added in red the total fit of the
SCPs (polynomial for the bkg + gaussian for the peaks, $f(x) + G(x)$).}
\label{fit7A}
\end{figure}

\begin{figure}
\plotone{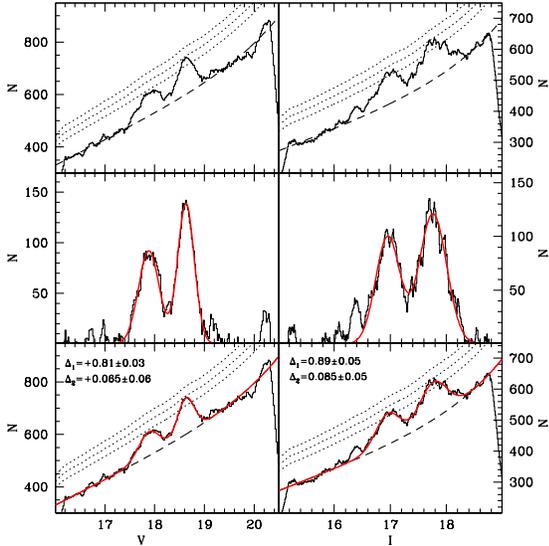}
\caption{As Fig.~\ref{fit7A}, but for a different field (field F5A). The
method is the same although this field shows more than one peak.}
\label{fit5A}
\end{figure}

\begin{figure}
\plotone{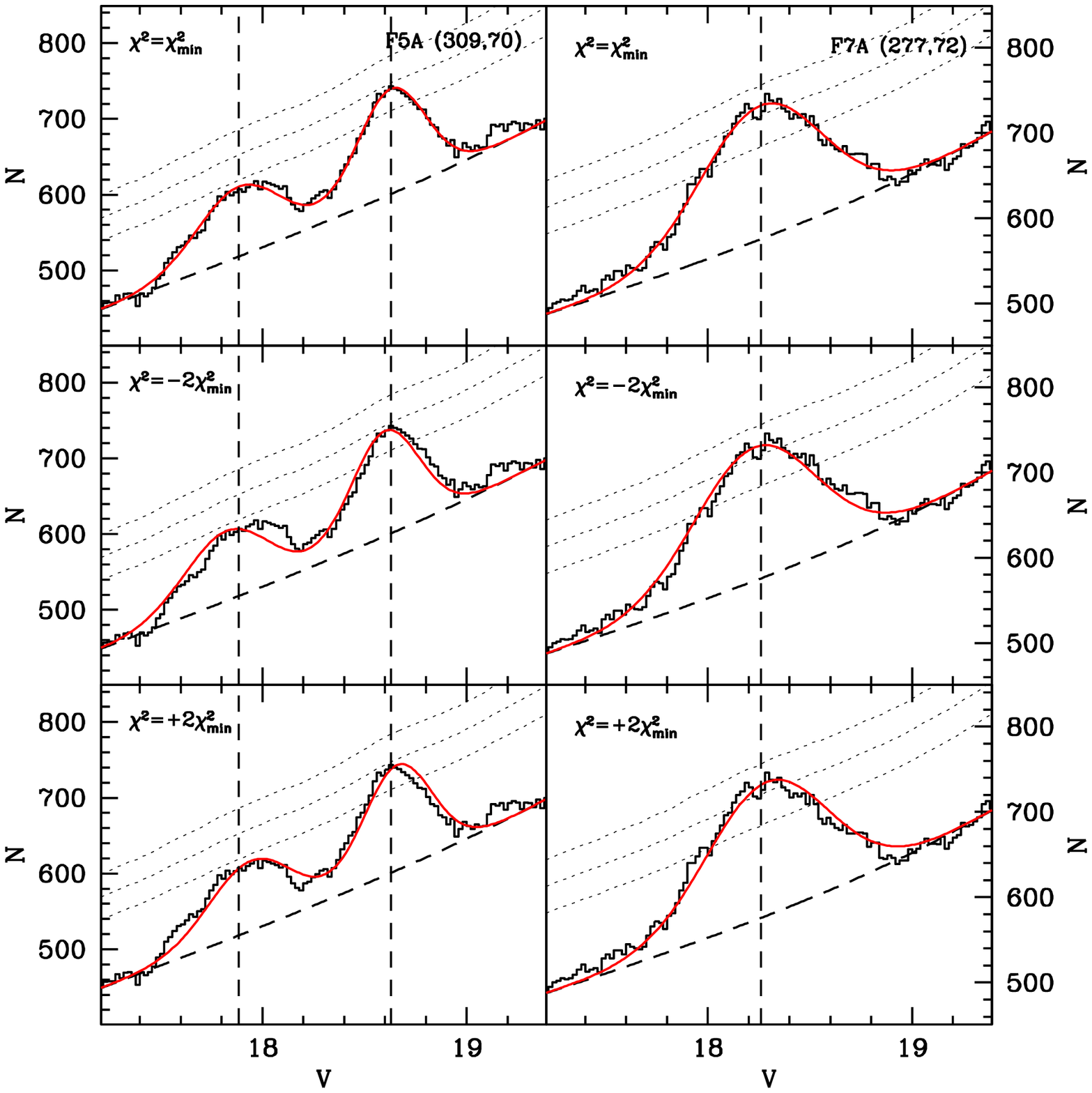}
\caption{Upper panels: de-reddened running histogram SCPs, zoomed in the region
of the peak(s), of color-selected RC candidates for the Stream field F5A (left
panel) and F7A (right panel). The meaning of the lines is the same as
Fig.~\ref{fit7A}, the red line represents the global model, $f(x) + G(x)$, with
the best fit value of $G(x)$ mean (value for which $\chi^2 = \chi^2_{min}$).
Middle and lower panels: as upper ones, with the exception that the values of
the $G(x)$ means are those that have $\chi^2 = \pm \chi^2_{min}$ ( $\chi^2 = -
2\chi^2_{min}$, middle panels and $\chi^2 = + 2\chi^2_{min}$, lower panels,
respectively). It is clearly visible that in these last two cases the fit is
totally unsatisfactory.}
\label{errfit}
\end{figure}

\subsection{The SCPs of Control Fields}
\label{scp_cf}

To verify empirically that the peaks we interpret as due to intersections of the
considered \los with Stream wraps are not due to Galactic structures,  we have
inspected all the color-selected SCPs of the Control Fields described in
Sect.~\ref{data}. The overall conclusion is that there is nothing similar to the
peaks we observe in the SCP of our on-Stream fields in generic Galactic fields at
similar distances from the plane and the center of the Galaxy.

\begin{figure}
\plotone{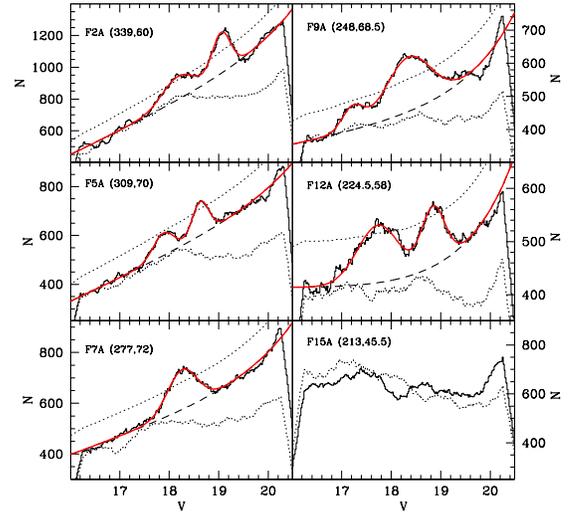}
\caption{Comparison between a sample of the on-Stream SCPs studied in this paper
(continuous histogram; best-fit model in red) and the SCP of the corresponding
Control Fields.}
\label{cfrV}
\end{figure}

In Fig.~\ref{cfrV} we show various examples: the SCPs of six on-Stream fields
(continuous lines) are compared to the SCPs of their corresponding CFs (dotted
lines, see Fig.~\ref{map}). The best-fit models for the on-Stream SCPs, together
with the background and the $3\sigma$ levels are also reported, using the same
symbols as in Sect.~\ref{modeling} and Sect.~\ref{detections} below. The two
SCPs are normalized by the ratio of the sampled areas, but any other reasonable
normalization (for example, by the ratio between the number of stars that fall
inside our color selection) does not significantly change the results.

The shapes of the SCPs are very similar in the range not affected by the peaks
associated with the Stream, as already observed when we have done the same
comparison in the main body (Fig.~\ref{fitSgr}). It is quite clear that the
strong and well defined peaks observed in on-Stream fields are lacking in the
SCPs of Control Fields (however there is no guarantee that genuine and yet
unrecognized structures are present also along these \los). It is also
reassuring to note that the models for the SCP of the contaminating
back/foreground population we have adopted for the on-Stream fields provide a
good description also of the CF SCPs, at least out to $V\simeq 18.5$. Beyond
this limit it is quite clear that in the on-Stream fields there is an additional
source of contamination, that has to be ascribed to RGB, SGB and MS stars from
the Stream population itself, as discussed in Sect.~\ref{SelCMD}. This provides
further support to the idea that the adopted approach of fitting the
back/foreground component directly on on-Stream SCPs is the most effective way
to get rid of this kind of self-containation from other species of Stream stars,
that would not have been possible if we merely subtracted the CF SCPs to the
on-Stream ones.

In the lower right panel we present the case of F15A, that will be discussed in
Sect.~\ref{detA}. It is interesting to note the close similarity between the two
considered SCPs for this field, where we do not detect any signal from the
Stream, and they are therefore expected to be (both) dominated by the generic
halo/thick disk Galactic population.

\subsection{Examples of on-Stream fits}
\label{exonstream}

In Fig.~\ref{fit7A} and Fig.~\ref{fit5A} we show two examples of application to
on-Stream fields, the fields F7A and F5A, respectively. In the first case a broad
peak with significance above 4$\sigma$ is detected in both the $V$ and $I$ SCPs.
The derived differential distances with respect to the main body are in good
agreement, within the uncertainties. The $f(x)+G(x)$ model provides an excellent
description of the observed SCPs.

Two significant peaks are detected in the SCPs of the field F5A
(Fig.~\ref{fit5A}), thus, in this case, we need a model with two gaussian
components. Both peaks are significantly narrower then that found in field F7A.
Nevertheless the model $f(x)+G_1(x)+G_2(x)$ provides an excellent representation
of the observed SCPs. The differential distances obtained from the $V$ and $I$
SCPs are in good agreement: there is no doubt that we are detecting {\em the same
structures in both SCPs}.

\begin{figure}
\plotone{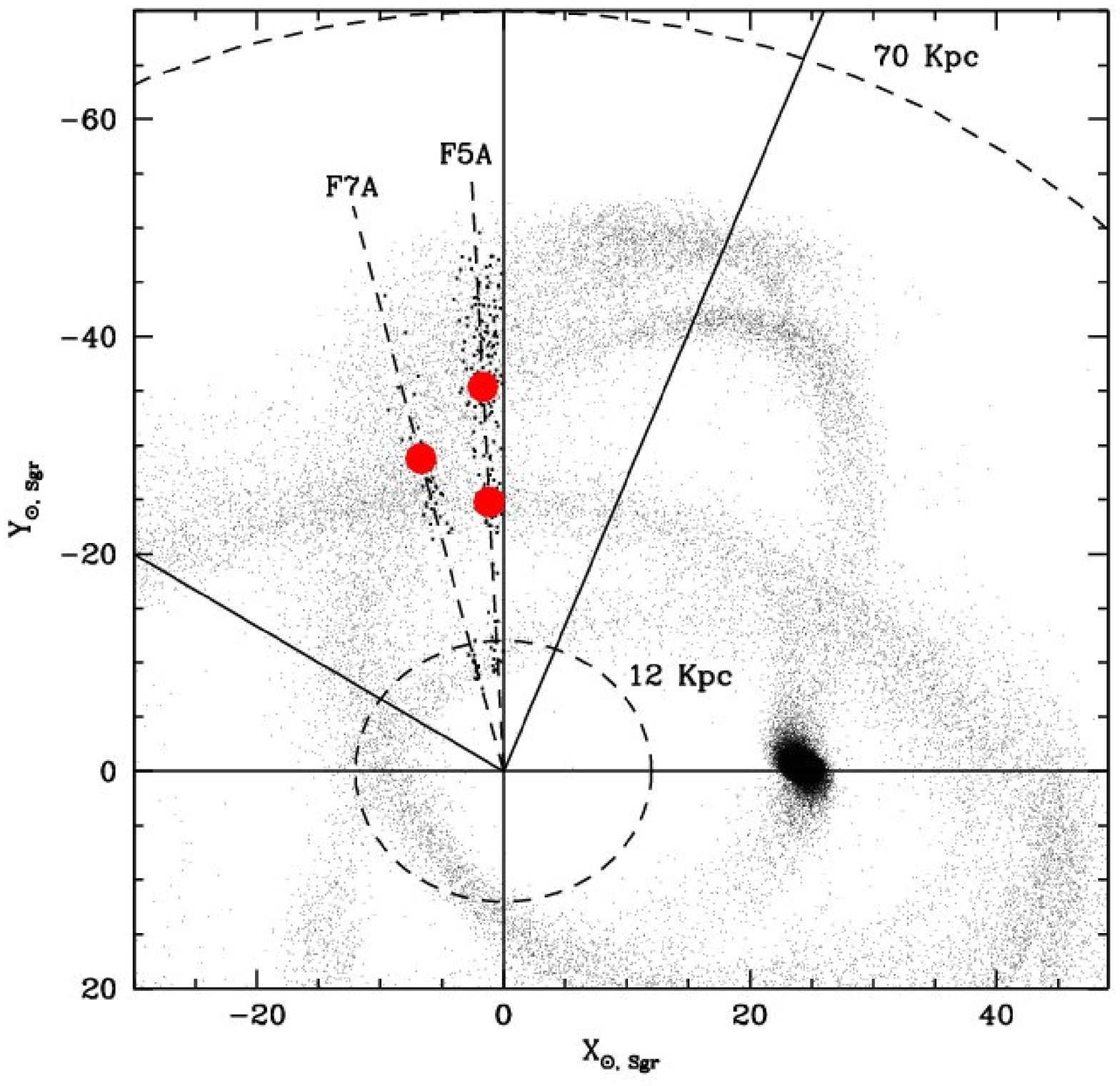}
\caption{N-body model of the tidal disruption of Sgr by \citet[][prolate Dark
Matter halo case]{law} in heliocentric Sgr coordinates \citep[see][]{maj03}. 
The heavily printed dots are the subset of the model particles that are
enclosed  in the two observed fields considered here (F5A and F7A). The observed
positions of the Stream in these fields, as estimated from the position of the RC
peaks in our SCPs, are plotted as filled (red) circles. 
}
\label{XY5A}
\end{figure}

To place the results shown in Fig.~\ref{fit7A} and Fig.~\ref{fit5A} into the
proper context, we plot the positions of the detected peaks into the
$X_{\sun,Sgr}$$Y_{\sun,Sgr}$ plane of the heliocentric Sgr coordinates as
defined by \citet{maj03}, in Fig.~\ref{XY5A}. This plane is defined to coincide
with the plane of the orbit of Sgr, hence the Stream is expected to be confined
within a few kpc about it. We compare our detections with one of the N-body
models of the tidal disruption of Sgr  by \citet{law}. In particular we plot in
Fig.~\ref{XY5A} the results of the evolution of the N-body model of Sgr within a
Galactic DM halo of prolate shape \citep[flatness q=1.25, see][for further
details on the models]{law,kat05}. To compare observations and model in a
consistent way we transformed our relative distances into absolute values by
adopting the same distance modulus for Sgr as \citet{law}, i.e. $(m-M)_0=16.9$
\citep{mateovar}. The points of the model that are encountered by the considered
F.o.V.s along the \los ~(F7A and F5A, from left to right, respectively) are
plotted as heavier dots.

Taking the considered model as a realistic representation of the actual Sgr
relic \cite[a very reasonable assumption, in first approximation;][]{law}, it is
clear that any \los ~around  the considered plane would cross one or more
different wraps of the Stream, at different distances (see Fig.~\ref{XY5A}).
The peak from F7A and the most distant peak from F5A seem to match
a distant portion of the leading arm. The nearest peak from F5A matches very
well with a wrap of the trailing arm that appears narrow and well defined and
that is crossed nearly perpendicular by the considered \los.   According to the
considered model, both \los should also cross a nearby wrap at a distance not
enclosed in our range of sensitivity, that is delimited by the two dashed
circles in Fig.~\ref{XY5A}. No wrap is expected to lie outside D=60 kpc in the
region sampled by our fields. 

\begin{figure}
\plotone{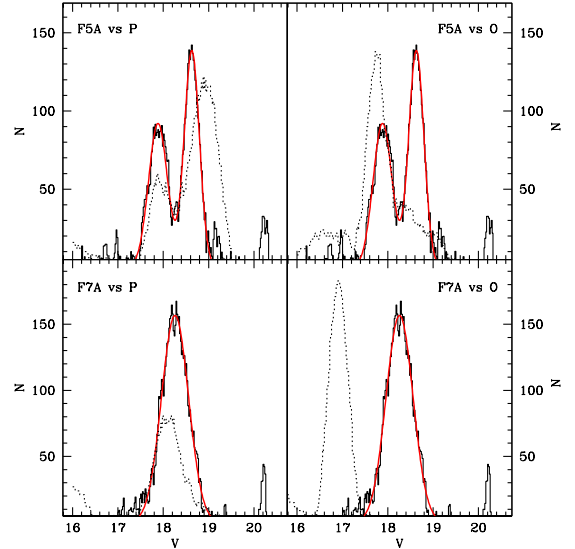}
\caption{Comparison between the observed field-subtracted RC SCPs from F5A 
(upper panels) and F7A (lower panels) and the predictions of the N-body models
by \citet{law} in the same directions (left panels: prolate-halo model, the same
as Fig.~\ref{XY5A}; right panels: oblate halo model). The observed SCPs are the
continuous running  histograms, the red continuous curves are the best-fit
gaussians, the dotted running histograms are the predictions of the models,
including the photometric errors. The histograms from the N-body models have
been multiplied by arbitrary normalization factors ($2\times$ and $3\times$, for
the left and right panels, respectively) to make the plots more readable.}
\label{compLF}
\end{figure}

Both the more distant F5A detection and the single F7A detection occur in
regions where the model predicts the confluence and crossing of different wraps.
At a first glance to Fig.~\ref{XY5A}, it may appear that the constraining power
of a single ``mean position'' of a Stream wrap, as derived with our method, is
not sufficient to describe the complex structure of the Sgr remnant along a
given \los. In Fig.~\ref{compLF} we compare the observed RC peaks of F5A and F7A
(and their best-fit gaussian models) with the peaks derived from the N-body
model shown in Fig.~\ref{XY5A} and from the oblate-halo model from the same set
\citep{law} by (a) selecting the model particles encountered by the FoV cone,
(b) assigning to each of them the absolute magnitude of the RC ($M_V^{RC}=
+0.72$ and $M_I^{RC}= -0.23$, according to Sect.~\ref{modeling}) and deriving
their apparent magnitude according to their distance, (c) adding gaussian
photometric errors as a function of the apparent magnitude similar to the
observed ones, and (d) producing the running histogram of the derived magnitudes
with the same settings adopted for the observed SCPs. To make easier the
comparisons shown in Fig.~\ref{compLF}, the synthetic SCPs have been multiplied
by an arbitrary normalization factor. The qualitative resemblance of the
observed and predicted structures for the prolate-halo model is striking. On the
other hand the oblate-halo model is clearly unable to reproduce the
observations, even in terms of number of Stream wraps encountered by the
considered \los. 

A more thorough comparison of our observations with theoretical models of the
disruption of Sgr will be presented in Sect.~\ref{models}. Here we are just
interested to demonstrate that our method allows a detailed comparison between
models and observations not only in terms of mean distances, but also in terms
of the actual {\em shape} of the structures along the {\em los} (see
Fig.~\ref{compLF}, left panels, in particular). In other words a fully
successful model of the Sgr Stream must reproduce the correct position  {\em and
shape} of the observed peaks: this provides the opportunity for a fruitful
detailed comparison between models and observations also in regions were
different Stream wraps cross each other. In the present study we provide the
position and the FWHM of  the peaks (as the most basic shape parameter, see
Sect.~\ref{sec.sigma}) but anyone interested in more detailed comparisons can
easily reproduce our results and obtain plots like Fig.~\ref{compLF}.

In Sect.~\ref{detections} the SCPs of all the considered on-Stream fields, with
the associated detections will be presented and briefly discussed. In agreement
with the qualitative predictions of the model shown in Fig.~\ref{XY5A}, in most
cases we will detect two peaks at different distances.

\subsection{Classification of the detections}
\label{classif}

We divided the detections in three categories, assigning a {\it flag} to each of
them, according to the following criteria:

\begin{itemize} \item $flag=1$: peaks that are above the $3\sigma$ threshold
	both in the $V$ and $I$ SCPs, and having the same $\Delta(Mag)$ in both
	passbands, within the errors. These are called {\it primary} peaks.
	
	\item $flag=2$: peaks having the same $\Delta(Mag)$ in both passbands,
	within the errors, but reaching the $3\sigma$ threshold only in one of
	the two SCPs. In all the $flag=2$ cases described in the following,
	the weaker peak is always just below  $3\sigma$. These are called {\it
	secondary} peaks.
	
	\item $flag=3$: clearly visible peaks having the same $\Delta(Mag)$ in
	both passbands, within the errors, but not reaching the $3\sigma$
	threshold in both the SCPs. These correspond to uncertain detections that
	we report just for completeness. In some case a weak peak in the SCP in
	one band is tentatively identified as it corresponds to a stronger peak
	in the other SCP. These peaks are called also {\it tertiary} peaks.
\end{itemize}

The observed SCPs and the detected peaks
will be briefly described and discussed in Sect.~\ref{detections}.

\subsection{Number of RC stars associated to each peak}
\label{sec.dens}

Our modeling of the observed SCP automatically provides also an estimate of the
total number of stars associated to any given peak. This gives a useful
additional constraint for theoretical models; we will illustrate this
possibility with an example in Sect.~\ref{densazi}. In Tab.~\ref{tab_dmag} we
provide the number of stars associated with a given peak normalized to an area
of 25 deg$^2$ (quite similar to the actual area of our fields). This number is
the weighted mean of the estimates obtained from the $V$ and $I$ SCPs, where the
assumed error on the estimate from each SCP is just the square root of the
observed number (hence it should be considered as a lower limit to the real
error).

Adopting the {\it Sgr34} field as a baseline \cite[having $N^{RC}_{Sgr34}\simeq 1500$
stars/deg$^2$, and $mu_V\simeq 25.5$ mag/arcsec$^2$;][]{lornuc}, and translating
our on-Stream estimates into stars/deg$^2$ units ($N^{RC}_{i}$, for the field
$i$)  we can transform our numbers into surface brightness, according to the
same formula used in \citet{mrc}:
\begin{equation}
\label{sbeq}
\mu_{V,i} = \mu_{V,Sgr34} -2.5log(\frac{N^{RC}_{i}}{N^{RC}_{Sgr34}}) + \Delta(m-M)_0
\end{equation}
derived from \citet{alviopix}, where $\Delta(m-M)_0 = (m-M)_0^{i}
-(m-M)_0^{Sgr34} = \Delta V = \Delta I$. Using the equation \ref{sbeq} we find
that the $V$ surface brightness of the portions of the Sgr Stream studied in the
present paper range between 30.6 mag/arcsec$^2$ and 33.6 mag/arcsec$^2$, quite
typical of tidal tails and in good agreement with previous results \cite[see,
for example,][and references therein]{sgrclus2}.

It is important to recall that the measured densities refer only to RC stars: in
the presence of a population gradient \citep[as is likely the case in the
Stream,][]{B06c,chou} they would trace different fractions of the total stellar
content at different positions along the Stream. Analogously, the derived
surface brightness estimates have been rescaled assuming the stellar mix of
{\it Sgr34} for all the considered portions of the Stream. For this reason these
estimates should be considered with caution: given the sense of the gradient it
is expected that they provide lower (faint) limits when converted into
luminosity or surface brightness. In this context, it is interesting to note
that if we convert our surface brightness into the same density units
($L_{\sun}$/kpc) adopted by \citet{nied}, we find that our results are fully
compatible with the trend of density as a function of RA derived by these
authors (see their Fig.~7), for both branches. On the other hand, our densities
are lower than theirs by a factor of $\sim 4-5$. It is reasonable to assume that
part of the difference may be accounted for by the effect of the population
gradient described above.

\subsection{Depth along the \los of the Stream wraps}
\label{sec.sigma}

Fig.~\ref{compLF} shows that the observed RC peaks contain also valuable
information on the characteristic size of the section of the Stream branches
crossed by our {\em los}, as peaks at similar distances display different
widths. To obtain a quantitative estimate of the linear width along the \los of
the structures identified here, we recurred to the synthetic population
described in Sect.~\ref{basti}. In particular we tried to reproduce the {\em
models} of the observed peaks\footnote{In this way we avoid any problem
associated with the (partial) overlap of {\em observed} peaks; adjacent
overlapping peaks are disentangled by our models as Point Spread Function
-fitting photometric packages disentangle the fluxes from two partially
overlapping stars on an image.} with smoothed histograms of the synthetic RC
population, properly including the effects of photometric errors. As done in
Sect.~\ref{basti}, we assign a distance along the \los to each star of  the
synthetic RC population according to a gaussian distribution having the same
mean and normalization as the observed peak, and we search for the value of
$\sigma_d$ giving the best match between the two models. In Tab.~\ref{tab_dmag}
we report the Full Width at Half Maximum (in kpc) of the adopted distributions,
FWHM=2.35$\sigma_d$. The best match is found by minimizing $\chi^2$, 
typical uncertainties are
$\sim 20$\%. The adopted procedure gets rid of the effects of the distance on the
width of the SCP peaks discussed in Sect.~\ref{basti}.

Since the synthetic population that we adopt is strictly single-age and
single-metallicity, the intrinsic luminosity width of its RC should be smaller
than the actual width of the RC of Sgr, as the latter hosts stars spanning a
range of ages and metallicities \citep[B06a,][]{gs,siegel,nuc08}. For this
reason the FWHM values we obtain in this way must be considered as strong {\em
upper limits} to the real values. Moreover, it has to be recalled that we report
FWHM along a given \los, that may have various incidence angles with respect to
the encountered Stream wraps. Applying the method to the {\it Sgr34} field we
obtain $FWHM\simeq 3$ kpc, not too far from the minor-axis FWHM in the plane of
the sky as obtained from the best-fit King (1962) model by \citet{maj03}, i.e.
$FWHM\simeq 1.1$ kpc, in particular if we take into account that the \los toward
the core of Sgr is (likely) not exactly perpendicular to the major axis of the
dwarf galaxy (see Fig.~\ref{XY5A}). Based on this test, it is reasonable to
assume that our FWHM overestimates the true values by a factor of $\ga 2$.

In any case, the ratio between the FWHM of two different \los/locations in the
Stream, or, equivalently, the differential trend of the FWHM as a function of
orbital azimuth along a given Stream wrap, can be directly compared to the
predictions of  theoretical models of the disruption of Sgr.

\section{On-Stream detections}
\label{detections}

In this section we present all the SCPs obtained from each analyzed field of 
branches A and B; we plot the SCPs in the $V$- and $I$-bands in Fig.~\ref{LFA}
for branch A, and Fig.~\ref{LFB} for branch B. Together with the observed SCPs,
we plot also the background model ($f(x)$, dashed lines), the threshold limits
for the detections (dotted lines, respectively $3,4,5\sigma$), which as
mentioned previously, include both the Poisson noise and the uncertainty in the
fit, and the global model that fits the observed SCPs ($f(x)+G(x)$, red lines).
Each field is labeled according to the names assigned in Table~\ref{coord} and
with its Galactic coordinates (l,b).

\begin{figure*}
\plotone{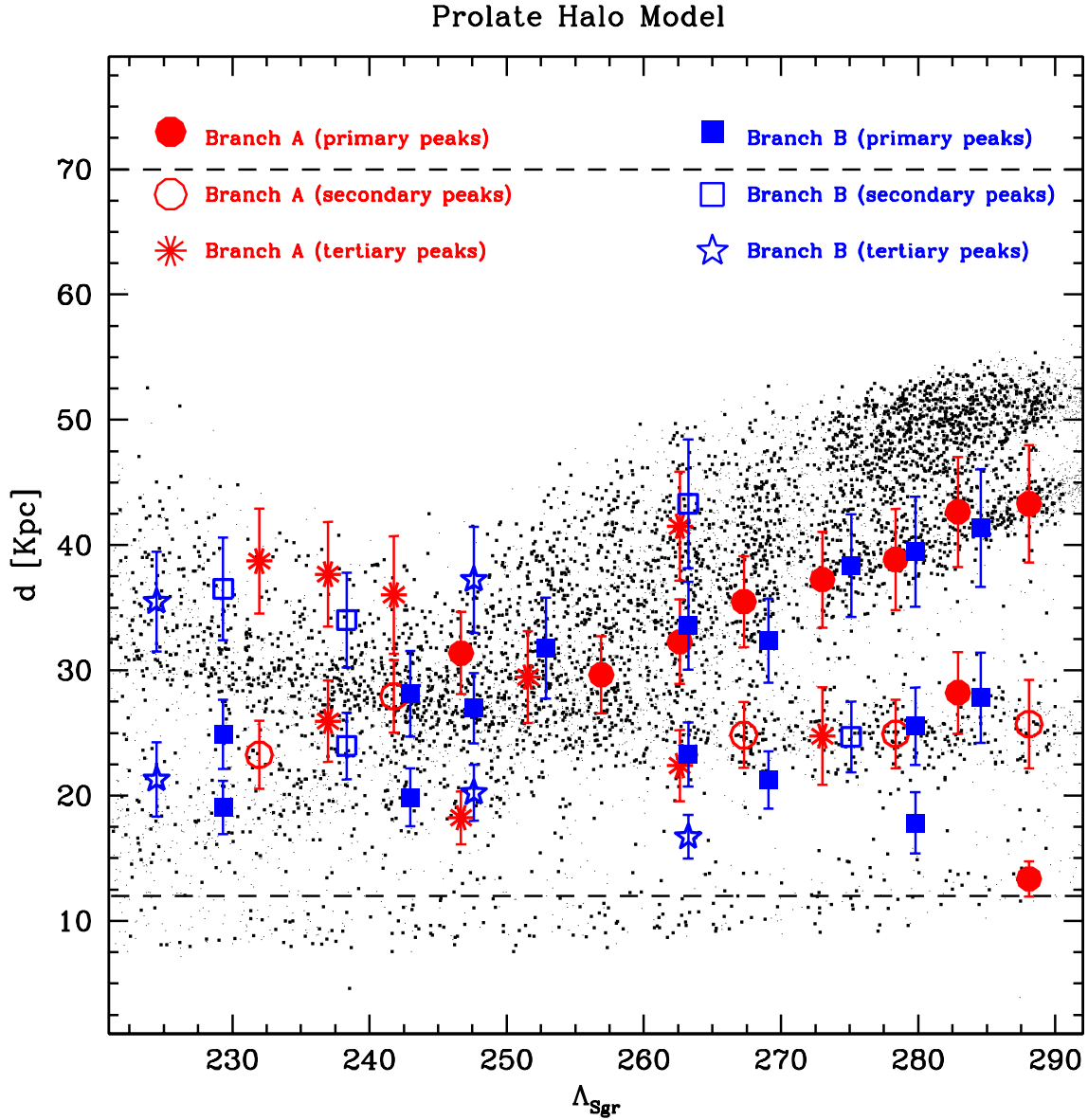}
\caption{Distribution of the {\em primary} (red filled circles - branch A -, and
blue filled squares - branch B), {\em secondary} (red open circles - branch A -,
and blue open squares - branch B) and {\em tertiary} detections (red starred
symbols - branch A, and blue stars - branch B) in the  $\Lambda_{Sgr}$ vs.
heliocentric distance plane (a true distance modulus of 16.90 has been adopted
here). The horizontal dashed lines enclose the range of sensitivity of our
method. The prolate-halo N-body model by \citet{law} is also reported (small
dots) as an aid for the interpretation of the plot. The heavier dots are those
enclosed in the cones of the considered FoVs.}
\label{ldist}
\end{figure*}

In summary, we detect 26 {\em primary} (flag=1) peaks, 10 {\em secondary}
(flag=2), and 14 {\em tertiary} (flag=3) peaks. Most of the considered SCPs show two
significant peaks, corresponding to subsequent crossings of different wraps of
the Stream along the \los.  The trend of peak distance as a function of Sgr
longitude ($\Lambda_{Sgr}$) shown in Fig.~\ref{ldist} can be useful to better
understand the morphology of the various SCPs presented below. Most primary peaks
appear to trace a wrap of the leading arm whose distance from the Sun steadily
decreases from $D\simeq 45$ kpc at $\Lambda_{Sgr}\simeq 290\degr$ to $D\simeq
20$ kpc at $\Lambda_{Sgr}\simeq 230\degr$. Both primary and secondary peaks
trace a more nearby filamentary structure at constant distance $D\simeq 25$ kpc,
from $\Lambda_{Sgr}\simeq 290\degr$ to $\Lambda_{Sgr}\simeq 260\degr$, that then
bends toward larger distances, reaching  $D\simeq 40$ kpc at
$\Lambda_{Sgr}\simeq 230\degr$. This feature is tentatively identified as a wrap
of the trailing arm (see Sect.~\ref{models}).  The two wraps cross at
$\Lambda_{Sgr}\sim 245\degr$ (see Sect.~\ref{models}). Some secondary and tertiary
peaks seem to trace more feeble distant or nearby wraps (see Sect.~\ref{models}
for a deeper discussion). The comparison with the considered model suggests that
most of the detected peaks can be associated with the Sgr Stream. The tertiary
peak at $\Lambda_{Sgr}\sim 263\degr$ and $D\simeq 18$~kpc, and the primary peak
at $\Lambda_{Sgr}\sim 280\degr$ and $D\simeq 19.5$~kpc, are possibly associated
to other overdensities in the Virgo constellation  \citep[see][and references
therein]{juric,duffvirgo,heidivirgo}, as discussed in some detail in
Sect.~\ref{models}.

\begin{figure}
\plotone{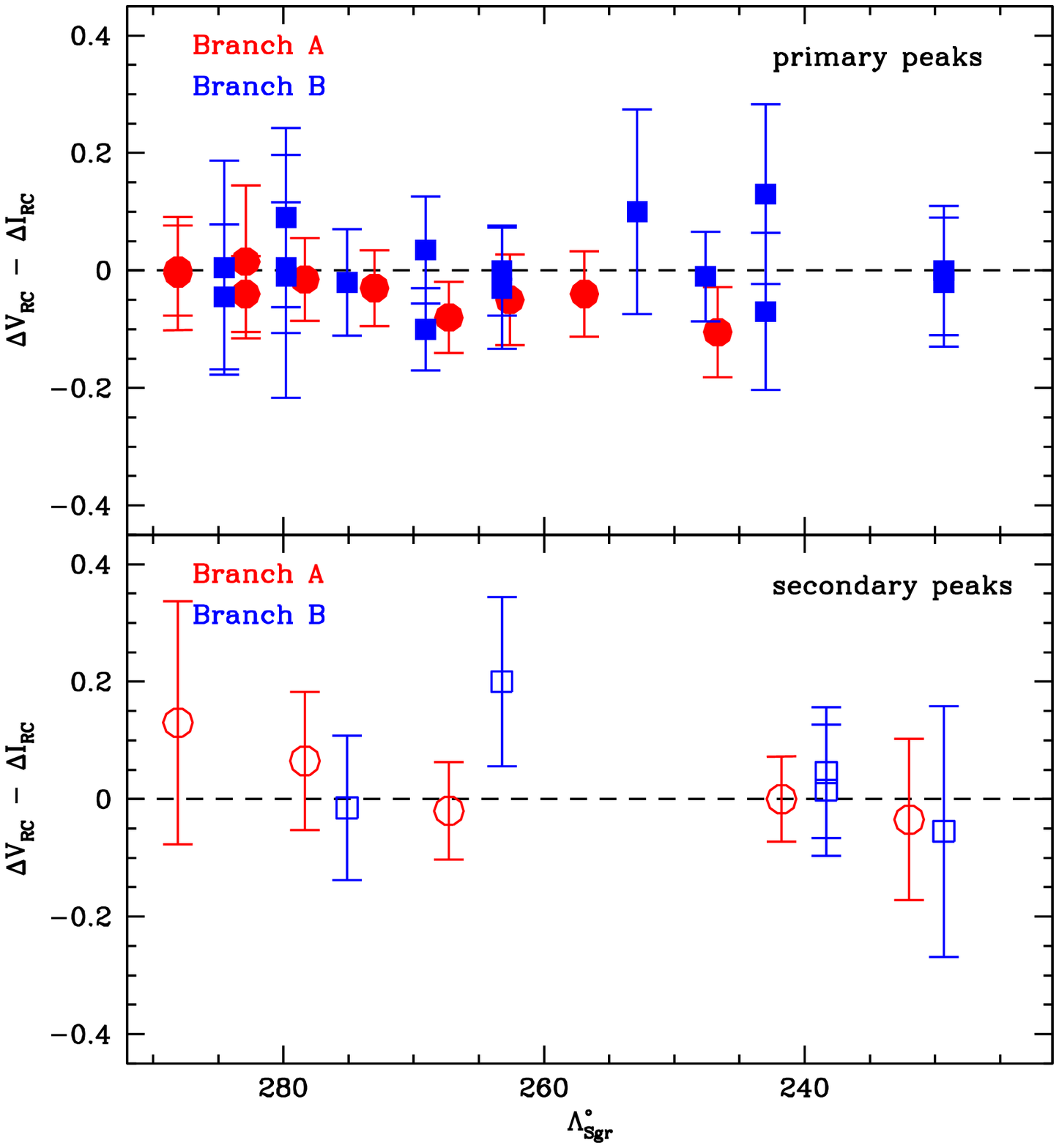}
\caption{Comparison between the differential distance moduli obtained from peaks
in the $V$ and $I$ SCPs for primary (upper panel) and secondary (lower panel)
peaks.}
\label{dmag}
\end{figure}

Before proceeding in the description of the various detections, we anticipate
that the differential distance moduli ($\Delta V=
V_{RC}^{Field}-V_{RC}^{Sgr34}$;  the analogous definition being valid also for
$\Delta I$), reported in Tab.~\ref{tab_dmag}, obtained from primary and
secondary peaks detected in the $V$ and $I$ SCPs are in {\em excellent} agreement, as
shown in Fig.~\ref{dmag}. This confirms the reliability of our detections and
distance estimates. For this reason, from Sect.~\ref{comp} on and in
Fig.~\ref{ldist} we adopt the mean of $\Delta V$ and $\Delta I$ as our final
differential distance moduli estimates.

\begin{table*}
\begin{center}
\caption{Distance, FWHM and density for the detected RC peaks.}
\tiny
\begin{tabular}{c|cc|cccc|cc|cc|c|cc|c}
\hline
field & $l^{\circ}$ & $b^{\circ}$ & $\Delta V$ & $\epsilon\Delta V$ & $\Delta I$  & $\epsilon\Delta I$ & d  &  $\epsilon$d   &  sign. (V)  & sign. (I) & FWHM  & N$_{RC}$  & $\epsilon$N & flag  \\
      &  deg  & deg  &    mag  &    mag  &    mag  &    mag  &  kpc  &  
kpc &        &        &  kpc  & [stars/25 deg$^2$]  & [stars/25 deg$^2$]  &  1 \\ 
&&&&&&&&&&&&&&\\
\hline
  1A  &  346  &  56  &  1.28  &  0.07  &  1.28  &  0.06  &  47.5  &  
5.1 &  $\geq 5\sigma$ &  $5\sigma$     &  9.8  &  241 &  11 &  1 \\ 
      &       &      &  0.21  &  0.14  &  0.08  &  0.15  &  28.2  &
3.9 &    $4\sigma$    &  $<3\sigma$    &  7.1  &  353 &  9  &  2 \\
      &       &      & -1.27  &  0.05  & -1.27  &  0.05  &  14.6  &
1.1 &    $3\sigma$    &   $3\sigma$    &  1.3  &  79  &  6  &  1 \\
\hline
  2A  &  339  &  60  &  1.23  &  0.05  &  1.27  &  0.04  &  46.8  &
4.8 &  $\geq 5\sigma$ & $\geq 5\sigma$ &  4.2  &  227 &  9  &  1 \\
      &       &      &  0.36  &  0.09  &  0.34  &  0.09  &  30.9  &
3.6 &    $4\sigma$    &   $3\sigma$    &  9.5  &  303 &  10 &  1 \\
\hline
  3A  &  331  &  63  &  1.04  &  0.04  &  1.05  &  0.05  &  42.6  &
4.4 &   $>5\sigma$    &  $\geq5\sigma$ & 11.3  &  345 &  13 &  1  \\
      &       &      &  0.06  &  0.07  &  0.00  &  0.09  &  26.6  &
3.0 &    $4\sigma$    &  $<3\sigma$    &  6.2  &  327 &  10 &  2  \\
\hline
  4A  &  320  &  66  &  0.94  &  0.04  &  0.97  &  0.05  &  40.8  &
4.2 &  $\geq 5\sigma$ &   $5\sigma$    &  7.5  &  158 &  8  &  1  \\
      &      &       &  0.08  &  0.09  &  0.05  &  0.15  &  27.1  &
4.3 &   $<3\sigma$    &  $<3\sigma$    & 14.6  &  226 &  10 &  3  \\
\hline
  5A  &  309  &  70  &  0.81  &  0.03  &  0.89  &  0.05  &  38.9  &
4.0 &    $4\sigma$    &   $4\sigma$    &  8.6  &  114 &  8  &  1  \\
      &       &      &  0.06  &  0.06  &  0.08  &  0.05  &  27.2  &
2.9 &   $<3\sigma$    & $\ge3\sigma$   &  5.1  &  116 &  7  &  2  \\
\hline
  6A  &  296  &  72  &  0.62  &  0.04  &  0.67  &  0.06  &  36.0  &
3.8 &    $4\sigma$    &   $4\sigma$    &  7.0  &  171 &  11 &  1  \\
      &       &      & -0.08  &  0.12  & -0.22  &  0.12  &  24.6  &
3.1 &   $<3\sigma$    &  $<3\sigma$    &  3.9  &   66 &  7  &  3  \\
      &       &      &  1.18  &  0.05  &  1.20  &  0.05  &  45.5  &
4.8 &   $<3\sigma$   &  $<3\sigma$     &  1.4  &  49  &  6  &  3  \\
 \hline
  7A  &  277  &  72  &  0.44  &  0.04  &  0.48  &  0.06  &  32.5  &
3.4 &  $\ge4\sigma$   & $\ge4\sigma$   & 10.9  &  188 &  10 &  1  \\
\hline
  8A  &  260  & 71.5 &  0.53  &  0.10  &  0.36  &  0.13  &  32.3  &
4.0 &   $<3\sigma$    &  $<3\sigma$    &  3.9  &  122 &  6  &  3  \\
\hline
  9A  &  248  & 68.5 &  0.53  &  0.05  &  0.63  &  0.05  &  34.4  &
3.6 &    $5\sigma$    &   $5\sigma$    & 17.0  &  231 &  10 &  1  \\                                      
      &       &      & -0.59  &  0.08  & -0.59  &  0.10  &  20.0  &
2.3 &  $<3\sigma$     &  $<3\sigma$    &  2.6  &  71  &  6  &  3  \\
\hline
 10A  &  238  & 65.5 &  0.33  &  0.06  &  0.33  &  0.04  &  30.6  &
3.2 &    $3\sigma$    &  $<3\sigma$    &  5.3  &  90  &  5  &  2  \\
      &       &      &  0.94  &  0.15  &  0.84  &  0.10  &  39.5  &
5.1 &  $<3\sigma$     &  $<3\sigma$    &  3.0  &  64  &  4  &  3  \\
\hline
 11A  & 230.5 &  62  &  0.93  &  0.08  &  1.03  &  0.07  &  41.3  &
4.6 &  $<3\sigma$     &  $<3\sigma$    &  9.5  &  42  &  6  &  3  \\
      &       &      &  0.15  &  0.15  &  0.18  &  0.07  &  28.4  &
3.6 &  $<3\sigma$     &  $<3\sigma$    &  6.2  &  37  &  5  &  3  \\
\hline
 12A  & 224.5 &  58  &  1.03  &  0.06  &  1.05  &  0.07  &  42.5  & 
4.6 &  $>3\sigma$     &  $<3\sigma$    &  4.2  &  103 &  7  &  3  \\
      &       &      & -0.08  &  0.09  & -0.05  &  0.10  &  25.5  &
3.0 &  $>3\sigma$     &   $3\sigma$    &  9.8  &  167 &  10 &  2  \\
\hline
\hline
  1B  &  353  &  61  &  1.16  &  0.08  &  1.20  &  0.09  &  45.3  &
5.2 &  $3\sigma$      &   $3\sigma$    &  6.9  &  172 &  8  &  1  \\
      &       &      &  0.32  &  0.10  &  0.32  &  0.15  &  30.5  &
3.9 &  $3\sigma$      &   $3\sigma$    &  12.5 &  352 &  11 &  1  \\
\hline
  2B  & 346.5 &  65  &  1.08  &  0.09  &  1.08  &  0.06  &  43.3  &
4.8 &  $\geq 5\sigma$ & $\geq 5\sigma$ &  9.3  &  187 &  10 & 1  \\
      &       &      &  0.18  &  0.12  &  0.09  &  0.09  &  28.0  &
3.4 &   $5\sigma$    &  $5\sigma$      &  8.8  &  374 &  11 &  1  \\
      &       &      & -0.65  &  0.14  & -0.64  &  0.15  &  19.5  &
2.7 &   $3\sigma$    &  $3\sigma$      &  3.5  &  124 &  8  &  1  \\
\hline
  3B  & 337.5 & 68.5 &  1.01  &  0.07  &  1.03  &  0.05  &  42.1  &
4.5 &  $>3\sigma$     &   $>3\sigma$   &  6.8  &  234 &  8  &  1  \\
      &       &      &  0.05  &  0.08  &  0.07  &  0.09  &  27.1  &
3.1 &  $3\sigma$      &   $<3\sigma$   &  6.8  &  170 &  8  &  2  \\
\hline
  4B  & 329.5 &  74  &  0.60  &  0.05  &  0.70  &  0.05  &  35.5  &
3.7 &  $>5\sigma$     &   $5\sigma$    & 15.0  &  271 &  11 &  1  \\
      &       &      & -0.24  &  0.05  & -0.28  &  0.07  &  23.3  &                      
2.5 &  $>4\sigma$     &   $3\sigma$    &  3.1  &  160 &  6  &  1  \\
\hline
  5B  & 313.5 & 78.5 &  0.73  &  0.05  &  0.73  &  0.05  &  36.8  &
3.8 &  $5\sigma$     &   $5\sigma$     &  6.5  &  125 &  7  &  1  \\
      &       &      & -0.08  &  0.07  & -0.05  &  0.07  &  25.5  &
2.8 &  $4\sigma$     &   $3\sigma$     &  4.1  &  133 &  7  &  1  \\
      &       &      &  1.38  &  0.10  &  1.18  &  0.10  &  47.5  &
5.5 &  $<3\sigma$    &  $>3\sigma$     &  6.9  &  56  &  7  &  2  \\
      &       &      & -0.80  &  0.05  & -0.77  &  0.05  &  18.3  &
1.9 &  $<3\sigma$    &  $<3\sigma$     &  2.2  &  61  &  5  &  3  \\
\hline
  7B  &  256  &  80  &  0.66  &  0.10  &  0.56  &  0.14  &  34.8  &
4.4 &  $3\sigma$     &  $>3\sigma$     &$\sim20.0$&201&  10 &  1  \\
\hline
  8B  &  234  &  77  &  0.25  &  0.05  &  0.26  &  0.05  &  29.6  &
3.1 &  $4\sigma$     &   $>4\sigma$    &  6.5  &  115 &  7  &  1  \\                 
      &       &      & -0.34  &  0.07  & -0.40  &  0.08  &  22.2  &
2.5 &  $<3\sigma$    &  $<3\sigma$     &  3.3  &  93  &  5  &  3  \\
      &       &      &  0.93  &  0.08  &  0.98  &  0.09  &  40.8  &
4.6 &  $<3\sigma$    &  $<3\sigma$     &  9.2  &  89  &  7  &  3  \\
\hline
  9B  &  224  &  73  &  0.41  &  0.09  &  0.28  &  0.12  &  30.8  &
3.7 & $\geq 3\sigma$ &   $3\sigma$     & 12.0  &  190 &  8  &  1  \\            
      &       &      & -0.44  &  0.07  & -0.37  &  0.11  &  21.8  &
2.5 &  $>3\sigma$    &  $3\sigma$      &  3.1  &  59  &  6  &  1  \\
\hline
 10B  & 216.5 &  69  &  0.76  &  0.10  &  0.75  &  0.04  &  37.3  &
4.1 &  $<3\sigma$    &  $>3\sigma$     &  7.0  &  111 &  8  &  2  \\
      &       &      &  0.02  &  0.102  & -0.02  &  0.04  &  26.3  &
2.9 &  $<3\sigma$    &  $>3\sigma$     &  4.4  &  74  &  8  &  2  \\
\hline 
 12B  & 208.5 & 60.5 &  0.88  &  0.15  &  0.94  &  0.15  &  40.0  &
5.6 &  $<3\sigma$    &  $3\sigma$      &  8.1  &  92  &  6  &  2  \\
      &       &      &  0.07  &  0.08  &  0.09  &  0.07  &  27.3  &
3.0 &   $3\sigma$    &  $3\sigma$      &  4.0  &  79  &  6  &  1  \\
      &       &      & -0.50  &  0.07  & -0.50  &  0.08  &  20.9  &
2.3 &   $3\sigma$    &  $3\sigma$      &  2.4  &  75  &  6  &  1  \\                                          
\hline
 13B  &  203  & 56.5 &  0.89  &  0.08  &  0.81  &  0.08  &  38.9  &
4.4 &   $<3\sigma$   &  $<3\sigma$     &  7.2  &  81  &  6  &  3  \\
      &       &      & -0.20  &  0.15  & -0.32  &  0.15  &  23.3  &
3.2 &   $<3\sigma$   &  $<3\sigma$     &  6.7  &  142 &  6  &  3  \\
\hline                          
\end{tabular}
\label{tab_dmag}
\end{center}
\end{table*}

\subsection{Branch A detections}
\label{detA}

\begin{figure*}
\plottwo{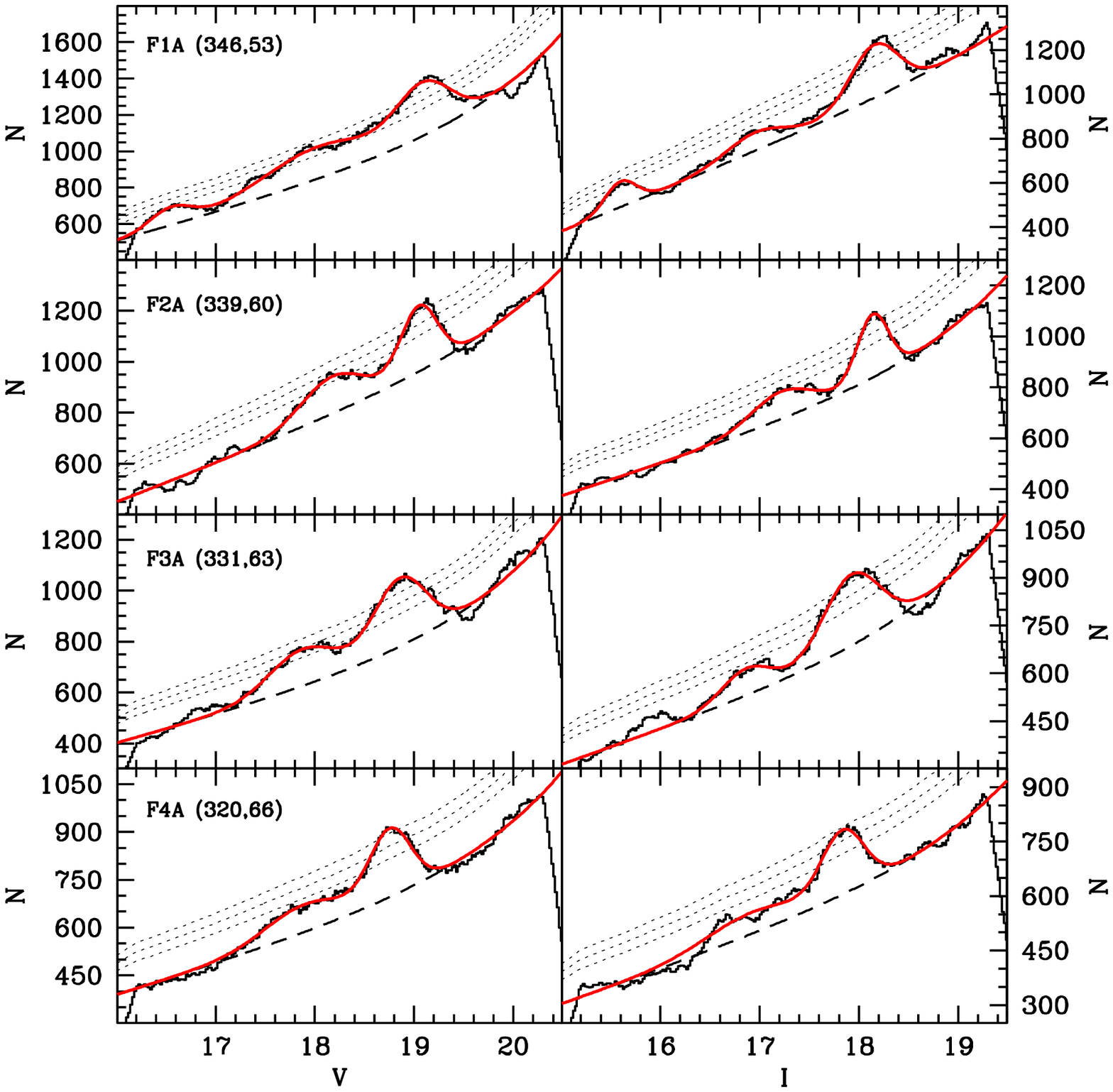}{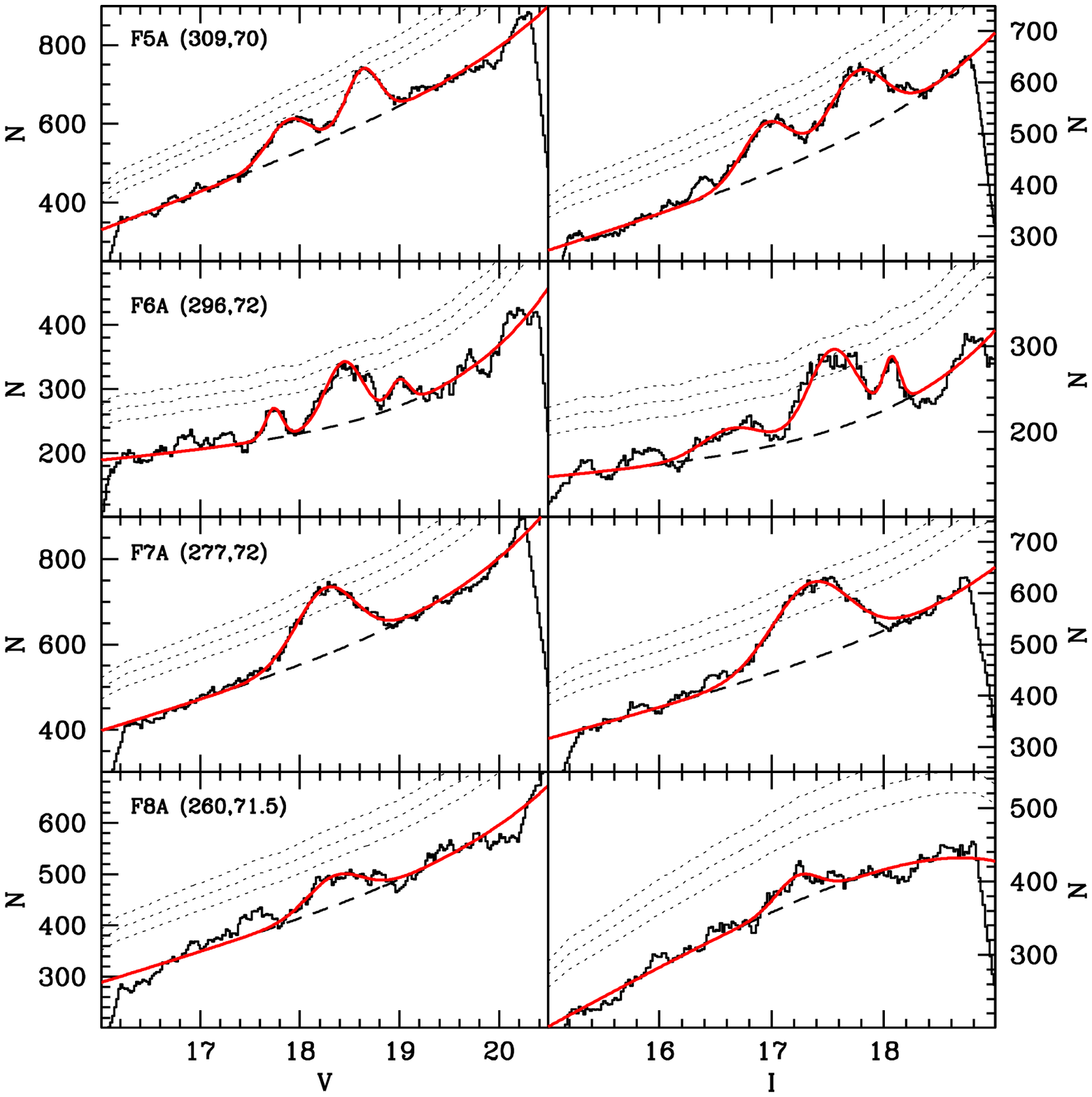}
\plottwo{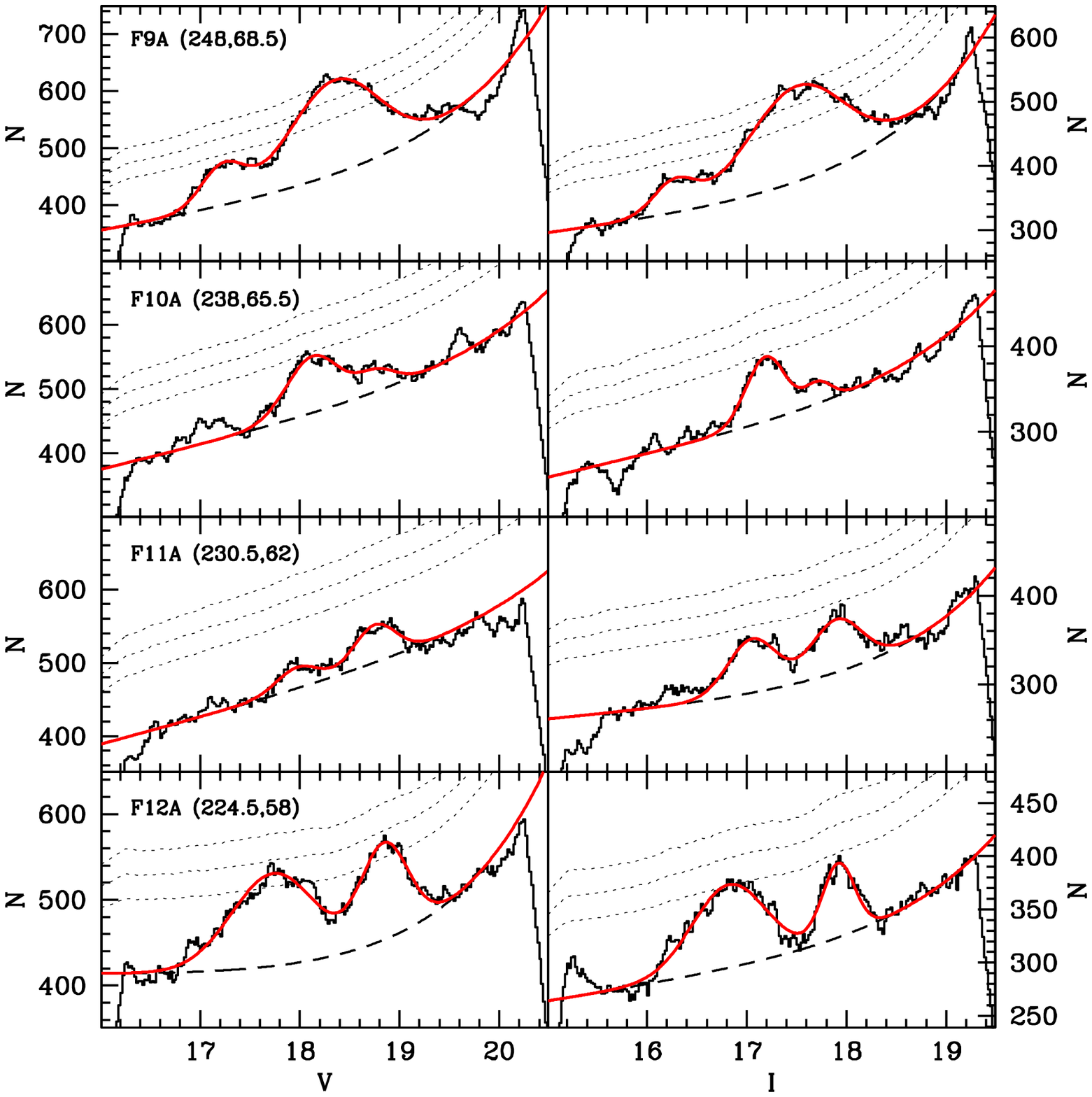}{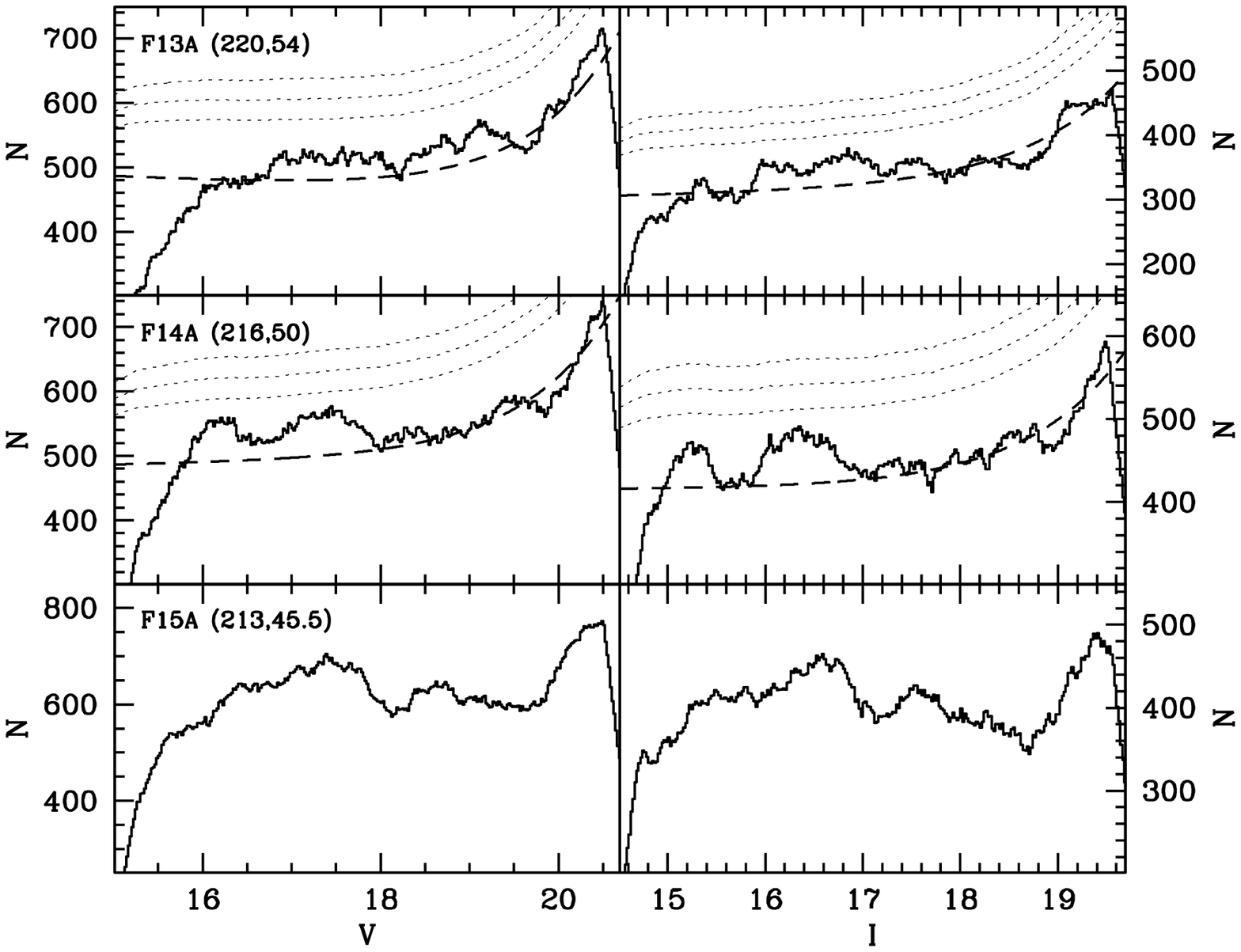}
\caption{Fits of the observed SCPs (in $V$ and $I$) for fields on Branch A of the Sgr
Stream. The numbers in parentheses are the Galactic longitude and latitude of
the center of the field, in degrees. The meaning of the symbols is the same as
in Figg.~\ref{fit7A} and \ref{fit5A}, above. No global fit has been attempted
for SCPs  that did not show significant RC peaks (F13A, F14A, F15A).}
\label{LFA}
\end{figure*}

In branch A we analyzed 15 fields, the corresponding observed SCPs and the
adopted best-fit models are shown in Fig.~\ref{LFA}. We obtained a total of 24
peak detections, with the following classification: 10 primary peaks, 5
secondary peaks and 9 tertiary peaks. The SCPs of the first five fields  (from
F1A to F5A, upper left panel and first two rows of the upper right panel of
Fig.~\ref{LFA}) display a common general behaviour: they present two main peaks,
the one at fainter magnitudes always being the strongest (a {\em primary} peak
in all cases), while the brighter ones are wider and span all the classes from
flag=1 to flag=3, depending on the specific field. It is quite clear that this
series of peaks traces two coherent structures placed at different distances
along the \los. The $I$-band SCP of F4A may suggest a splitting of the
brighter/weaker peak into two separate components: we consider this
interpretation as unlikely, nevertheless the result obtained with a three  peaks
model is briefly discussed in Sect.~\ref{detS}. The only exception is a primary
peak detected at $V\sim 16.5$ in F1A: this likely corresponds to the nearest
wrap of the Stream that emerges from the $d\la 12$~kpc circle (where our method
is blind), which is the most Eastern of all the considered \los (see
Sect.~\ref{models}). 

The SCPs of F6A (second upper right panel of Fig.~\ref{LFA}) present an overall
structure similar to those described above. However we identified additional
(fainter) peaks, and finally we adopted a three peak solution, whose validity is
confirmed by the inspection of SCPs obtained with reduced bin width (i.e., higher
resolution\footnote{We note that this is the only case in which a change in the
bin width produced a change in the interpretation of the SCPs.}; see
Sect.~\ref{detS} for an alternative). The newly-resolved third peak, at
magnitude $V \sim 19$ ($I \sim 18$), corresponds to a distance $D \simeq 42$ kpc
at $\Lambda_{Sgr}\simeq 265\degr$; this detection seems related to a very
distant wrap of the leading arm (see for example Fig.\ref{xy_pro}). F6A is the
only branch-A field in which we detect a peak related to this distant wrap of
the leading arm, that was observed also by Bel06 (see Sect.~\ref{comp}); the
same structure is detected in branch B along the same \los,  as well as along an
additional one (F8B).

F7A is one of two fields in branch A that presents only one detection: SCPs
(third upper right panels of Fig.~\ref{LFA}) show a single, very prominent {\em
primary} peak. As discussed above, this \los intercepts a region where two or
three wraps of the Stream cross each other. Fig.~\ref{compLF} shows that the
presence of a single peak is nevertheless consistent with model predictions.  In
both the SCPs of the adjacent field F8A, a remarkably weak peak appears at
similar position as in F7A, hence we obtain only a tentative flag=3 detection.
We have no convincing explanation for the weakness of the peak detected in this
field: it may be related to the complex structure of the various Stream wraps or
to a local dip in the density along the Stream. However, the derived distance is
in good agreement with the trend observed in others Branch-A fields. In the SCPs
of F9A field we identify again two peaks, the faintest one being very prominent
and wide; also this \los intercepts a region of crossing wraps, thus superposed
structures may contribute to the production of a strong and remarkably wide
primary peak.  The weaker/brighter peak is more interesting: it is clearly
identified in both SCPs, even if below the $3\sigma$ threshold, at $V \sim 17.20$
($I \sim 16.30$): as discussed later in Sect.~\ref{models}  this feature may
trace a  near wrap of the Stream that was never detected before. 

SCPs of fields from F10A to F12A show two peaks at similar positions, with a
remarkable variety of absolute and relative strengths. This may reflect the
highly structured morphology that is suggested by models in this region (see
Sect.~\ref{models}). F11A is crossed by the Orphan Stream
\citep[Bel06,][]{belorp}. While the distance of this structure ($\sim 30$ kpc
toward this \los) does not match with the detected peaks we cannot rule out some
contamination from Orphan Stream stars in this field.

We did not find any convincing signal in F13A, F14A and F15A; the overall shape
of the SCPs appear quite different from the other cases and, in the case of F15A,
the polynomial model did not provide a satisfactory fit to the background
population. In particular, the SCPs present a strong excess at bright magnitudes
($V\la 18$, $I\la 17$) with respect to those in the previously-discussed fields,
such that they appear flat or even decreasing with increasing magnitude.  These
are the fields at the lowest galactic latitude, hence we attribute these
features to contamination by (relatively) nearby stars from the Galactic thick
disk (and, possibly, the Monoceros structure, see Fig.~1 of Bel06) that
overwhelms the signal from the Sgr Stream RC. This seems confirmed by the
comparison with the corresponding CFs (the case of F15A is shown in
Fig.~\ref{cfrV}, in Sect.~\ref{scp_cf}), that display SCPs essentially
indistinguishable from those of the on-Stream fields. This implies that the
adopted technique can be used successfully only at large distances from the
Galactic plane. Given the above reasons we preferred not to consider for further
analysis the possible peaks at V(I)$\sim 16(15)$ and V(I)$\sim17.5(16.5)$ in
F14A.

\subsection{Branch B detections}
\label{detB}

In branch B we analyzed 13 fields, obtaining a total number of 26 detections,
with the following classification: 16 primary, 5 secondary and 5 tertiary peaks.
The main structures found in branch A are mirrored also in branch B, as clearly
shown in Fig.~\ref{ldist}. In all the fields (except F7B) we detect at least two
peaks; in two cases (F8B and F12B) we also detect a third peak; in F8B this is
likely tracing the more distant Stream wrap running nearly parallel to the main
wrap of the leading arm (see above and Fig.~\ref{ldist}); in another case (F5B),
in addition to three peaks analogous to those in F8B, we found an additional
nearby peak (see Sect.~\ref{models} for a discussion). An alternative
interpretation for the SCPs of F12B is presented in Sect.~\ref{detS}. Quite
suprisingly, the  SCPs of F6B appear completely smooth and featureless. In this
case we were not able to find an explanation for this behavior (but see above
for the discussion of the similar case of F8A).

In analogy with F7A (and F8A) F7B is the only case of branch B SCPs fitted with a
single peak model. The morphology of the $I$-band SCP and the comparison with the
adjacent F8B field suggest that two, or possibly three peaks may be merged
together in this SCP. However we were unable to resolve the peak into separate
components even in SCPs with smaller bin width (as for the case of F6A). We
caution the reader that this primary detection is likely concealing significant
--- but as yet undetected --- substructure.

\begin{figure*}
\plottwo{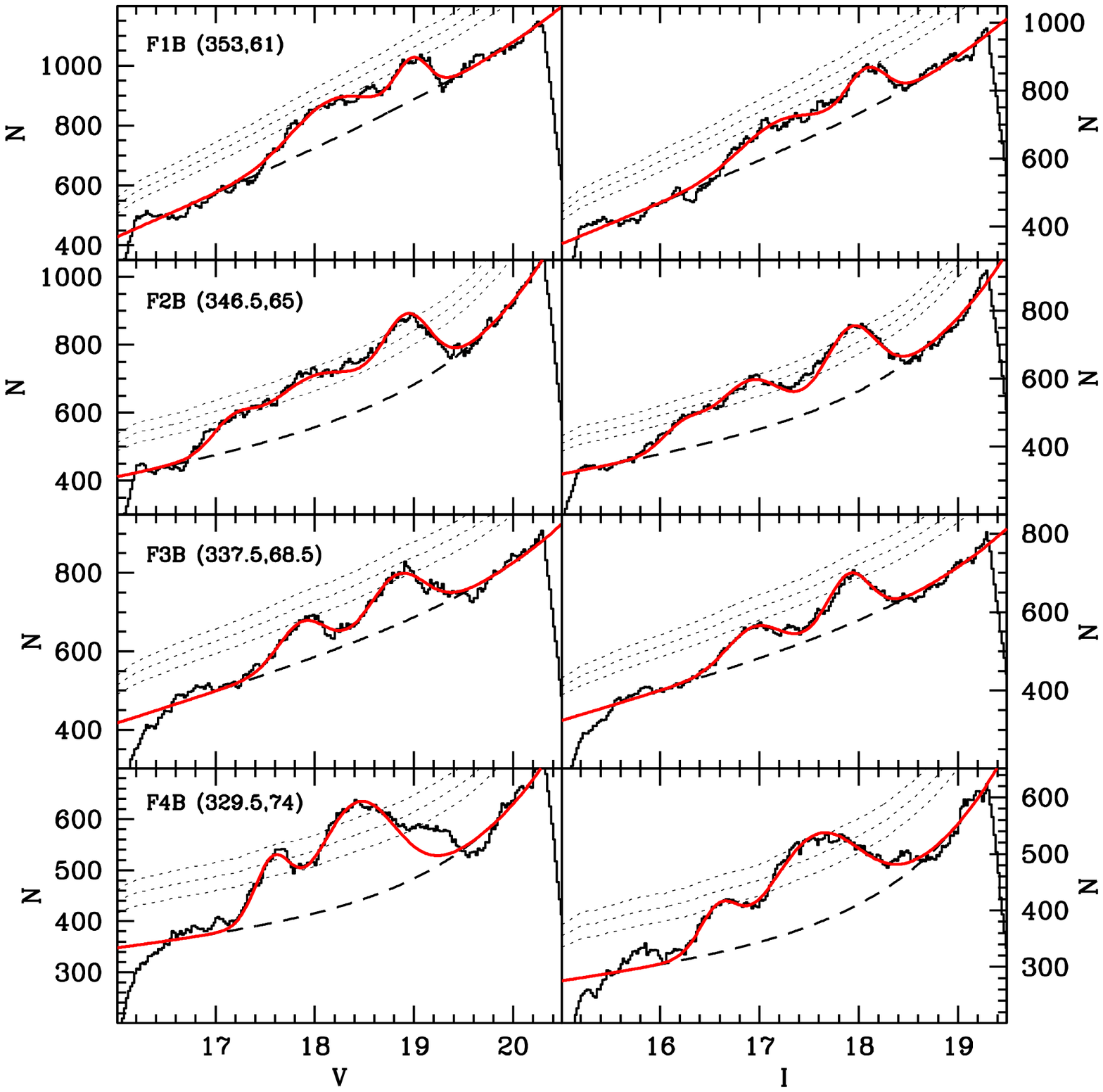}{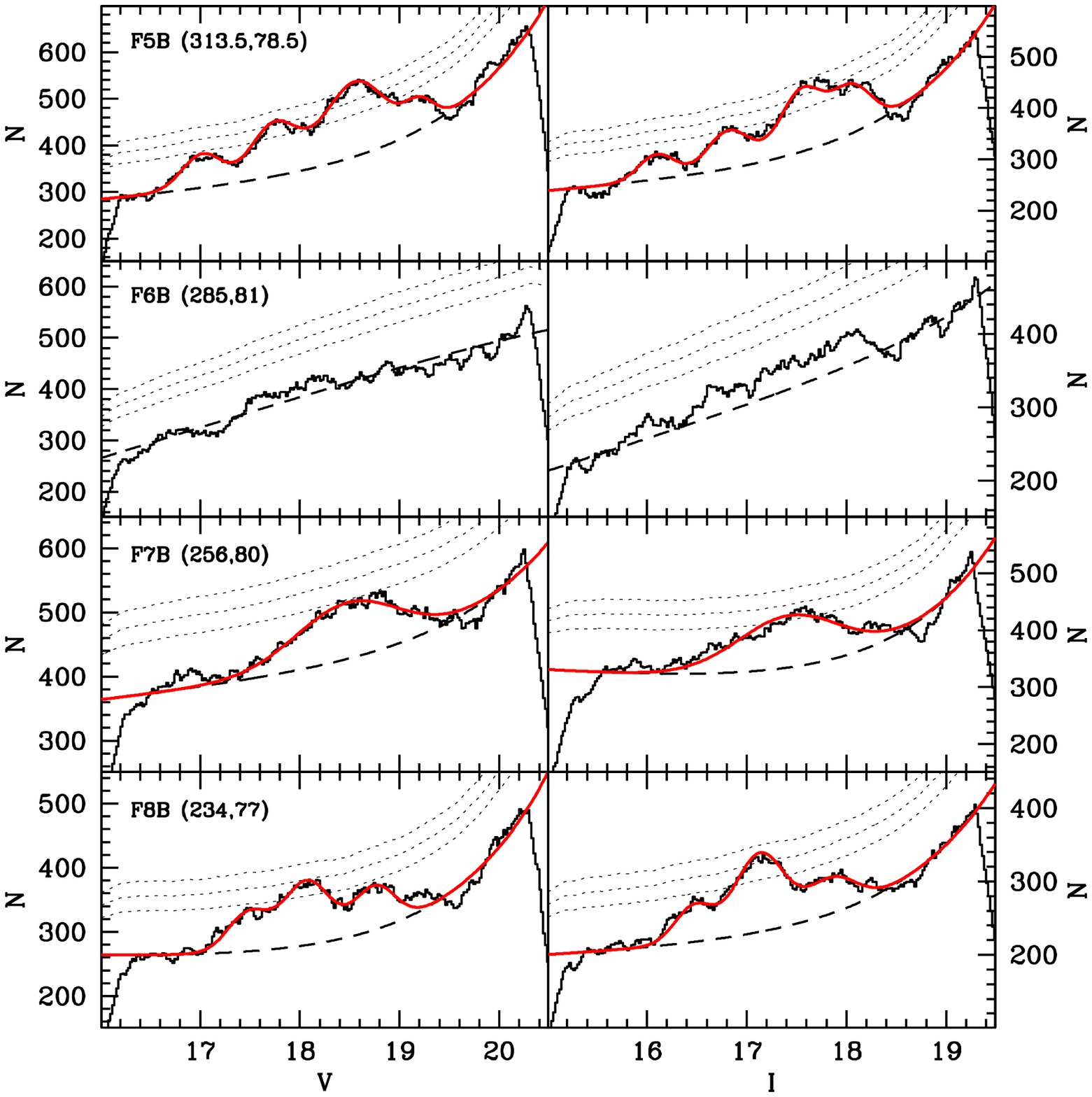}
\plottwo{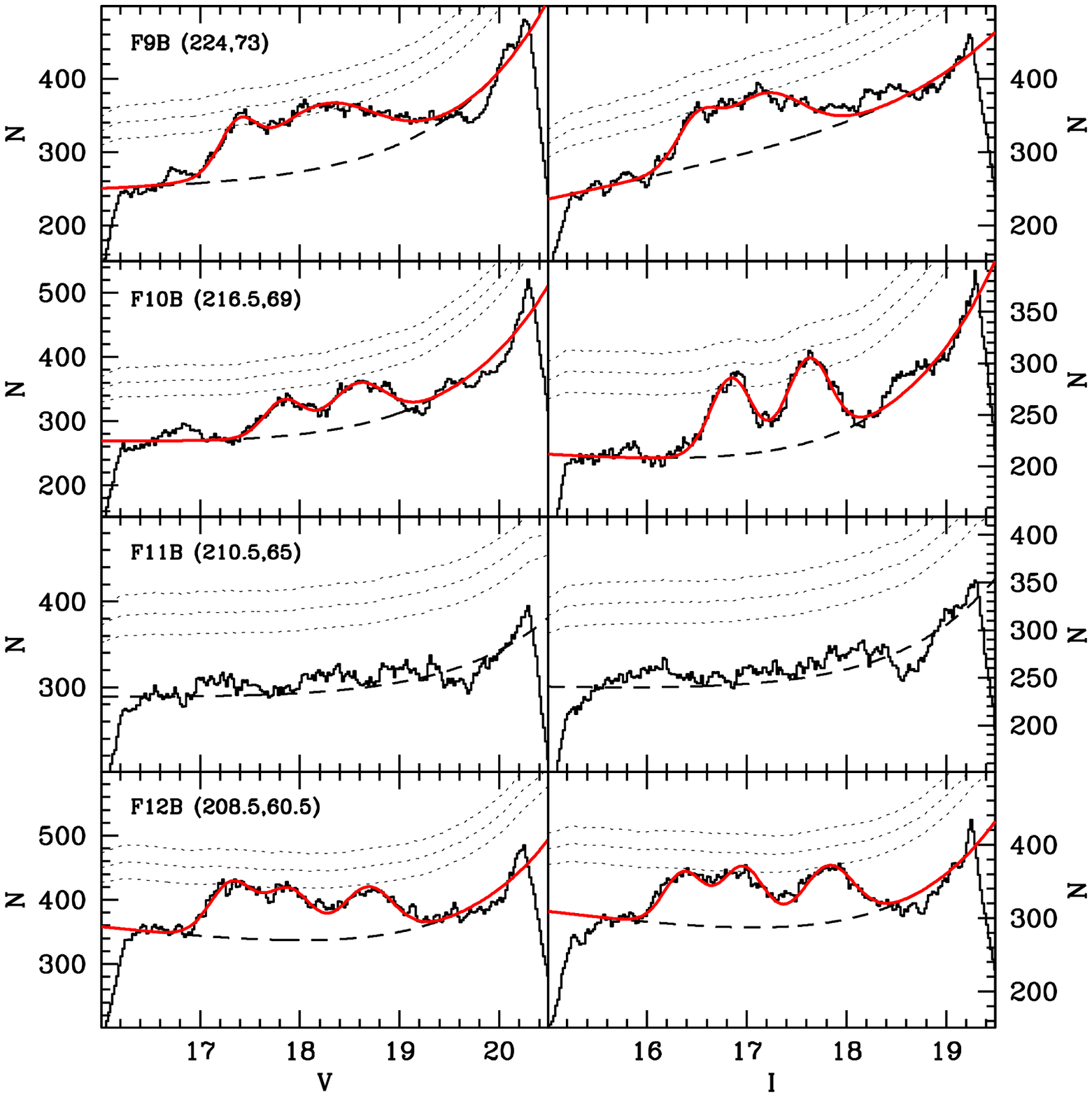}{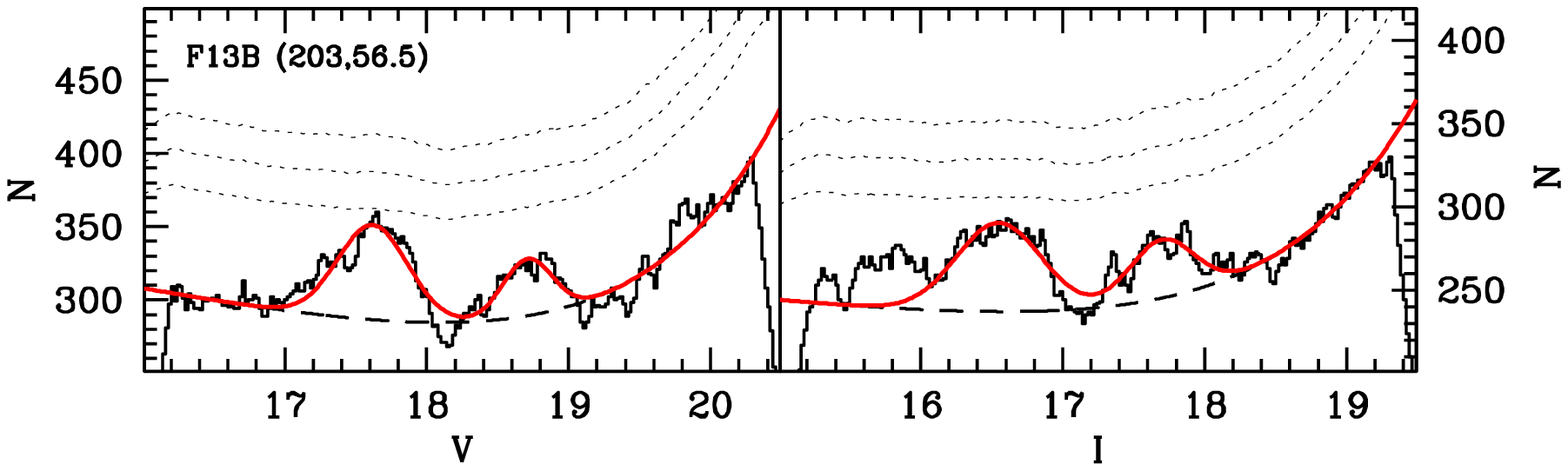}
\caption{The same as Fig.~\ref{LFA} but for fields on Branch B of the Sgr
Stream.  No global fit has been attempted for SCPs  that did not show significant
RC peaks (F6B, F11B).}
\label{LFB}
\end{figure*}

\subsection{A few special cases}
\label{detS}

\begin{figure}
\plotone{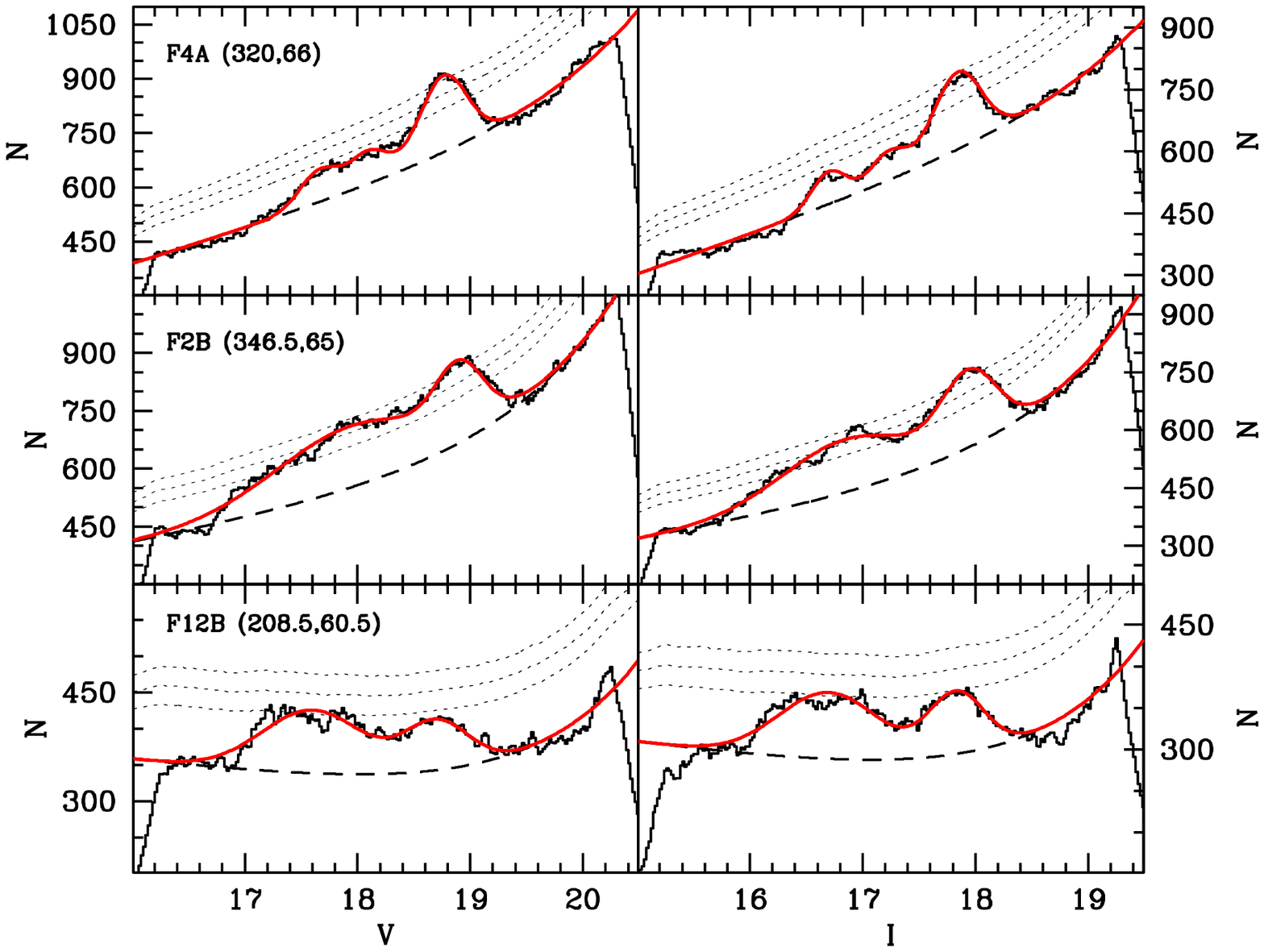}
\caption{Fits of the observed SCPs (in $V$ and $I$) for fields of the branch A
and B of Sgr Stream for which we present acceptable alternative models, with respect to
the solutions shown in Figs.~\ref{LFA} and Figs.~\ref{LFB}. The numbers in
parentheses are the Galactic longitude and latitude of the center of the field,
in degrees. The meaning of the symbols is the same as in Figg.~\ref{fit7A} and
\ref{fit5A}, above.}
\label{lf_spc}
\end{figure}

There are a few cases in which the observed SCPs do not provide unequivocal
indications for the model to be adopted, in particular concerning the number of
$G_i(x)$ functions to be included in the model, i.e. the number of detected
peaks. F7B, briefly discussed above, is the only case in which we feel that the
observed peak is due to the merging of two (or, more likely, three) adjacent
peaks that we cannot resolve. In Fig.~\ref{lf_spc} we present acceptable
alternative models (with respect to the solutions shown in Figs.~\ref{LFA} and
\ref{LFB}, and listed in Tab.~\ref{tab_dmag}) for the three cases in which our
preference for the adopted models (Tab.~\ref{tab_dmag}) is only marginal, and is
also supported by the continuity within a large scale structure (a Stream wrap).
In Tab.~\ref{tab_spc} we report the corresponding alternative solutions, that
can be replaced with those of Tab.~\ref{tab_dmag} by those readers who may use
our values to constrain models of Sgr, if they judge them more appropriate,  for
some reason.

\begin{table*}
\begin{center}
\caption{Alternative solutions for three \los.}
\tiny
\begin{tabular}{c|cc|cccc|cc|cc|c|cc|c}
\hline
field & $l^{\circ}$ & $b^{\circ}$ & $\Delta V$ & $\epsilon\Delta V$ & $\Delta I$  & $\epsilon\Delta I$ & d  &  $\epsilon$d   &  sign. (V)  & sign. (I) & FWHM  & N$_{RC}$  & $\epsilon$N & flag  \\
      &  deg  & deg  &    mag  &    mag  &    mag  &    mag  &  kpc  &  
kpc &        &        &  kpc  & [stars/25 deg$^2$]  & [stars/25 deg$^2$]  &  1 \\ 
&&&&&&&&&&&&&&\\
\hline
  4A  &  320  &  66  &  0.94  &  0.04  &  0.97  &  0.05  &  40.8  &
4.2 &  $\geq 5\sigma$ &   $5\sigma$    &  7.5  &  158 &  8  &  1  \\
      &       &      &  0.28  &  0.07  &  0.36  &  0.10  &  30.5  &
3.5 &   $<3\sigma$   &  $<3\sigma$     &  3.2  &  119 &  7  &  3  \\
      &       &      & -0.16  &  0.09  & -0.17  &  0.09  &  24.4  &
2.8 &   $<3\sigma$   &  $<3\sigma$     &  1.9  &  107 &  6  &  3  \\
\hline
  2B  & 346.5 &  65  &  1.08  &  0.09  &  1.08  &  0.06  &  43.3  &
4.8 &  $\geq 5\sigma$ & $\geq 5\sigma$ &  9.3  &  187 &  10 &  1  \\ 
      &       &      &  0.13  &  0.15  &  0.08  &  0.12  &  27.6  &
3.7 &   $5\sigma$     &  $5\sigma$     & 14.0  &  499 &  14 &  1  \\
\hline
 12B  & 208.5 & 60.5 &  0.88  &  0.15  &  0.94  &  0.15  &  40.0  &
5.6 &  $<3\sigma$    &  $3\sigma$      &  8.1  &  92  &  6  &  2  \\ 
      &       &      & -0.21  &  0.15  & -0.17  &  0.15  &  24.1  &
3.3 &  $3\sigma$     &  $3\sigma$      & 12.0  &  155 &  9  &  1  \\
\hline
  \end{tabular}
\label{tab_spc}
\end{center}
\end{table*}

\subsection{Intra-Branch fields}
\label{intra}

In the present analysis we do not consider the structure of the Stream in the
Dec direction. We fully adopted the view of \citet{FoS}, where the leading arm
of the Stream as seen from TO stars in the SDSS bifurcates into branch A and
branch B around RA=220$\degr$ and the separation between the branches increases
with decreasing RA. We proceeded to a basic verification of this scenario by
looking at the SCPs of a few Intra-Branch (I) fields (not shown here, for
brevity), located at intermediate Dec with respect to the A and B fields F5, F7,
F10 and F12. In agreement with the results of \citet{FoS}, we find that the SCPs
of F5I and F7I mimic the structure of the SCPs of the corresponding A and B
fields, showing peaks at the same position and with similar shape, but weaker
than in the on-Stream fields (i.e. tracing a lower stellar density). In the SCPs
of F10I the peaks seen in the A and B SCPs are just barely visible and they
completely disappear in F12I. Hence, these limited set of tests confirm the
reality and the morphology of the Stream bifurcation as observed by \citet{FoS}.

\subsection{The color of the RC peaks}
\label{color}


\begin{figure}
\plotone{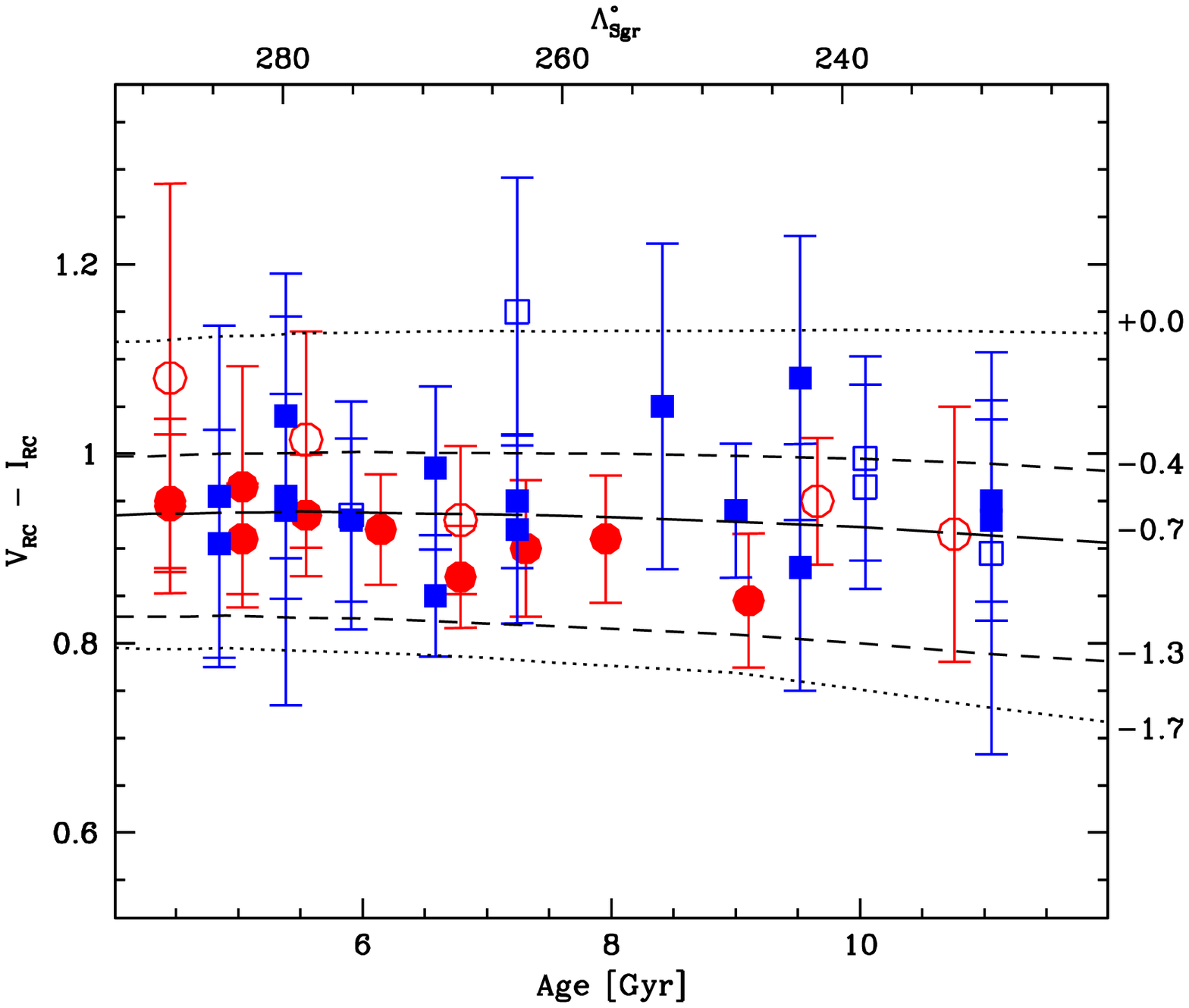}
\caption{De-reddened color of the observed RC peaks as a function of
$\Lambda_{Sgr}$. The observed points are compared with the predictions of the
\citet{gs} theoretical models for different metallicities, plotted as a function
of age of the population in the range 4-12 Gyr. The scale on the lower
horizontal axis (age) refers to models, that on the upper horizontal axis
($\Lambda_{Sgr}$) refers to observed points.}
\label{viRC}
\end{figure}

The color of the RC peak is known to be very sensitive to metallicity and weakly
sensitive to age, in the range of ages relevant for the present study
\citep[4-12 Gyr,][]{gs}. As our procedure of independent peak detections in $V$-
and $I$-band SCPs automatically provides the colors of the RC peaks, it is worth
checking if there is any hint of a color (metallicity) gradient along the
Stream. In Fig.~\ref{viRC} the colors of the observed peaks are compared to the
theoretical models by \citet{gs}. All of the peak detections shown in
Fig.~\ref{viRC} are compatible with having the same color within the
uncertainties (that are quite large for some non-primary peaks). It is
interesting to note that the large majority of points cluster around the
$[M/H]=-0.7$ model, in good agreement with the results by \citet{B06a} and
\citet{carrell} (see also Sect.~\ref{SelCMD}). No significant trend of color
(metallicity) of the RC population with orbital azimuth is apparent and the few
points showing the larger color difference from the mean locus are always among
those having the most uncertain color estimates.

\section{Comparison with previous analyses}
\label{comp}

Before discussing in detail the comparison between our distance estimates and
the findings from previous works, it is worth considering the difference between
the performances of the various adopted tracers. The intrinsic stability (and
ubiquity along the Stream) of our standard candle (RC stars),  the adopted
analysis, best suited for the detection and location of RC peaks, and the purely
differential nature (Stream vs. main body) of our measures, make our distance
estimates the most comprehensive, accurate and {\em homogeneous} set publicly
available (even if limited to the region of sensitivity described above). The
uncertainties associated with our estimates are lower than any previous work,
with a typical values of $\leq 5\%$, raising to $\leq 10\%$ in the worst cases.
For example, \citet{maj03} reports that the characteristic uncertainty of their
photometric parallaxes based on M-giant is $\simeq 20$\%; \citet{cma} showed
that uncertainties in the age/metallicity of the considered populations may lead
to systematics of order $\sim 30$\% in the distance scale based on M giants. F
stars (assumed to be TO stars of Sgr) proved to be an excellent mean to trace
even very feeble substructures \citep{FoS,heidivirgo}. However the assumption of
a common absolute magnitude for all color-selected F stars implies large
uncertainties, as these stars span a range of luminosities much larger than RC
stars.  For example, if we consider the distribution in $V$ magnitude of (a) the
RC selected with our color window, and (b) the MSTO stars selected in color as
done by Bel06 (and limited to $V>20.0$) in the photometry of the {\it Sgr34} field, we
found two obvious single peaked distributions, but while the FWHM of the RC peak
is $\simeq 0.3$~mag, the MSTO star peak has $FWHM\simeq 2$~mag.   Indeed,
\cite{cole08}, in their pilot project on stripe 82, showed that the assumption
of a fixed magnitude for these stars may lead to very large errors. Blue
Horizontal Branch stars are easier to select against the Galactic
fore/background, but are rarer than RC stars. Moreover, even if selected in a
color range where the Horizontal Branch is really nearly horizontal, the
distribution in magnitude of these stars is not expected to be as clearly peaked
as the RC (see B06c). In this sense, the Sub Giant Branch (SGB), used by Bel06
and \citet{kellerev}, is more promising, as it is a very narrow feature in CMDs
of metal rich populations. However it should be much more sensitive to
metallicity and age variations than the RC (see, for example, B06a, and
references and discussion therein), and being much (intrinsically) fainter, its
use is limited to a lower distance range, for any given dataset. Finally, RR
Lyrae stars \citep{ive00,vivas,prisag,kellsag} can provide distances with even
superior accuracy with respect to our method; well sampled light curves can also
give indications on physical properties of individual stars (metallicity, for
example) that cannot be obtained from RC stars. However RR Lyrae are (likely)
less frequent than RC stars over most of the Stream extension and, above all,
they need time series information to be safely identified and to obtain a
reliable apparent magnitude averaged over the pulsation period: for this
reason   the available data cover a much smaller region of the sky with respect
to generic ``single epoch'' standard candles. 

\subsection{Comparison with specific detections}

\citet{yan00} were the first to interpret a stellar over-density in the halo as
possibly due to the Sgr Stream. In the first available (equatorial) stripe of
the SDSS they identified an excess of A-type stars around $\Lambda_{\sun} \sim
295\degr$, adjacent to our field F1A.  The heliocentric distance inferred is of
48 kpc, in good agreement with our estimate for the main wrap of the leading arm in
this direction ($D\simeq 45$ kpc at $\Lambda_{Sgr}\simeq 290\degr$). This
result was later confirmed by the more thorough study by \citet{heidi}, that
used F stars as main tracers. Although they do not comment on it, the
\citet{yan00} data also showed an excess of A-type stars less than 20 kpc away
along the same \los (see their Fig.~18 and 19). This may be more easily
identified with the constant-distance coherent structure we see at $d\simeq
25$~kpc than to the nearest wrap that we (possibly) detect at $\Lambda\simeq
287\degr$ and $d\simeq 13$ kpc.

A similar detection of two density enhancements toward the Northern Loop was
reported by \cite{ive00}, from the study of RR Lyrae in the same SDSS stripe
studied by \citet{yan00}, and by \citet{vivas}, also using RR Lyrae  from the
QUEST RR survey,  which explored nearly the same region of sky ($\Lambda_{\sun}
\sim 270\degr-290\degr$). Both studies comment primarily on an excess of RR
Lyrae stars at 45-50 kpc (corresponding to the main wrap of the leading arm);
however a structure around $\sim 20$ kpc is also noted. 

\citet{maj03}, provided a clear panoramic view of the Sgr Stream using M giants
as standard candles; they were able to trace very neatly the trailing tail all
over the Southern Galactic hemisphere, as well as part of the leading arm closer
to the main body of the galaxy, up to $RA\simeq 190\degr$. They report two cases
of M giants excess along the \los in common with the present analysis. The most
evident at a distance  $D \sim 45$ kpc, compatible with our estimates, and the
other one, less pronounced, at a distance $D \sim 20-25$ kpc, for which the
interpretation is not so clear as in the case of A stars and RR Lyrae
detections. 

All the detections mentioned above, as well as others toward specific
directions, also not included in the range considered here
\citep{david01,david,sgrclus2,vivas}, are collected and reported in Fig.~17 of
M03. This figure, as well as Fig.~19 in \citet{law10}, clearly illustrates how
it may be difficult and misleading to put results from different sources (and on
different distance scales) all together. In this sense, it is more fruitful to
compare our results with other data sets providing homogeneous distance
estimates for significant portions of some wrap in common with those detected
here.

For instance, Bel06, who used A-F dwarf stars from the SDSS to trace the Stream,
detected a distant gradient along the main wrap of the leading arm that is in
good agreement with our results (for both branches). More interestingly, Bel06
found a double detection in a few branch A \los (from F5A to F7A): in addition 
to the main wrap of the leading arm, they found also a more distant structure,
$\sim 15$ kpc behind. This finding is also  in excellent agreement with our
results (see Sect.~\ref{detA}). The only difference is that we detect this
structure, at similar distance, also in the corresponding branch B fields.

\begin{figure}
\plotone{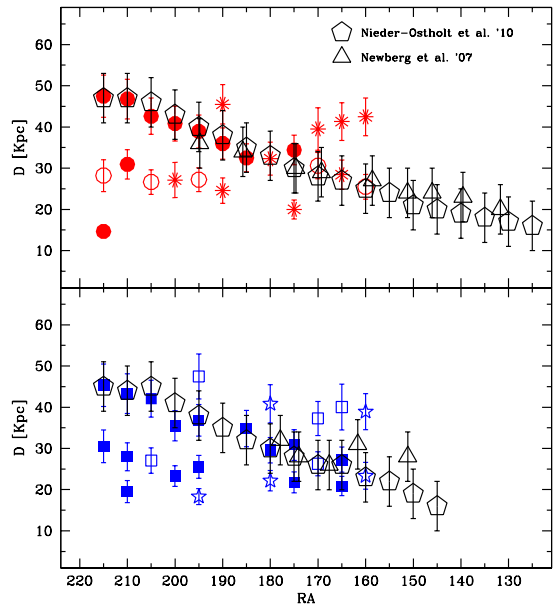}
\caption{Comparison between our distance estimates (same symbols as in
Fig.~\ref{ldist})  and those by \citet[][open triangles]{heidivirgo} and by
\citet[][open diamonds]{nied}. The match between the different indicators is
remarkably good, for the portions of the leading arm considered by the three
studies. The trailing arm is detected only in the present study.}
\label{comp_dist}
\end{figure}

\citet{heidivirgo}  investigated the relationship between several
previously-identified  substructures in the direction of Virgo and the Sgr
Stream using imaging and spectroscopic observations of F stars and BHB stars
from SDSS and SEGUE. In their Tab.~1, they reported the detections associated to
the Sgr Stream, providing also estimates of the distance of these structures.
This allowed us to perform the direct comparison with our results that is
presented in Fig.~\ref{comp_dist}. The agreement for the structure detected in
both studies (main wrap of the leading arm) is very good, both for branch A and
B. 

\cite{nied} investigated the leading arm of Sgr Stream in the same region of the
``field of stream'' analyzed by Bel06, using BHB candidates from the SDSS.  The
distances to BHBs are calculated assuming an absolute magnitude $M_g=0.7$ and
are in good agreement with the results of Bel06. In Fig.~\ref{comp_dist} we
compare our estimates with those obtained by \cite{nied}, reported in their
Tab.~1, in the region of sky going from $RA \geq 160\degr$ to $RA \leq
220\degr$. Our distances are calculated using a true distance modulus of 
$(m-M)_0=17.10$, roughly the same adopted by \cite{nied} (the $g$ magnitude of
the BHB in the main body is $g \sim 17.80$). In all the regions where the
data-sets overlap the match is very good, both for branch A and B detections.
The trailing arm of Sgr Stream is not sampled by \cite{nied} that concentrated
their analysis on the detection of the leading arm and on an accurate distance
estimate for this wrap of the Stream.  

In conclusion, the overall agreement with previous detections of the leading arm
is very good. The situation for the other coherent structures detected here is
more difficult to judge; in our view the only firm conclusion that can be drawn
is that several independent studies found evidence for some structures located in
front of the main wrap of the leading arm, in the considered range of $\Lambda$.
It is unclear if some of these detections can be associated with the constant
distant (putative) wrap of the trailing arm detected here or to even more nearby
wraps. In this sense it is interesting to note that a similar coherent
structure, at a similar distance, is detected also by \citet{kellerev}, using
SGB stars (see his Fig.~7).

\section{Comparison with models}
\label{models}

As soon as it was realized that Sgr was likely undergoing tidal disruption,
several authors attempted to model the process by means of N-body simulations,
to establish the plausibility of proposed models and to infer the properties
(mass, orbit) of the original system
\citep{velaz,kat95,iba97,edel,ibamodel,gomez,kat99,jiang,helmi01}. It is
interesting to note that \citet{velaz} were able to provide estimates of
perigalactic and apogalactic distances and orbital period remarkably similar to
those obtained in the most recent studies, just one year after the discovery of
Sgr  \cite[$R_{peri}\simeq 10$ kpc, $R_{apo}\simeq 52$ kpc and $P_{orb}\simeq 0.76$ Gyr, to compare, for instance, with $R_{peri}\simeq 15$ kpc,
$R_{apo}\simeq 60$ kpc and $P_{orb}\simeq 0.85$ Gyr, from][]{law}. The possible
r\^ole of Sgr in the formation of the Galactic Disk warp was studied by
\citet{razum} and \citet{bailin}.

However, since the Sgr Stream appears as a remarkably-coherent structure crossing
a large part of the Galactic halo on a nearly-polar orbit, it seems the ideal
tracer to study the overall shape and the degree of clumping of the Galactic
halo as a whole. For this reason, the most recent N-body modeling efforts have
focused on constraining the shape of the DM halo of the Milky
Way \citep{iba01b,david,aminashape,law,kat05,fell,law2,law10}. However it turned out that
the conclusions of these studies depended on the specific set of observational
constraints considered, and it is now generally accepted the idea that none of
the static-potential axisymmetric halo models considered is able to reproduce
simultaneously all the available positional and kinematic data \cite[see][for
references and discussion]{yan09,law2}. In a recent contribution \citet{law2}
anticipated that the adoption of triaxial halo models can help to solve this
problem: in Sect.~\ref{tri} we briefly consider the N-body model 
they produced as a follow-up of that analysis \citep{law10}.  In any case, it is quite clear that currently available models are far
from perfect, and more detailed simulations are needed to extract all the
possible information on the Galactic DM halo from the Sgr Stream, as more (and
more accurate) observational constraints become available. For example,
\citet{fell} interpreted the bifurcation of the trailing arm giving rise to the
A and B branches considered here as produced by the precession between two
subsequent orbits. As the implied amount of precession is relatively small,
this, in turn, requires that the potential felt by Sgr should be nearly
spherical. However the similarity between the two branches (in terms of
distance, kinematics and stellar content) led \citet{yan09} to suggest that in
fact the two branches are composed by stars lost at the same epoch, i.e. they
are in the same orbital phase. In this case the separation between the two
branches would not be related to orbital precession and would have nothing to
say about the shape of the potential. In their recent analysis, \citet{nied} adopt the same view as \citet{yan09}\footnote{See also the discussion in \citet{law10}.}.

\begin{figure}
\plotone{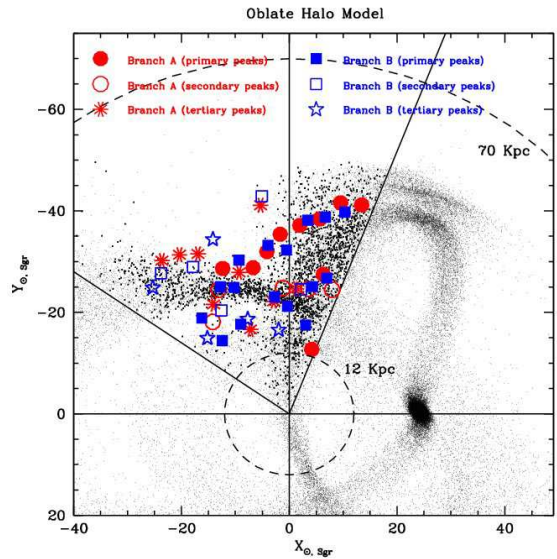}
\caption{Comparison between the positions of the observed RC peaks and the
predictions of Law et al.'s N-body model of the disruption of Sgr within an {\em
oblate} DM halo in the $X_{\sun, Sgr}$ vs. $Y_{\sun, Sgr}$ plane (i.e. the
approximate plane of Sgr orbit seen face-on). Heavier dots indicate model
particles enclosed in our FoVs cones.  To be consistent with \citet{law} we
converted our differential distances into absolute ones by assuming a true
distance modulus of 16.90 for the main body of Sgr. }
\label{xy_obl}
\end{figure}

\begin{figure}
\plotone{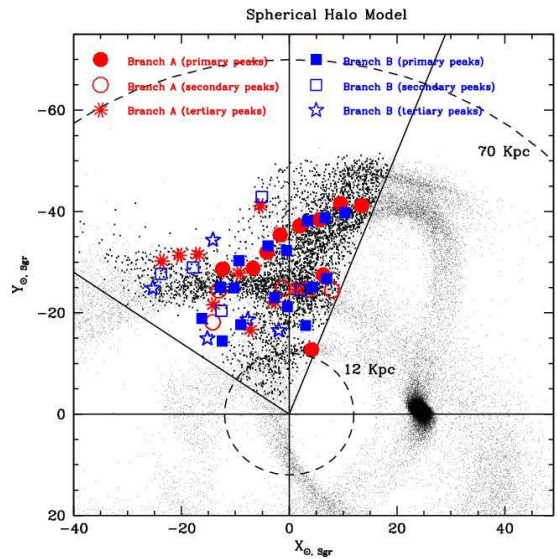}
\caption{The same as Fig.~\ref{xy_obl}, but for a N-body model within a {\em
spherical} DM halo.}
\label{xy_sph}
\end{figure}

\begin{figure*}
\plotone{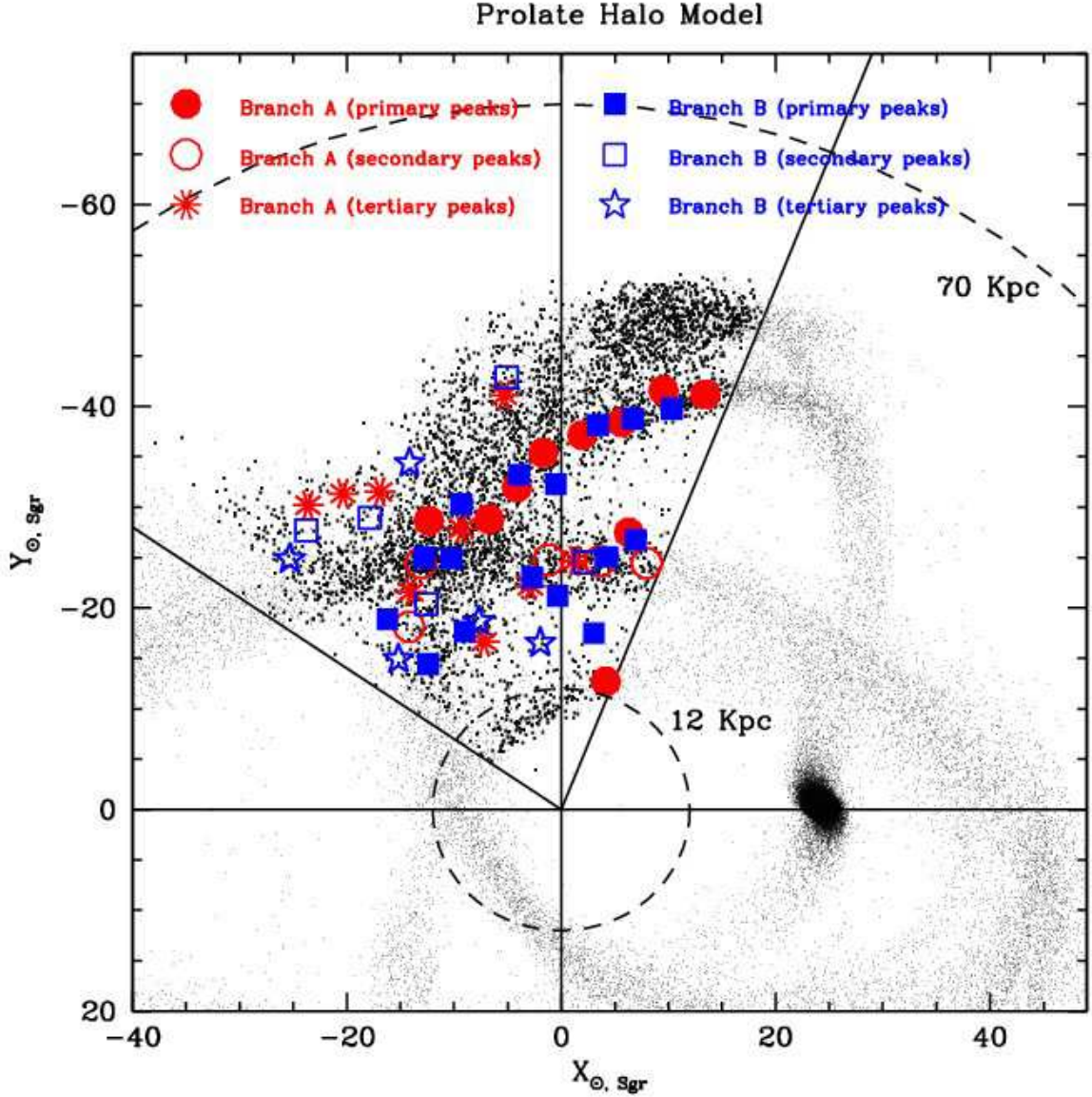}
\caption{The same as Fig.~\ref{xy_obl} and \ref{xy_sph}, but for a N-body model
within a {\em prolate} DM halo.}
\label{xy_pro}
\end{figure*}

\begin{figure}
\plotone{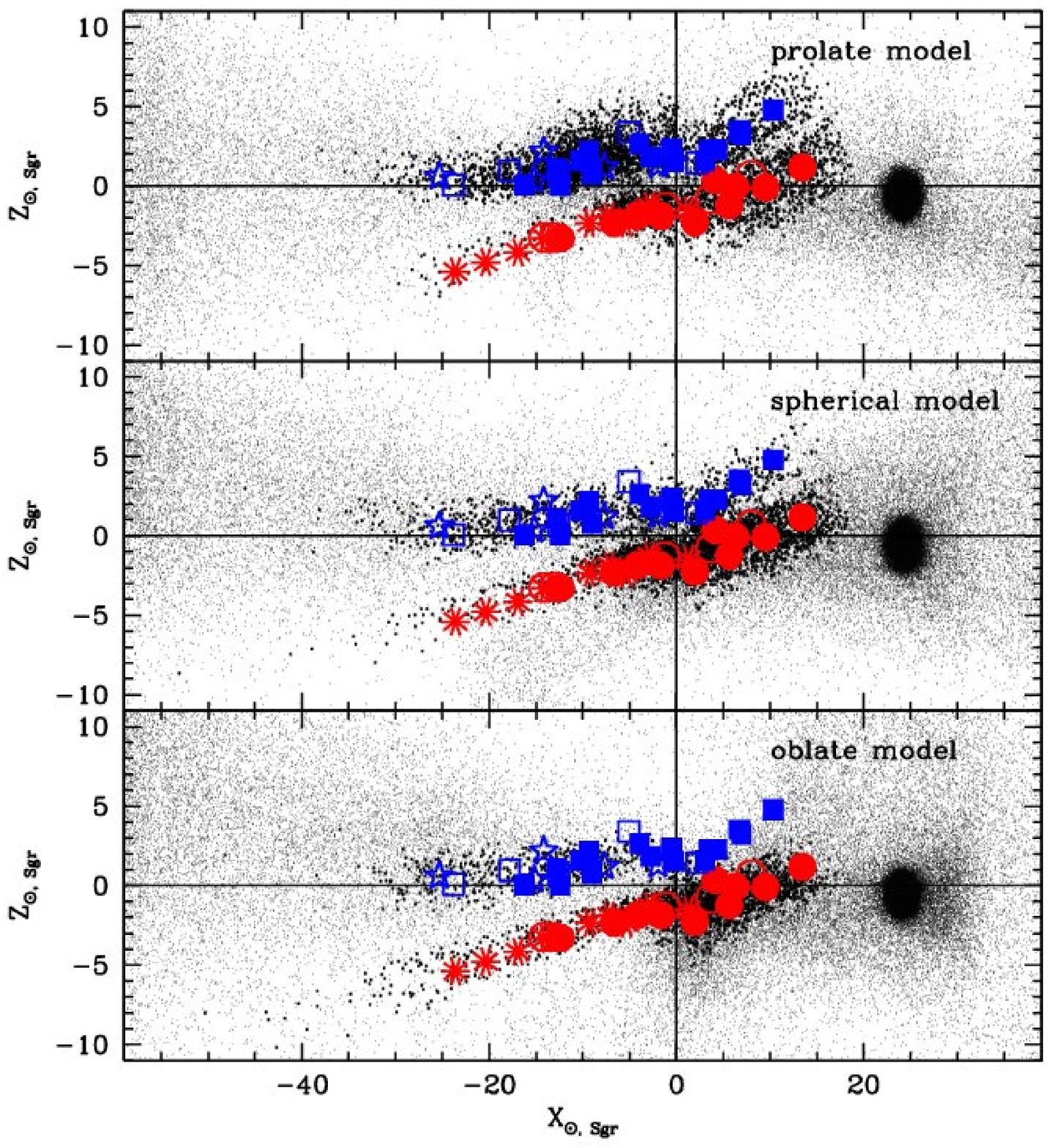}
\caption{Comparison of observed peaks positions with Law et al.'s models in the
$X_{\sun, Sgr}$ vs. $Z_{\sun, Sgr}$ plane (i.e. perpendicular to the orbital
plane of Sgr, that is seen edge-on). The symbols are the same as in
Fig.~\ref{xy_obl}.}
\label{xz}
\end{figure}

In the present contribution we provide very accurate distance estimates along
the northern branches of the Sgr Stream as powerful constraints for future
generations of Sgr disruption models that will include effects like halo
triaxiality, dynamical friction, time-evolving Galactic potential, etc. In this
section we discuss our findings in  comparison with the predictions of the three
models by \citet[][L05 hereafter]{law}, just to show how powerful accurate
distance constraints can be in distinguishing between different models (some
examples of such comparisons have already been presented in
Sect.~\ref{exonstream} and Fig.~\ref{ldist}). One of the main aims of the
studies by L05 and  \citet{kat05}, was to use the existing observations on the
Sgr Stream to constrain the shape of the Galactic halo. For this reason the
three models they provide\footnote{Publicly available at {\tt
http://www.astro.virginia.edu/~srm4n/Sgr}} describe the final state of the
evolution of a realistic progenitor of Sgr after a few orbits within a Galactic
potential having a {\em prolate}, {\em spherical}, or {\em oblate} DM halo. For
sake of simplicity, in the following we will refer to these models as to the
Oblate (O), Spherical (S) and Prolate (P) models, respectively. In all the
models, each particle is flagged according to the peri-galactic passage in which
it become unbound from the main body of the galaxy. Here we refer to stars still
bound or lost during the current peri-galactic passage as having $p=0$; $p=-1,
-2, -3, -4$ refers to particles lost one, two, three and four peri-galactic
passages ago, respectively. $p=0$ stars are out of the range accessible with the
fields considered here, according to the L05 models. When we speak of ``young''
and ``old'' wraps of the Stream we refer to portions of the Stream whose
population is dominated by particles lost in the most recent or less recent
peri-galactic passages, respectively, on an age-scale encompassing the last
$\sim 5$ orbits, i.e. $\sim 3-4$ Gyr.

In Figg.~\ref{xy_obl}, \ref{xy_sph} and \ref{xy_pro} the three models are
compared with the positions of the observed RC peaks in the $X_{\sun, Sgr}$ vs.
$Y_{\sun, Sgr}$ plane\footnote{For brevity, in the following we will drop the
${\sun, Sgr}$ index any time we found this convenient; $X_{\sun, Sgr}$,
$Y_{\sun, Sgr}$ and $X$, $Y$ are interchangeable. For the same reason, the
values of $X$, $Y$ must be always intended as expressed in kpc, even if not
explicitly stated.}, as in Fig.~\ref{XY5A}, above. It is immediately apparent
from Fig.~\ref{xy_obl} that the trend traced by our primary peaks rules out the
O model, that fails to reproduce the most prominent branch of the Stream seen in
SDSS data, i.e. the portion of the leading arm descending from the North
Galactic Pole (while there is some agreement for the - putative - nearby portion
of the trailing arm).  The case of the spherical model is similar, even if the
disagreement between observations and model prediction is less severe
(Fig.~\ref{xy_sph}). 

\begin{figure*}
\plottwo{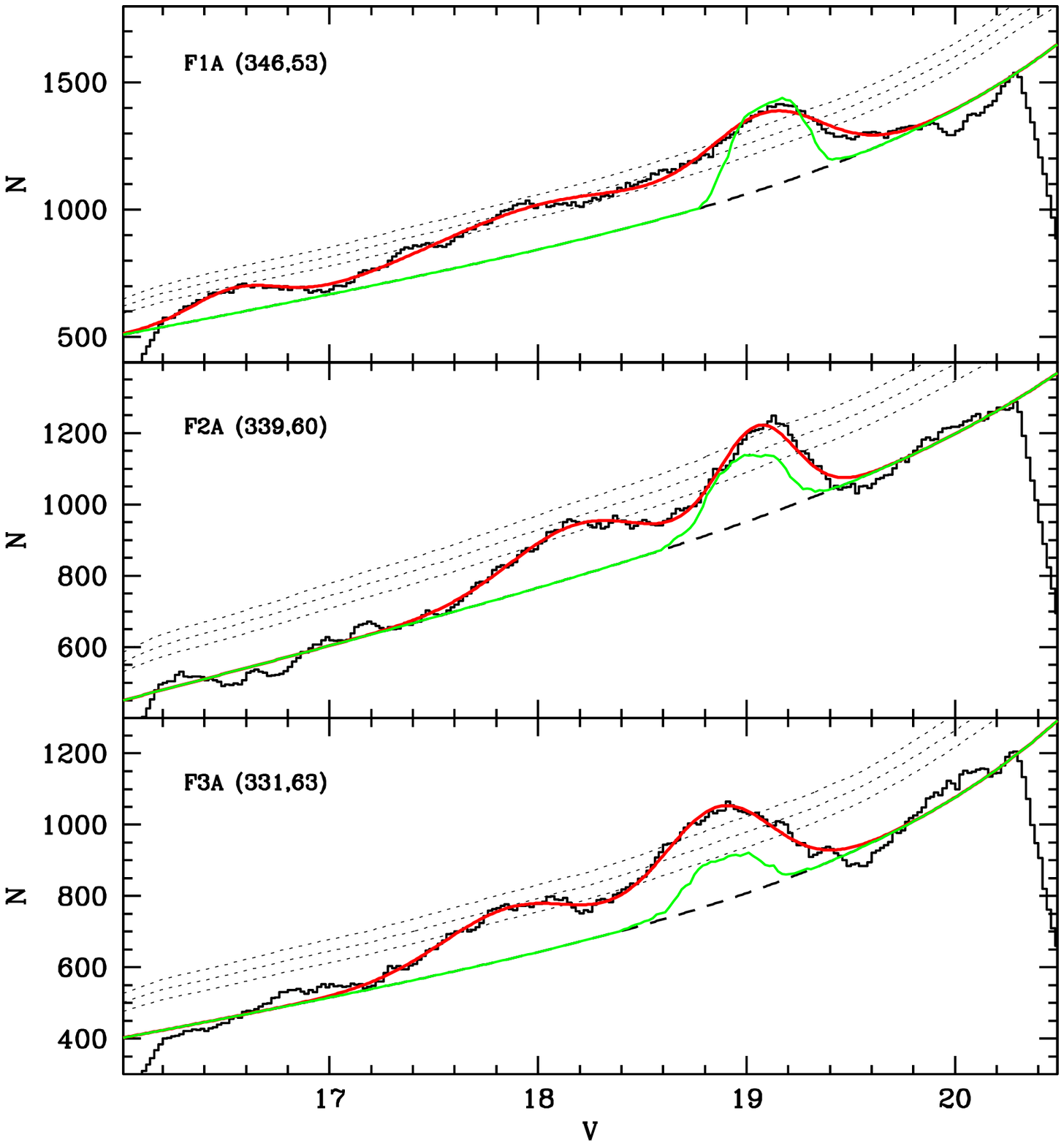}{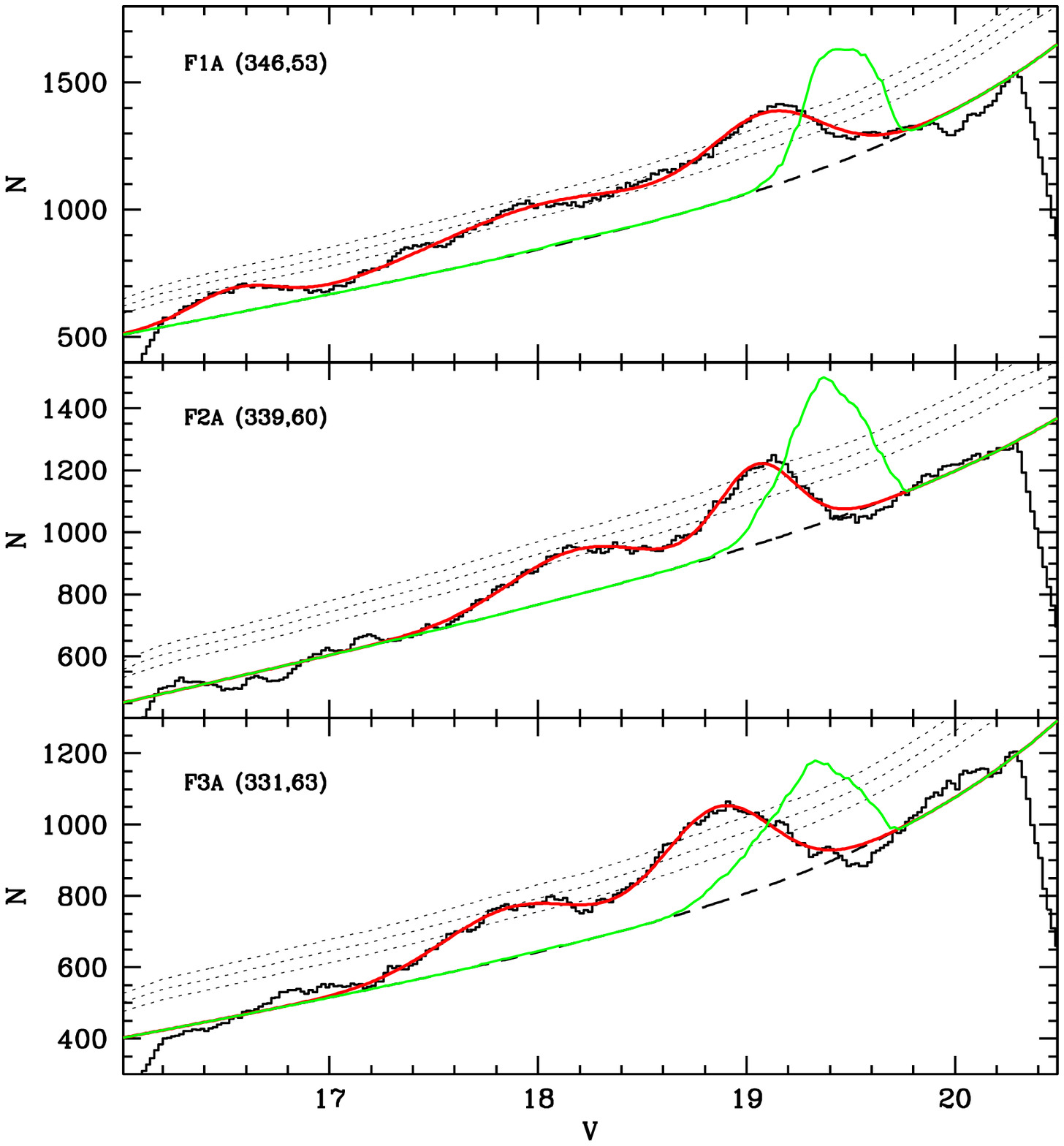}
\plottwo{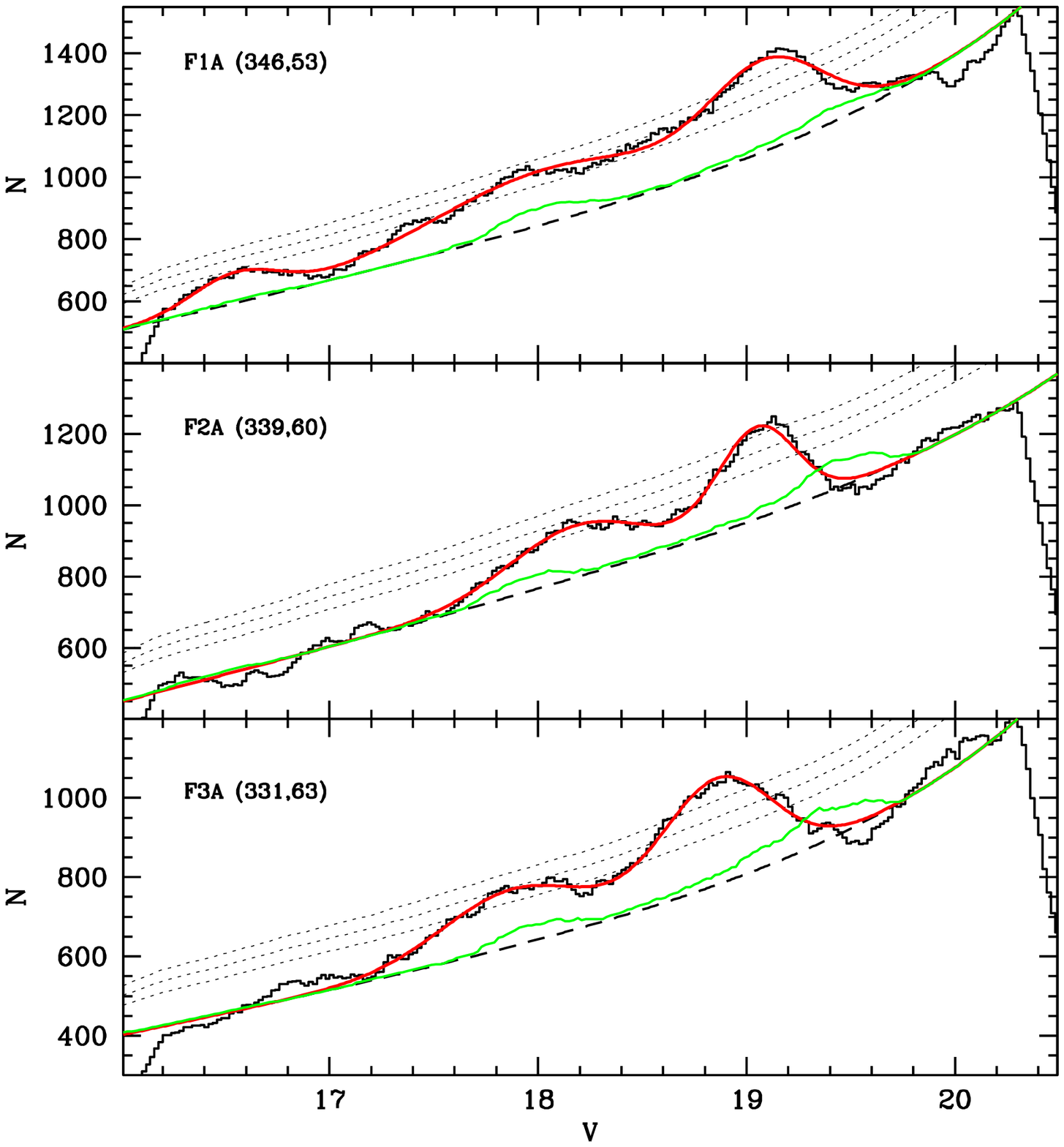}{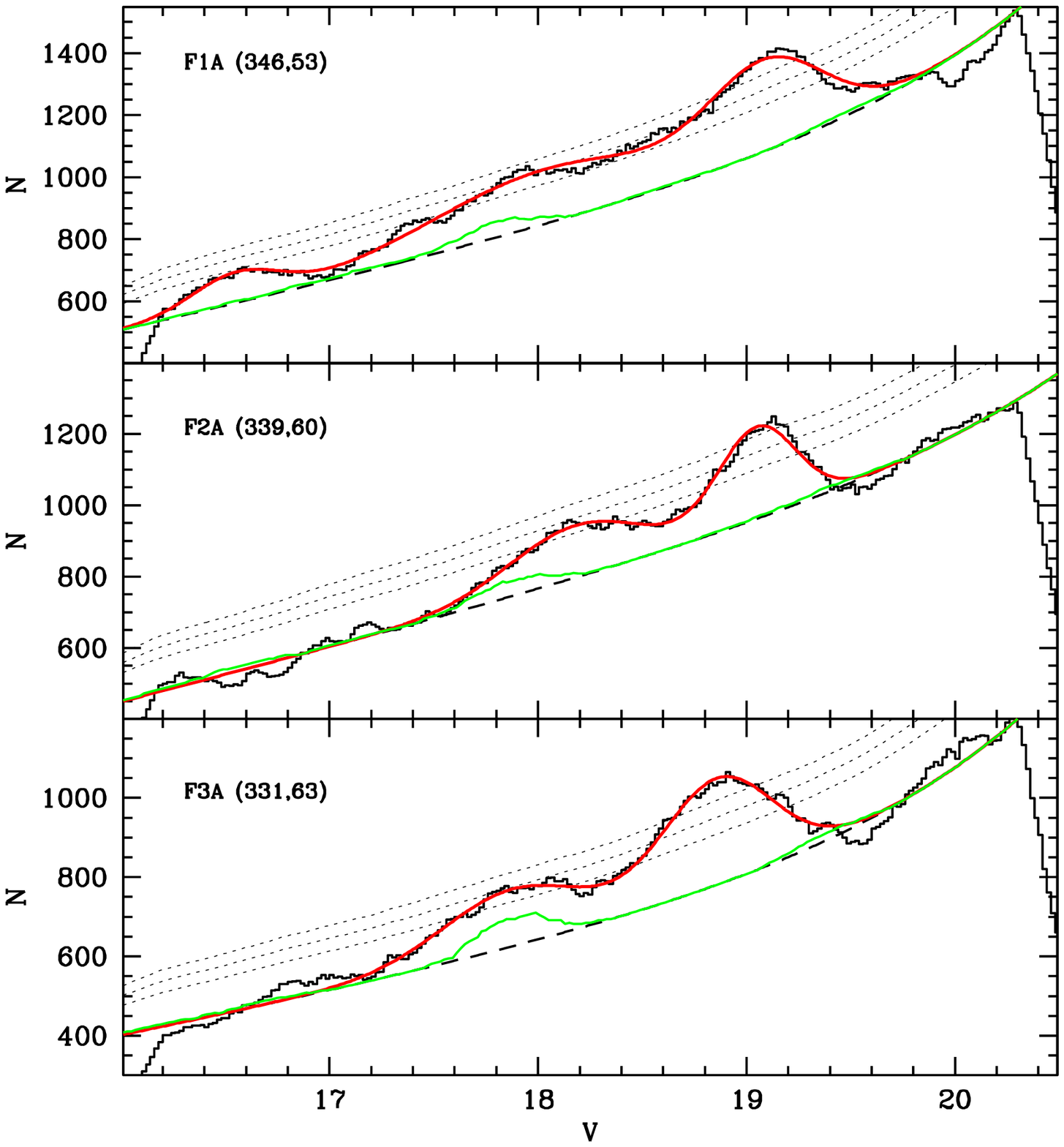}
\caption{Comparison of the observed SCPs in the field F1A, F2A, F3A with the
synthetic SCPs obtained by adding the (arbitrarily normalized) \los histogram from
the prolate N-body model (as in Fig.~\ref{compLF} above) to the best-fit
background model of the considered field. Each of the four (triple) panels
report the synthetic SCPs including only N-Body debris stripped from the main body
one, two, three or four orbits ago, going from the upper left, to the upper
right, to the lower left and to the lower right panels, respectively.}
\label{dircomp}
\end{figure*}

The comparison with the prolate model is the most interesting and we take it
also as the occasion to describe the trends found in our data in a deeper
detail. For a more fruitful discussion we provide Fig.~\ref{xy_pro} in a larger
format with respect to its analogs for the oblate and spherical models (see also
Fig.~\ref{ldist}).  It should be stressed that, in the following, we interpret
the coherent structures we have detected using this specific model as a
guideline. For an example of a different interpretation see Sect.~\ref{tri},
below. There are several features worth noticing in Fig.~\ref{xy_pro}:

\begin{enumerate}

\item For $X_{\sun, Sgr}\ga -7$~kpc the agreement between the positions of our
      primary branch-A peaks and the portion of the leading arm going from
      $(X,Y)\simeq(16,-40)$\footnote{In the following we drop the $\sun, Sgr$
      indices, for brevity. The unity is always kpc.} to $(X,Y)\simeq(-7,-28)$
      is {\em excellent}. Also primary branch-B peaks follow the same trend  
      thus confirming that the two structures lie at {\em the same distance}
      \cite[see also Fig.~\ref{ldist}, and][]{FoS,fell,yan09}. According to the
      considered model, this part of the leading arm is dominated by $p=-1$
      particles up to $X_{\sun, Sgr}\simeq 0$, and by a mix of $p=-2$ and $p=-3$
      particles for $X_{\sun, Sgr}\la 0$.

\item Several detections seem to extend the path of the arm down to
      (X,Y)=(-15,-15), possibly suggesting a slightly less elongated shape of
      the arm with respect to the model predictions. The coherence of the
      structure is less   clear in this region: the models predict that various
      wraps cross here and this may be source of some confusion.

\item The model predicts the presence of a more ancient (mostly populated by
      $p=-2$ and $p=-3$ particles) and wider wrap running nearly parallel to the
      portion of the leading arm described above, but behind it. This structure
      has been detected in branch-B, where one flag=2 point, at
      $(X,Y)\sim(-5,-44)$, in coincidence with a branch-A detection, and a
      flag=3 detection at   $(X,Y)\sim(-14,-35)$]. These points appear to trace
      the outer edge of this wrap , as depicted by the considered model. On the
      other hand, there is no detection (in any branch) for $X\ga 0$, i.e. where
      the detection of the second wrap should be easier, according to the model,
      as the separation from the inner wrap increases with X and the feature is
      denser and narrower in that region. This lack of detection seems confirmed
      by the independent results of \citet{FoS}, that, however, detect the most
      distant wrap at $X<0$ only in the direction of Branch A. To have a deeper
      insight into this problem in Fig.~\ref{dircomp} we provide a direct
      comparison between observations and model at the SCP level, as done in
      Fig.~\ref{compLF}. Here we compare the observed SCP of the F1A, F2A and F3A
      fields with the SCPs obtained from the model in the considered \los for
      particles lost one, two, three and four peri-galactics ago. From the
      upper-right panel it is clear that the dense $X>0$ part of the outer wrap,
      produced by $p=-2$ particles in the model, has no counterpart in the
      observed SCPs and would be easily detected if actually there.  On the other
      hand, the sum of the relics having $p=-1,-3$, and $p=-4$ provide a
      satisfactory match to all the observed peaks. This suggests that there is
      a real mismatch between the L05 P model predictions and our observations
      in this part of the halo.  We note that the spherical model suffers from
      the same problem, while the oblate model does not predict a strong signal
      at that position, but it fails to match all the observations at $X<0$ for
      this wrap.

\item A coherent series of detections lying at nearly constant $Y\sim 25$  kpc,
     traced from $X\sim 8$ to $X\sim -14$ in both branches,  traces a
     filamentary structure that is identified here for the first time. Isolated
     detections with M giants and RR Lyrae were previously reported at $\Lambda
     \simeq 295\degr$ \citep{maj03,vivas,ive00}.  This feature matches quite
     well a wrap of the trailing arm that is present in all the L05 models; it
     can be appreciated from Fig.~\ref{ldist} and Fig.~\ref{xy_pro} that the
     agreement with the P model is very good. For $X< -14$ however the positions
     of the peaks do not trace the model prediction anymore. This apparent
     discontinuity along this branch cannot be (only) due to the distance
     effects discussed in Sect.~\ref{basti} as the distance is expected to
     increase and the sensitivity of the method should increase accordingly.
     Moreover we are able to detect peaks both more and less distant than the
     position predicted by the model along these {\em los}. This feature has no
     counterpart in the triaxial halo model discussed in Sect.~\ref{tri}, below.
     It is clear that additional information is needed to understand better the
     nature of this structure, from, for example, the kinematics of member
     stars.

\item There are primary and tertiary branch-B detections, plus one tertiary
      branch-A detections, tracing a feeble (but coherent) spur from an ancient
      ($p=-3,-4$) wrap, predicted by the model to arch between
      $(X,Y)\sim(-10,-15)$ and $(X,Y)\sim(-3,-22)$. As far as we know this is
      the first detection of this nearby portion of the Stream. A couple of
      primary branch-B  detections (and a tertiary branch-A detection) may trace
      similar substructures on the near side of the constant-distance portion of
      the trailing arm (see Sect.~\ref{tri} for an alternative interpretation).

\item There are a couple of other cases of slight distance mismatches between
      branch-A and branch-B detections, occurring, however in the region around
      $(X,Y)\sim(-10,-25)$ where different wraps of the Stream cross each other.
      It may be challenging to disentangle the various contributions based on
      distances alone. A more interesting case is provided by the two pairs of
      detections around $(X,Y)\sim(-23,-30)$, a region where the model predicts
      only feeble structures and branch-B detections are clearly more nearby
      than branch-A ones. It is intriguing to note that the few particles of the
      model lying in this region are not uniformly distributed but appear to
      form two approximately parallel tiny bridges that reasonably reproduce the
      observed pattern. Also in this case this is the first detection of such
      structures. 
      
\item Both the P and S models by L05 predict the presence of a fairly dense and
      narrow wrap composed by $p=-3$ and $p=-4$ particles crossing the
      accessible range of the X,Y plane from $(X,Y)\sim(-7,-7)$ to
      $(X,Y)\sim(5,-12)$, where it emerges from the $d\la 12$~kpc zone of
      insensitivity of our method (the triaxial model briefly discussed in
      Sect.~\ref{tri} also displays a similar feature). Here we have a primary
      detection from the SCP of F1A, that currently is the first detection of
      this nearby wrap of the leading arm\footnote{While the distinction between
      {\em leading} and {\em trailing} arms is easy and sensible for particles
      lost in the latest two peri-galactic passages, it becomes increasingly
      blurred for Stream wraps dominated by more ancient relics, as a particle
      can reach the same position in these parts of the Stream both from the
      leading and from the trailing sides of the tidal tails.}. A detailed
      exploration of the $d\la 12$~kpc zone would require a different kind of
      analysis, hence it is postponed to a future contribution. We note, however, that
      \citet{mon_grad} studied the chemical composition and the kinematics of a small sample of    M giants that can be attributed to this nearby wrap.

\item Two Branch-B detections, located at $(X,Y)\sim(-3,-16)$ and
      $(X,Y)\sim(+3,-18)$, F5B and F2B, respectively, do not seem to match any
      significant structure of the spherical and prolate L05 models; the primary
      one (that with positive X) is marginally consistent with the part of the
      leading arm plunging toward the Sun of the oblate model (but see
      Sect.~\ref{tri}, below). As anticipated in Sect.~\ref{detA}, their
      position [(RA,Dec)=($195\degr,+16\degr$) $d\sim 18$ kpc, and
      (RA,Dec)=($190\degr,+18.5\degr$) $d\sim 19.5$ kpc, respectively] is fully
      compatible with the outer fringes (i.e., the high Galactic latitude edge)
      of the nearby overdensity S297+63-20.5, discovered by \citet{heidi} and
      discussed in detail in \citet{heidivirgo}. It is unclear why we do not
      detect the structure in other adjacent fields, or in the corresponding
      branch A fields. This may be due to the intrinsic weakness of the RC
      signal from these nearby features, or it may reflect a high degree of
      complexity of the sub-structures, as suggested in the analyses by
      \citet{kellvir} and \citet{vivasvirgo}. \citet{heidivirgo} provided
      positional and kinematic evidence arguing against the association of
      S297+63-20.5 with the Sgr Stream, that was originally proposed by
      \citet{davirgo} and cannot be completely ruled out at the present stage
      \citep[see also the discussion in][]{law10}. However Fig.~\ref{xy_pro}
      provides further support for the conclusions by \citet{heidivirgo}: the
      peaks detected here do not present any continuity with the main branches
      of the leading and trailing arms of the Stream as traced in the present
      analysis (but see also Sect.~\ref{tri}, below). Our data suggest that the
      leading arm crosses the Galactic plane at $\sim 10$~kpc from the Sun,
      toward the Anticenter, in agreement with \citet{heidivirgo} and
      \citet{seab}.  On the other hand, the identification of S297+63-20.5 with
      the Virgo Stellar Stream \citep[VSS,][]{duffvirgo,vivasvirgo} seems likely,
      while the relationship between VSS and the Virgo Over Density
      \cite[VOD,][]{juric,heidivirgo,kellvir} is less certain
      \citep[see][]{heidivirgo,kellerev}. We are currently following up these
      possible detections of S297+63-20.5/VSS in F5B and F6B (also looking for
      the structure at lower latitudes). If confirmed, they would provide the
      first detection of RC stars in these structures, in analogy with the cases
      of Boo~III discussed in \citet{boo3}. RC stars may provide new insights on
      the nature of complex series of structures recently identified in the
      direction of Virgo \citep{kellvir,kellerev}.

\end{enumerate}

All the features and correlations with the P model described above can be seen
even more clearly and directly in Fig.~\ref{ldist}, that provides the most
natural way to compare our measures with models. For example, the match between
two weak model structures described at point 6, above, and our detections can be
very clearly appreciated in that plot, at $230\degr\la \Lambda \la 245\degr$ and
$d\simeq 37$~kpc. The linear trend of increasing distance with decreasing
$\Lambda$ of the two parallel sets of observed points is very nicely matched by
corresponding filaments of particles in the model.

Fig.~\ref{xz} shows that the overall morphology of the three models is
remarkably similar in the $X_{\sun,Sgr}$ vs $Z_{\sun,Sgr}$ plane and reproduces
the general trends of the data (except for the Oblate model, that predicts a
total lack of particles for branch-B detections at $X_{\sun,Sgr}>0$, at odds
with observations). A more detailed analysis is beyond the scope of the present
paper. On the other hand, we must conclude, from the results summarized above,
that the {\em prolate} model by \citet{law} is the one  (among those considered
here) providing the best match to the positional data considered here. It should
be stressed that with this {\em we do not intend to say that a prolate halo
model is favored by our data}, as the comparison was limited to just three very
specific models that are already known not to be able to fit all the positional
and kinematical observational constraints available \citep{law2}.  In particular
it should be recalled that the available radial velocities of Stream stars seems
to favor prolate models \citep{aminashape}, while the angular precession of the
leading arm with respect to the trailing arm favor spherical or slightly oblate
models \citep[L05;][and references therein]{kat05,heidivirgo,prisag}. We simply
note that any future model intended to fit all the observed characteristics of
the Sgr Stream must have a spatial structure {\em very} similar to that of the
{\em prolate} model by \citet{law}, at least in the portion of space sampled by
our study, unless an alternative origin is assumed for the $d\sim 25$~kpc
structure we tentatively interpreted as the trailing arm.

\subsection{Trends of depth as a function of orbital azimuth}
\label{depazi}

\begin{figure}
\plotone{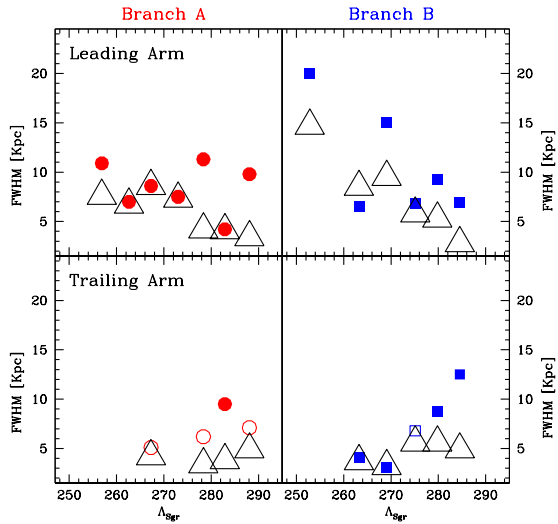}
\caption{Comparison between the trends with $\Lambda_{Sgr}$ of the observed FWHM
along the \los (the same symbols as Figg.~\ref{xy_obl}, \ref{xy_sph} and
\ref{xy_pro}, above, and the predictions of the Prolate model by L05 (open
triangles). The comparison is limited to the \los from F1 to F7 (i.e., those
providing the cleanest tracing of the leading and trailing arms). The upper row
of panels refers to detections in the leading arm, the lower row to detections
in the trailing arm. The left and right columns refer to branch A and
branch B, respectively. The field numbers increase (from F1 to F7) from right to
left, as in Fig.~\ref{ldist} and Figg.~\ref{xy_obl}, \ref{xy_sph},
\ref{xy_pro}.}
\label{sig7}
\end{figure}

In line with the above discussion, in Fig.~\ref{sig7} we compare the FWHM along
the \los described in Sect.~\ref{sec.sigma} with those measured from the
distribution of particles of the Prolate model of L05, along the same \los. The
following discussion is mainly intended to illustrate the possible use of the
derived FWHM. It should be considered that there are additional sources of
uncertainty affecting this comparison, associated with the the measure of FWHM
in models. For example, the  measured width depend on the actual number of
particles of the model, the limited number of particle may lead to
underestimates of the actual width. This expected effect is clearly confirmed in
Fig.~\ref{sig7}, where observed FWHM are always equal or larger than their model
counterparts. The disentanglement of overlapping structure may be also
problematic, as it is unavoidably performed in different ways in the observed
SCPs and in the N-body models.

To minimize the possible ambiguities associated with the collapse of complex
structures along the \los into a single FWHM measure (see Sect.~\ref{modeling}),
especially in regions where different wraps cross one another,   in
Fig.~\ref{sig7} we limit our comparison to $X>-10$ kpc peaks tracing the two
main wraps (leading and trailing arms) that are 30 kpc apart at 
$\Lambda_{Sgr}=290\degr$ and cross each other at $\Lambda_{Sgr}=265\degr$, and
we consider only primary and secondary peaks.


The most interesting and sensible comparison is between the trends of the FWHM
as a function of orbital azimuth. The upper panels of Fig.~\ref{sig7} show that
the observed and predicted trends for the leading arm are indeed similar, both
in direction and in amplitude, for both branches. The agreement of the absolute
values of the FWHM is also satisfying (within a factor of $\sim 2$), with four
Branch A and one Branch B detections closely matching the model predictions. The
FWHM of the considered detections from F1A and F3A give some reason of concern,
as they break the continuity of the observed trend: this may suggest that there
may be some unresolved structure in these peaks. Alternatively we have to accept
variations of a factor of $\sim 2$ as due to the uncertainty inherent to the
adopted method of estimating FWHM. The overall agreement is reasonable also for
the putative trailing arm.

It is interesting to note that the different trends observed in the two branches
of the leading arm are reproduced by the P model that do not present any
bifurcation (see also Fig.~\ref{dens}, for a similar behavior in the $\Lambda$
vs. density trend in the leading arm).

\subsection{Trends of density as a function of orbital azimuth}
\label{densazi}

\begin{figure}
\plotone{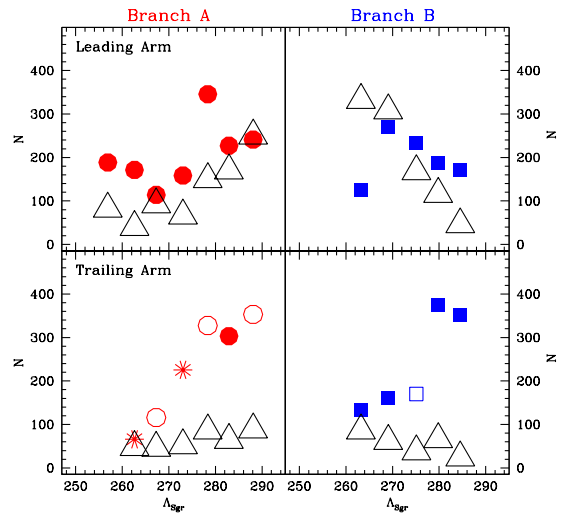}
\caption{The same as Fig.~\ref{sig7} for the density of RC stars as a function
of $\Lambda$. An arbitrary normalization factor of 2 is applied to all the
density values from the theoretical model. }
\label{dens}
\end{figure}

In strict analogy with the analysis described in the previous subsection, in
Fig.~\ref{dens} we present the comparison of the observed and predicted trends
of the stellar density (see Sect.~\ref{sec.dens}) as a function of $\Lambda$.
The measured density is compared with the density of particles in the same wrap
of the P model. The density scale of the model has been multiplied by the
arbitrary factor 2.5, to achieve a reasonable normalization with the observed
values. As in Sect.~\ref{depazi} the comparison presented is just intended as
illustrative of the possible use of these numbers, and it is limited to the
cleanest portions of the leading and trailing arms, at $X>-10$ kpc. 


For the leading arm the match between the overall observations and the models is
acceptable, in particular for branch A. The highly discrepant point at $\Lambda
\sim 263\degr$ is associated with an especially complex SCP model, with three
overlapping peaks (F5B): for this reason we are inclined to ascribe the
discrepancy to an erroneous density estimate.
 
On the other hand, while the model predicts low or even negative gradients of
density with increasing $\Lambda$, the observations show a very strong positive
gradient, similar in both branches\footnote{It has to be recalled that the
bifurcation in two branches is an observed property of the main wrap of the
leading arm. There is no reason to discuss other wraps as divided into two
branches: here this is merely incidental, due to adopted distribution of the
observed fields that were chosen to trace the bifurcation.}. This is an obvious
example of the kind of constraints that can be achieved with these data: in
principle, any fully successful model of the disruption of Sgr must also
reproduce a density gradient similar to the observed one. However it has to be
taken into account that the available models are intended to describe the dark
matter halo in which the baryonic part of the galaxy is embedded. While, for
example, stars and DM particles in the Stream should not greatly differ in their
kinematical and positional properties, their density would follow the same
trends only if mass strictly follows light also in tidal tails, which is very
unlikely to be the case \citep[see][and references therein]{pena}.

\subsection{The triaxial model by Law \& Majewski 2010}
\label{tri}

When the present manuscript was ready for submission, a preprint was posted
\citep{law10}, following up the preliminary analysis by \citet{law2}. In that
study a new N-body model of the disruption of Sgr within a triaxial Galactic
potential, is shown to provide a reasonable match to most of the existing
observational constraints. In particular, the new model reproduces the distance
$vs.$ $\Lambda$ trend reported by Bel06 for the main wrap of the leading arm,
the precession between the leading and trailing arms, and it matches the
existing sets of kinematic measures.   


\begin{figure}
\plotone{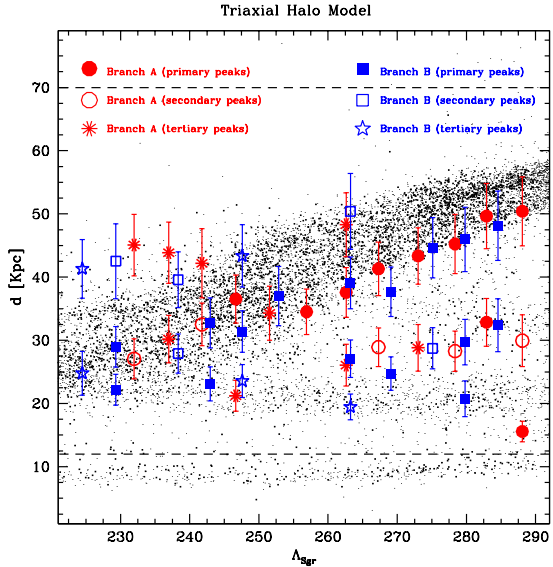}
\caption{Comparison with the new ``triaxial halo'' (T) model by \citet{law10}:
distance vs. orbital azimuth. Symbols are the same as in Figg.~\ref{xy_obl},
\ref{xy_sph} and \ref{xy_pro}. Our measures have been rescaled to match the
distance scale adopted by \cite{law10}, that implies a distance to the main body
of Sgr $d=28$ kpc (i.e. $(m-M)_0=17.23$, instead of $(m-M)_0=16.90$, as adopted
by L05).}
\label{trid}
\end{figure}


\begin{figure}
\plotone{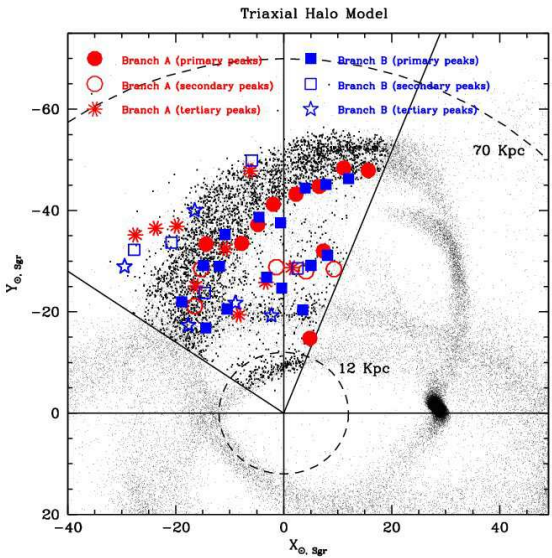}
\caption{Comparison with the new ``triaxial halo'' (T) model by \citet{law10}:
the $X_{\sun, Sgr}$ vs. $Y_{\sun, Sgr}$ plane, with the same assumptions and
symbols as Fig.~\ref{tri}.}
\label{trixy}
\end{figure}

As we have stressed before, it is far beyond the scope of the present analysis
to find out which is the best available model. However it is worth showing the
comparison between our results and this new model, for completeness and (above
all) as a very instructive example of how the interpretation  of observed
features may depend on the considered model (see Sect.~\ref{conc} for
discussion). From the inspection of Fig.~\ref{trid} and Fig.~\ref{trixy} it can be
concluded that the triaxial halo model provides a reasonable match to distance
gradient of the main wrap of the leading arm, over the whole range of $\Lambda$
covered by our data. This is not unexpected as the model is found to fit the
observations by Bel06 in this regard, and we are in good agreement with these
authors. The same is true for the sparse detections behind the main wrap, that
were also found by Bel06. It is interesting to note that, in a similar fashion
to the P and S models, the T model predicts a remarkable increase of the density
of this wrap for $\Lambda \ga 275\degr$ ($X\ga 0$) that is not observed, neither
in the present work or by Bel06. Moreover, the T model does not seem to display
the narrow and dense structure of the main wrap of  the leading arm that in the
P model appears to match so well our coherent set of primary detections in that
region. We postpone a detailed comparison between the observed structure along
the \los and the predictions of the P and T model to a future contribution: here
we limit the discussion to the main features of the models (i.e. trends of
distance with orbital azimuth). 

The new model makes predictions very similar to
those of the P and O models also regarding the nearest wrap of the Stream,
running across the whole range of $\Lambda$ sampled by our data. However it should be noted that it predicts this wrap to lie below our sensitivity limit at any $\Lambda$, in the observed range, thus it is unable to match the observed points at $\Lambda\simeq 287\degr$ and $d\simeq 13$ kpc, at odds with the P model.
The T model presents a very coherent narrow wrap of the trailing arm running at
nearly constant $d\simeq 20$~kpc from $\Lambda\simeq 290\degr$ to $\Lambda\simeq
250\degr$, then it begins to bend gently toward $d\simeq 25$~kpc from
$\Lambda\simeq 235\degr$ where it crosses with the leading arm. This feature
matches {\em very nicely} the nearby ($d\le 25$ kpc) detections that  we
tentatively attributed to S297+63-20.5/VSS and to an ancient spur of the leading
arm, in the comparison with the P model described in detail in
Sect.~\ref{models}, above. 

On the other hand, the coherent structure we detect from  $\Lambda\simeq
260\degr$ to $\Lambda\simeq 250\degr$, that we interpreted as a wrap of the
trailing arm, is not present in the T model. The same is true for the $d\ge 40$
kpc structures at $\Lambda \le 245\degr$.  The orbital path of the simulated Sgr
galaxy matches also these structures, so it is not excluded that they may
correspond to very ancient wraps. However,  it has to be noted that the T model
is the remnant of the evolution of a Sgr progenitor for $\simeq 8$ orbits (not
just $\sim 4$ as for S,O, and P) models, thus it should include wraps populated
from more ancient stripping events than the S, O, and P models.

In conclusion, while the P model still appears to provide a more thorough match
of the observed structures, the T model provides a promising alternative that
deserves to be  investigated in further detail. Not surprisingly, the mere
comparison with our own (limited) data-set shows that both models need to be
refined.

\section{Summary and conclusions}
\label{conc}

We have used RC stars to trace the long tidal tail of the Sgr dSph galaxy in the
portion of the Northern sky sampled by the SDSS-DR6. Structures along the
line-of-sight are identified as peaks in the (otherwise smoothly increasing) $I$-
and $V$-band SCPs of color-selected samples of candidate RC stars, from $\sim
5\degr \times 5\degr$ fields covering the whole extension of the two main
branches (A and B) of the Sgr Stream identified by \citet{FoS} in the same
dataset. Any other part of the Stream in addition to these branches is expected
to lie (approximately) in the same plane, i.e. it should be visible in the
considered fields. The analysis was focused on obtaining the most accurate and
reliable distances to all the wraps of the Stream that we were able to detect. 

Many significant peaks were consistently found in both the SCPs of several
fields. The observed SCPs were modeled as a series of Gauss curves (one for each
peak) superposed to a polynomial accounting for the smooth fore/background
population. For each significant peak we derived a purely differential estimate
of its distance (with uncertainties $\le$ 10\%), an estimate of the FWHM along
the \los, and an estimate of the associated density of RC stars attributable to
the considered structure. All the derived quantities are provided in
Tab.~\ref{tab_dmag} as powerful constraints for the new generations of models of
the disruption of the progenitor of Sgr dSph within the Milky Way halo.

To illustrate the potential of our measures in that context we compared them
with the three models made publicly available by L05. These provide a realistic
realization of the present epoch configuration of particles that were originally
bound to a progenitor similar to Sgr that was evolved for $\simeq 4$ orbital
periods within a static Galactic potential with different degrees of flattening
(a spherical, oblate and prolate halo, respectively). The models (and in
particular the Prolate halo one, that matches well most of our observations) are
also used as guidelines for the interpretation of our results. The great
complexity of a structure like the Sgr Stream, multiply-wrapped around the
Galaxy, requires a process of convergence between models and observations: the
latter must constrain models but the former are indispensable to re-conduce such
a complexity to a single structure (see Sect.~\ref{tri}).

Our technique resulted in higher-accuracy distance estimates with respect to
previous studies, and demonstrated high sensitivity to feeble structures.
However, the sensitivity is easily destroyed by contamination from Galactic
sources: for these reasons we had to limit our survey to $b>50\degr$, while
other (more abundant) tracers are able to follow the Stream down to $b>30\degr$.
The overall agreement with previous analyses is good (see Sect.~\ref{comp}).
Finally, and most importantly, our method proved especially efficient in the
detection of (relatively) nearby structures.  In the following we summarize and
briefly discuss the main conclusions of the present study, taking
Fig.~\ref{ldist} and Fig.~\ref{xy_pro}, as references.

\begin{itemize}

\item For $\Lambda\ga 255\degr$ ($X\ga -10$ kpc) the leading arm of the Stream
      is cleanly and coherently detected in both branches, going from $d=43$ kpc
      at $\Lambda\simeq 290\degr$ to $d=30$ kpc at $\Lambda\simeq 255\degr$.
      This is in full agreement with the results obtained with other tracers
      \citep{heidivirgo,nied}. This portion of the leading arm is the most
      unambiguous and robustly constrained.

\item In the same range of $\Lambda$ (and $X$) a remarkably coherent structure
      is also very clearly detected at nearly constant distance from us,
      $d\simeq 25$ kpc. According to the S and P models  by L05 this can be
      interpreted as a wrap of the trailing arm, while it has no obvious
      counterpart in the recently presented T model \citep{law10}. The P model
      matches the observed structure very well. Previous detections of this wrap
      were reported only around $\Lambda=295\degr$ \cite[see][for discussion and
      references]{maj03}. 

\item The comparison with the L05 models strongly suggest that the run of the
      relative distance as a function of $\Lambda$ of the two wraps described above
      has a strong power in discriminating between different models of the Stream. In
      particular the S and O models by L05 clearly fail to reproduce the observed
      trends. On the other hand the P model reproduces the trend nearly perfectly.

\item Weak detections of a further, more distant wrap (running parallel to the
      leading arm, in the same range of  $\Lambda$ as above) were also obtained. These
      support similar results by \citet{FoS}. An enhancement of the density of this
      wrap at $\Lambda\ga 275\degr$, predicted by the S, P and T models, seems to be
      excluded by the present analysis (in agreement with Bel06). 

\item Turning to the $\Lambda\la 255\degr$ ($X\la -10$ kpc) portion of the
      survey, this is characterized by a very complex structure, partly due to
      the crossing of multiple wraps predicted to occur in this region by all
      the models. Hence the interpretation of these structures is less
      straightforward, and must be considered as tentative. However, the P model
      appears to provide a reasonable match to all the detections in this
      region: for these reasons we  adopt it as a guideline for our best-effort
      interpretation of the data (see Sect.~\ref{tri} for an alternative view).

\item The leading arm seems to be traced beyond $\Lambda\simeq 255\degr$,
      continuing its trend of linear decrease of its distance down to  $d\simeq
      20$ kpc at $\Lambda\simeq 220\degr$. Extrapolating from the observed trend
      one would expect the arm to cross the Galactic disk at $\sim 10$~kpc from
      the Sun, in agreement with the conclusions by
      \citet{heidivirgo,seab,law10}. The degree of coherence of the detections
      in this portion of the leading arm is lower, suggesting the possible
      presence of further (unresolved) substructure or due to higher
      uncertainties associated to weaker and overlapped structures.

\item In the same region, the continuation of the trailing arm is coherently
      traced where predicted by the P model up to $\Lambda\ga 240\degr$ ($X\ga
      -15$ kpc).  For $\Lambda\la 235\degr$, in particular, we lack any
      detection corresponding to the well defined structure predicted by the
      model (the same is true for the T model). On the other hand, coherent
      detections are obtained {\em behind} the main wrap of the trailing arm as
      predicted by the P model for $\Lambda\la 240\degr$. These detections may
      indicate a different shape for that portion of the trailing arm. However,
      as discussed above, they match  two more feeble structures running
      parallel to the main arm. It is obvious that the P model is not adequate
      to fit all our observations, in spite of the good overall match.

\item The most nearby detections are the more difficult to interpret robustly.
      However the single primary detection at $d\simeq 13$~kpc and $\Lambda
      \simeq 287\degr$ (just beyond the $d\le 12$~kpc ``zone of avoidance'' of
      our technique) matches the prediction of all the three L05 models, as well
      as for the model by \citet{law10}. For this reason we are quite confident
      to have detected for the first time the nearest wrap of the leading arm.
      We are currently following up this finding, to check if the predicted
      $d\sim 10$ kpc wrap can be detected also in other \los.

\item The three detections at $d\sim 20$~kpc and  $\Lambda \simeq 245\degr$ are
      matched by a spur of the P model. The two detections at $d\simeq 18$ kpc and 
      $\Lambda \ga 260\degr$ have been tentatively ascribed to the S297+63-20.5/VSS
      overdensity. The T model matches very well {\em all} of these detections with a
      single narrow wrap of the Sgr trailing arm. However \citet{law10} confirm that
      the kinematics predicted by their model toward VSS is markedly different from
      what observed by \citet{duffvirgo} and \citet{heidivirgo}.

\item The overall trends of FWHM along the \los as a function of $\Lambda$ of
      the P model provide a reasonable match to our primary detections of the
      leading arm. It is especially interesting to note that the model
      reproduces the different trends encountered in the two branches, even if
      it does not produce the observed bifurcation. This seems to provide
      further support to the view \citep[adopted by][]{segue,nied} that branch~A
      and branch~B are substructures within the same wrap of the Stream, and not
      different wraps as proposed by \citet{fell}. However, a limited set of
      tests performed on intra-branches fields suggests that the bifurcation in
      the Dec direction shown by \citet{FoS} is real. Probably a deeper,
      thorough and independent analysis of the Dec structure of this wrap of the
      Stream is warranted (see \citet{law10} for possible alternative
      explanations).
 
\item The observed trends of density as a function of RA along the leading arm
      (branches A and B) are in fair agreement with those by \citet{nied}. Our
      estimates of the total luminosity per kpc at any given RA are lower than
      theirs by a factor of $\sim 4-5$.

\item Kinematic follow up of the newly identified structures is clearly urgent.
      \citet{carrell} recently demonstrated that this can be carried on using
      exactly the same tracer stars, i.e. RC stars.
  
\end{itemize} 

\acknowledgments
This research has been financially supported by INAF through the PRIN 2007 grant CRA 1.06.10.04 ``The local route to galaxy formation...'' and by ....

This research make use of SDSS data. 
Funding for the SDSS and SDSS-II has been provided by the Alfred P. Sloan Foundation, the 
Participating Institutions, the National Science Foundation, the U.S. Department of Energy, the National Aeronautics and Space Administration, the Japanese Monbukagakusho, the Max Planck Society, and the Higher Education Funding Council for England. The SDSS Web Site is http:\/\/www.sdss.org\/.
The SDSS is managed by the Astrophysical Research Consortium for the Participating Institutions. The Participating Institutions are the American Museum of Natural History, Astrophysical Institute Potsdam, University of Basel, University of Cambridge, Case Western Reserve University, University of Chicago, Drexel University, Fermilab, the Institute for Advanced Study, the Japan Participation Group, Johns Hopkins University, the Joint Institute for Nuclear Astrophysics, the Kavli Institute for Particle Astrophysics and Cosmology, the Korean Scientist Group, the Chinese Academy of Sciences (LAMOST), Los Alamos National Laboratory, the Max-Planck-Institute for Astronomy (MPIA), the Max-Planck-Institute for Astrophysics (MPA), New Mexico State University, Ohio State University, University of Pittsburgh, University of Portsmouth, Princeton University, the United States Naval Observatory, and the University of Washington.

\bibliographystyle{apj} 

\end{document}